\documentclass[
  epj,
  nopacs,
]{svjour}
\usepackage{graphicx}
\usepackage[square,comma,sort,compress,numbers]{natbib}
\usepackage[percent]{overpic}
\usepackage{pict2e}
\usepackage{siunitx}
\usepackage[table,usedvipsnames]{xcolor}
\usepackage[T1]{fontenc}
\usepackage{textcomp}
\usepackage{footnotehyper}
\usepackage{flushend}
\usepackage{enumitem}
\usepackage{dialogue}

\usepackage{xmpincl}
\includexmp{metadata}

\newcommand{\orcid}[1]{%
  ~\href{https://orcid.org/#1}{\includegraphics[width=8pt]{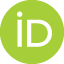}}%
  }

\newcommand{\linkedorcid}[1]{\href{https://orcid.org/#1}{#1}}

\usepackage[hang,flushmargin]{footmisc}

\makeatletter
\renewenvironment{dialogue} {%
    \begin{list}{} {%
      \setlength\itemsep{-0.4ex plus 0.1 ex}
      \setlength\parsep{0.6ex plus 0.4ex} 
      \setlength\rightmargin{0pt}
      \defcommand\speak [1] {\item[{##1}]}%
      
    }%
    \PreDialogue\relax
  }{%
  \end{list}%
}
\makeatother

\makeatletter \def\NAT@def@citea{\def\@citea{\NAT@separator\,}} \makeatother

\makeatletter
\def\makeheadbox{{%
}}
\makeatother
\DeclareSIUnit{\au}{{a.u.}}

\newcommand{\narrator}[1]{\noindent\emph{#1}}

\usepackage[
  bookmarks=true,
  colorlinks,
  linkcolor=blue,
  urlcolor=blue,
  citecolor=blue,
  plainpages=false,
  pdfpagelabels,
  final,
  breaklinks=true
]{hyperref}
\hypersetup{
  pdftitle={Dialogue on analytical and ab initio methods in attoscience}, 
  pdfauthor={Gregory S. J. Armstrong, Margarita A. Khokhlova, Marie Labeye, Andrew S. Maxwell, Emilio Pisanty, and Marco Ruberti}
}

\begin{document}
\title{Dialogue on analytical and {\em ab initio} methods in attoscience}

\author{
Gregory S. J. Armstrong\orcid{0000-0001-5949-2626}\inst{1}
\and Margarita A. Khokhlova\orcid{0000-0002-5687-487X}\inst{2,3}
\and Marie Labeye\inst{4}
\and Andrew S. Maxwell\orcid{0000-0002-6503-4661}\inst{5,6}
\and Emilio Pisanty\orcid{0000-0003-0598-8524}\inst{2,5}
\and Marco Ruberti\orcid{0000-0003-0424-3643}\inst{3}
}                     
\institute{
Centre for Theoretical Atomic, Molecular, and Optical Physics, Queen's University Belfast, Belfast BT7 1NN, United Kingdom
\email{gregory.armstrong@qub.ac.uk}
\and
Max Born Institute for Nonlinear Optics and Short Pulse Spectroscopy, Max-Born-Stra{\ss}e 2A, Berlin 12489, Germany
\and
Department of Physics, Imperial College London, South Kensington Campus, London SW7 2AZ,
United Kingdom
\and
PASTEUR, D\'{e}partement de chimie, {E}cole Normale Sup\'{e}rieure, PSL University, Sorbonne Universit\'{e}, CNRS, 75005 Paris, France
\and
Institut de Ciencies Fotoniques, The Barcelona Institute of Science and Technology, 08860 Castelldefels (Barcelona), Spain
\and
Department of Physics and Astronomy, University College London, Gower Street, London WC1E 6BT, United Kingdom
}

\date{\today}
%

\abstract{
The perceived dichotomy between analytical and \emph{ab initio} approaches to theory in attosecond science is often seen as a source of tension and misconceptions.
This Topical Review compiles the discussions held during a round-table panel at the `Quantum Battles in Attoscience' \textsc{cecam} virtual workshop, to explore the sources of tension and attempt to dispel them.
We survey the main theoretical tools of attoscience---covering both analytical and numerical methods---and we examine common misconceptions, including the relationship between \emph{ab initio} approaches and the broader numerical methods, as well as the role of numerical methods in `analytical' techniques.
We also evaluate the relative advantages and disadvantages of analytical as well as numerical and \emph{ab initio} methods, together with their role in scientific discovery,
told through the case studies of two representative attosecond processes:
non-sequential double ionisation and resonant high-harmonic generation.
We present the discussion in the form of a dialogue between two hypothetical theoreticians, a numericist and an analytician, who introduce and challenge the broader opinions expressed in the attoscience community.
\\[1em]
\small
Accepted Manuscript for
\href{%
  https://doi.org/10.1140/epjd/s10053-021-00207-3%
  }{%
  \color[rgb]{0,0,0.55}%
  \textit{Eur.\ Phys.\ J.\ D} \textbf{75}, 209 (2021)%
  }, 
available as %
\href{%
  https://arxiv.org/abs/2101.09335%
  }{%
  \color[rgb]{0,0,0.55}%
  arXiv:2101.09335 %
  }%
under %
\href{%
  %
  https://creativecommons.org/licenses/by/4.0/%
  }{%
  \color[rgb]{0,0,0.55}%
  CC BY%
  }.
} 


%
\maketitle

\makeatletter
\global\setbox\authrun=\hbox{\small\ignorespaces 
  G.\,S.\,J. Armstrong et al.: Dialogue on analytical and {\em ab initio} methods in attoscience
  }
\makeatother
        
\section*{Introduction}
\label{intro}

Modern developments in laser technologies have kick-star\-ted the attosecond revolution, which formed the field of attoscience, dealing with dynamics on the attosecond ($\SI{e-18}{s}$) timescale~\cite{brabec2000, krausz2009, calegari2016}. 
Attosecond science was born with the study of above-threshold ionisation (ATI) and high-order harmonic generation (HHG) driven by strong laser pulses.
As it has matured over the past three decades, attoscience has given us access to phenomena which were previously thought to be inaccessible---%
including the motion of valence electrons in atoms~\cite{Goulielmakis2010Aug}, 
charge oscillations in mo\-le\-cules~\cite{Calegari2014}, 
as well as the direct observation of the electric-field oscillations of a laser pulse~\cite{Goulielmakis2004direct}---%
and it has also spurred advances in ultrafast pulse generation which have opened a completely new window into the dynamics of matter.

The meteoric progress of attoscience has been fuelled, on the one hand, by formidable experimental efforts, and, on the other hand, it has been supported by a matching leap in our theoretical capabilities.
These theoretical advances have come in a wide variety, forming two opposing families of analytical and numerical approaches.
While these two families generally work together, the dichotomy between analytical and numerical
methods is sometimes perceived as a source of tension within the attoscience community.

In this paper we present an exploration of this dichotomy, which collects the arguments presented in the panel discussion `Quantum Battle 3 –- Numerical vs Analytical Methods' held during the online conference `Quantum Battles in Attoscience' \cite{battle}.
Our main purpose is to resolve the tension caused by this dichotomy, by 
identifying the critical tension points,
developing the different viewpoints involved,
 and finding a common ground between them.

This process forms a natural dialogue between the analytical and numerical perspectives.
We delegate this dialogue to two hypothetical `combatants'---
\begin{dialogue}
\speak{Analycia}
Hi, I'm Analycia Formuloff, and I am an attoscience theorist working with analytical approaches.
\speak{Numerio}
Hello, my name is Numerio Codeman, and I'm a computational scientist working on \emph{ab initio} methods. 
\end{dialogue}
---who will voice the different views expressed during the panel discussion.

We follow the dialogue between \textsc{Analycia} and \textsc{Numerio} through three main questions.
First, in Section~\ref{sec:1}, we explore the scope and nature of analytical and numerical methods, including the interchangeability of the terms `numerical' and `\emph{ab initio}'.
We then analyse, in Section~\ref{sec:Advantages_Disadvantage}, the relative advantages and disadvantages of the two approaches, using non-sequential double ionisation (NSDI) as a case study.
Finally, in Section~\ref{sec:discovery}, we examine their roles in scientific discovery, via the case study of resonant HHG.
In addition, in Section~\ref{sec:discussion}, we present some extra discussion points, 
as well as our combatants' responses to the questions raised by audience members,
and a summary of the responses to several polls taken during the live session.

\section{\texorpdfstring{`\emph{Ab initio}'}{'Ab initio'} and analytical methods}
\label{sec:1}

A constructive discussion is always based on a good knowledge of the subject. 
To this end, in this section we tackle the 
 subtleties in the definitions of `\emph{ab initio}', `numerical' and `analytical' methods:
we detail their differences, and we present a rough classification of the various theoretical methods used in attosecond science. 
We first concentrate on \textsc{Numerio}'s speciality, \emph{ab initio} methods, then we move to \textsc{Analycia}'s forte, analytical theories. 
For each combatant, we first introduce their theoretical approach, and then list the main methods in the corresponding toolset. 
After these presentations, \textsc{Numerio} and \textsc{Analycia} discuss the friction points they have with each other's methods.

\subsection{\emph{Ab initio} and numerical methods}
\label{sec:ab-initio}
In its dictionary sense, \emph{ab initio} is Latin for `from the beginning'.
Thus, a theoretical method can be defined to be \emph{ab initio} when it tackles the description of a certain physical process starting from first principles, i.e., using the most fundamental laws of nature that---according to our best understanding---govern the physics of the phenomena that we aim to describe.

Within an \emph{ab initio} framework, the inputs of the theoretical calculation should be limited to only well-known physical constants, with any interactions kept as fundamental as possible.
This means that no additional simplifications or assumptions may be made on top of what we believe are the established laws of nature. 
In other words, the specific aspects of the physical process of interest need to be approached without using specially-tailored models.

\vspace{1ex}

\narrator{%
We now bring our combatants to the stage, to discuss the consequences of this definition.
}

\begin{dialogue}

\speak{Analycia}
This is an extremely stringent definition, which will substantially limit the number of methods that can be classified in the \emph{ab initio} category.
But, more importantly, it just delays the real question: what does `fundamental' mean in this context?

\speak{Numerio}
The answer to this question is, in essence, a choice of `reference frame', within theory-space, which will frame our work.
This choice is tightly connected to the physical regime that we want to describe.
We know that attoscience, as part of atomic, molecular and optical physics, is ultimately grounded on the Standard Model of elementary interactions in particle physics, which gives---in principle---the `true' fundamental laws.
However, much of this framework is largely irrelevant at the energies that concern us.
Instead, we are only interested in the quantum mechanics of electrons and atomic nuclei interacting with each other and with light, and this gives us the freedom to restrict ourselves to quantum electrodynamics (QED), or with its `friendlier' face as light-matter interaction~\cite{Cohen-Tannoudji1997}.

\speak{Analycia}
What does this mean, precisely?
If QED is the right framework, that means we must retain a fully relativistic approach as well as a full quantisation of the electromagnetic field.

\speak{Numerio}
For most problems in attoscience, this would be overkill, as relativistic effects are rarely relevant.
Instead, it is generally acceptable to work in the context of non-relativistic quantum mechanics, and to introduce relativistic terms into the Hamiltonian at the required level of approximation.
These are the basic laws responsible for `a large part of physics and the whole of chemistry', as recognised by Dirac as early as 1929~\cite{dirac1929}. 

\speak{Analycia}
I can see how this is appropriate, so long as spin-orbit and inner-core effects are correctly accounted for.
However, what about field quantisation?

\speak{Numerio}
We normally deal with strong-field settings where laser pulses are in coherent states comprising many trillions of photons, which means that a classical description for electromagnetic radiation is suitable.

\speak{Analycia}
That typically works well, yes, but it is also important to keep in mind that it can blind you to deep questions that lie outside of that framework~\cite{Lewenstein2020quantum}.
In any case, though: as `fundamental', would you be satisfied with a single-electron solution of the Schrödinger equation?

\speak{Numerio}
No, this would not be appropriate---cutting down to a single electron is generally going too far.
While this can be very convenient for numerical reasons, restricting the dynamics to a single particle invariably requires adjusting the interactions to account for the effect of the other electrons in the system, via the introduction of a model potential.
This approximation can be validated using a number of techniques which can make it very solid, but it always entails a semi-empirical step and, as such, it rules out the `\emph{ab initio}' label in its strict sense.

\speak{Analycia}
That leaves a many-body Hamiltonian of for\-mi\-da\-ble complexity.

\speak{Numerio}
It does! Let me show you how it can be handled.
\end{dialogue}

\vspace{-5mm}

\noindent
\paragraph{The `\emph{ab initio}' toolset}
The complexity of simulating the time-dependent Schrödinger equation (TDSE) with the multi-electron Hamiltonian of atomic and molecular systems can be tamed using a wide variety of approaches.
Most of these are inherited from the field of quantum chemistry, and they differentiate from each other via the level of approximation they employ, and by the ranges of applicability where they are accurate.

However, every approach in this space must face a challenging trade-off between accuracy in capturing the relevant many-body effects, on the one hand, and the computational cost that it requires, on the other.
The key difficulty here is the handling of electron-correlation effects, which are difficult to manage at full rigour.
Because of this, many methods adopt an `intermediate' approach, which allows for lower computational expense, while at the same time limiting the accuracy of the physical description.

\begin{figure}
    \centering
    \includegraphics[width=\columnwidth]{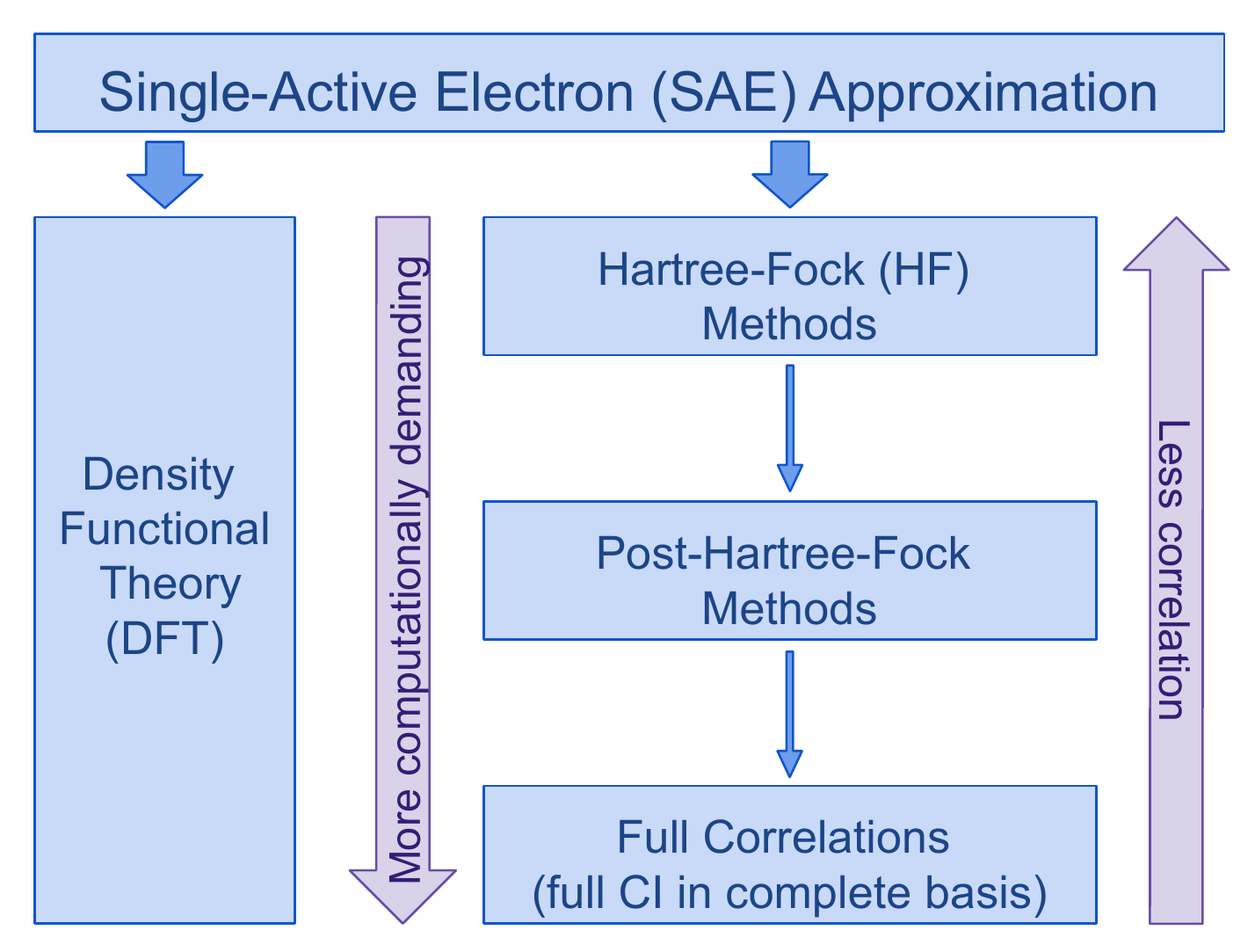}
    \caption{Schematic representation of the hierarchy of methods to describe electron correlation.}
    \label{fig:corr}
\end{figure}

Numerical methods thus form a hierarchy, schematised in Fig.~\ref{fig:corr}, with rising accuracy as more electron correlation effects are included:

\begin{itemize}

\item
\textbf{Single-Active-Electron} (SAE) approaches are the simplest numerical approaches to the TDSE~\cite{scrinzi2014}, though they are only \emph{ab initio} for atomic hydrogen, and require model potentials to mimic larger systems.
Nevertheless, they can be used effectively to tackle problems where electron correlation effects do not play a role, 
and their relative simplicity has allowed multiple user-ready software packages that offer this functionality in strong-field settings~\cite{Bauer2006, Tulsky2020, Patchkovskii2016, Fritzsche2019}.

\item
\textbf{Density Functional Theory} (DFT) allows an effective single-particle description~\cite{Ullrich2012}, widely considered as \emph{ab initio}, which still includes electron correlation effects through the use of a suitable `exchange-corre\-la\-tion functional'~\cite{Kohn1965}.
Within attoscience, examples of approaches which specifically target attosecond molecular ionisation dynamics include 
real-time time-dependent (TD)-DFT \cite{DeGiovannini2013, Wopperer2017, DeGiovannini2018}
and time-dependent first-order perturbation theory static-exchange DFT \cite{Calegari2014,Lara-Astasio2018}.
More broadly, TD-DFT approaches are robust enough that they appear in several user-ready software packages~\cite{Tancogne2020, Noda2019, Apra2020, Garcia2020Siesta} suitable for attoscience.%
\footnote{For a detailed examination of the status (and shortcomings) of DFT and TD-DFT, see the talks of N.~Maitra, A.~Schild and K.~Lopata at Ref.~\cite{BIRS-CMO-Workshop}}

\item
\textbf{Non-equilibrium Green's function theory} also allows one to describe the many-body problem from first principles by using effectively-single-particle approaches~\cite{Perfetto2018, Perfetto2018JPCL}.

\item
\textbf{Quantum-chemistry approaches} go beyond the SAE approximation and DFT to include, directly, the effects of electron correlation~\cite{szabo1996}.
The starting point for this is generally the Hartree-Fock (HF) mean-field approach, though this is rarely sufficient on its own. 
Because of this, quantum-chemistry methods climb the ladder all the way to the full Configuration Interaction (CI) limit, a complete description of electron correlation (which is generally so computationally-intensive that it is out of reach in practice).

Most of the standard approaches of quantum chemistry were developed to describe bound states of molecular systems~\cite{szabo1996,Schirmer2018}, 
and they have also proven to be highly successful for modelling band structures in solid-state systems~\cite{Evarestov2012}. 
Nevertheless, they often require significant extensions to work well in attoscience, particularly regarding how the ionisation continuum is handled.
Recent examples of these extensions include 
\emph{ab initio} methods based on the algebraic diagrammatic construction (ADC)~\cite{Ruberti2014JCP, Simpson2016, Averbukh2018, Ruberti2018PCCP, ruberti2018} 
and its restricted-correlation-space extension (RCS-ADC)~\cite{Ruberti2019, Ruberti2019PCCP, Ruberti2021, schwickert2020}, 
multi-reference configuration interaction (MRCI)~\cite{majety2015hacc, Majety2015PRL},
and multi-configuration time-dependent Hartree~\cite{Beck2000} and Hartree-Fock \cite{sukiasyan2009} methods,
as well as restricted-active-space self-consistent-field (RAS-SCF)~\cite{Marante2017, Marante2017PRA, Klinker2018} approaches. 

\item
\textbf{Basis-set development} is another crucial element of the numerical implementation work for \emph{ab initio} methods in attoscience, since the physics accessible to the method, as well as its computational cost, are often determined by the basis set in use.
Recent work on basis sets includes the development of
B-spline functions, both on their own~\cite{Ruberti2014JCP, Simpson2016, Averbukh2018, Ruberti2018PCCP, ruberti2018, Ruberti2019, Ruberti2019PCCP, Ruberti2021, schwickert2020, Toffoli2016}, and in hybrid combinations with Gaussian-type orbitals (GTOs)~\cite{Marante2017, Marante2017PRA, Klinker2018}, 
as well as
finite-difference approaches~\cite{clarke2018, brown2020, Benda2020}, finite-element discrete-variable-representation functions~\cite{Miyagi2013, Miyagi2014}, 
grid-based methods~\cite{sato2013, Greenman2010}, 
and simple plane-waves~\cite{NguyenDang2014}. 

\end{itemize}

\noindent
A more extensive (though still non-exhaustive) list of methods is shown in Table~\ref{tab:numerical-methods}. Here we focus on methods for the description of ultrafast electron dynamics, happening on the attosecond time-scale. For numerical methods tackling the (slower) nuclear motion in attoscience, we refer the reader to Ref.~\cite{Vrakking2018}.  

\begin{table*}[t]
\centering
\definecolor{headercellbackground}{rgb}{0.92,0.95,1}
{\renewcommand{\arraystretch}{1.2}%
\begin{tabular}{lll}
\multicolumn{3}{l}{
  \cellcolor{headercellbackground}%
  \rule{0pt}{1.2em}%
  Fully \textit{ab initio} methods%
  }
\\
Time-Dependent B-spline Algebraic Diagrammatic Construction  
 & \cite{Ruberti2014JCP, Simpson2016, Averbukh2018, Ruberti2018PCCP, ruberti2018} 
  & TD B-spline ADC
\\
Time-Dependent B-spline Restricted-Correlation-Space ADC
 & \cite{Ruberti2019, Ruberti2019PCCP, Ruberti2021, schwickert2020}
  & TD B-spline RCS-ADC
\\
Multi-Reference Configuration Interaction: \textsc{TRECX}  
 & \cite{trecx,majety2015hacc,Majety2015PRL}
  & MRCI: \textsc{TRECX}
\\
Restricted-Active-Space Self-Consistent-Field: XCHEM 
 & \cite{Marante2017,Marante2017PRA,Klinker2018} 
  & RAS-SCF: XCHEM
\\
Multi-Configuration Time-Dependent Hartree-Fock 
 & \cite{Carette2013,Sawada2016}
  & MCTDHF
\\
Time-Dependent Complete-Active-Space Self-Consistent-Field 
 & \cite{sato2013}
  &  TD CAS-SCF
\\
Time-Dependent Restricted-Active-Space Self-Consistent-Field 
 & \cite{Miyagi2013,Miyagi2014} 
  & TD RAS-SCF
\\
Time-Dependent Configuration Interaction Singles 
 & \cite{Greenman2010,Toffoli2016,NguyenDang2014}
  & TD CIS 
\\
$R$-matrix with time dependence  
 & \cite{clarke2018,brown2020,Benda2020}
  & RMT
\\
Real-time Non-Equilibrium Green's Function
 & \cite{Perfetto2018,Perfetto2018JPCL}
  & Real-time NEGF
\\
\multicolumn{3}{l}{
  \cellcolor{headercellbackground}%
  \rule{0pt}{1.2em}%
  DFT/hybrid/non-\textit{ab-initio} methods%
  }
\\
Real-time Time-Dependent Density Functional Theory 
 & \cite{ DeGiovannini2013,Wopperer2017,DeGiovannini2018}
  & Real-time TDDFT 
\\
Time-Dependent First-Order Perturbation Theory static-exchange DFT  
 & \cite{Calegari2014,Lara-Astasio2018}
  & 
\\
Single-Active Electron Time-Dependent Schrödinger Equation 
 & \cite{Bauer2006, Tulsky2020, Patchkovskii2016, Fritzsche2019}
  & SAE TDSE
\\
\end{tabular}
}
\caption{
  Rough survey of numerical methods for attoscience and strong-field physics.
  }
\label{tab:numerical-methods}
\end{table*}

\subsection{Analytical methods}

The use of analytical methods to describe strong-field phenomena has a long-storied pedigree dating back to the 1960s~\cite{keldysh1965}, before laser sources could reach sufficient intensities to drive quantum systems beyond the lowest order of perturbation theory.
As a general rule, analytical methods are approaches for which the governing equations can be solved directly, under suitable approximations, and the solutions can be written down in `exact' or `closed' form.

\vspace{1ex}
\narrator{We return to our combatants on the stage, where \textsc{Numerio} is dissatisfied with this definition.}

\begin{dialogue}
\speak{Numerio}
That just seems like it is kicking the can down the road. What does `closed form' mean?

\speak{Analycia}
As it turns out, when the term `closed form' is placed under examination, its precise meaning turns out to be rather elusive and ultimately quite ambiguous~\cite{berry2007, Borwein2013closed}.
That is to say: 
  which `forms' does the term `closed form' actually include? 
  Which ones does it exclude?
  Does it stop at elementary functions, i.e., exponentials and logarithms?
  Or must it cover special functions, like the Bessel functions?
  And if we do intend to include special functions as part of the toolbox of analytical methods, which special functions should be included?
  Do hypergeometric functions or Meijer $G$ functions make the cut?
  What about newly-minted functions expressly defined to encapsulate some hard numerical problem?
  (As one example of this, take the recent proof of the integrability of the Rabi model of quantum optics~\cite{Braak2011}---should the functions defined in that work be considered special functions?)
  More importantly: what does it really mean for a function to be a `special' function?

\speak{Numerio}
But hasn't this question been answered long ago?

\speak{Analycia}
Well, this is the kind of question where one could hope that we could look to the mathematicians to provide an answer---say, by supplying an objective classification of functions, from elementary through exponentials to the `higher' transcendentals---as to where this class should stop.
Unfortunately, however, when such objective classifications are attempted, they run into a bog of vague answers and incomplete taxonomies which leave out large classes of useful functions.
Ultimately, as Paul Tur\'an put it~\cite{andrews1999special}, `special' functions are simply \textit{useful} functions: they are a shared language that we use to encapsulate and communicate concepts and patterns~\cite{berry2007}, and their boundaries (and with it, the boundaries of analytical methods in general) are subjective and a product of tradition and consensus.

\speak{Numerio}
This distinction seems like trivial semantics to me, to be honest.

\speak{Analycia}
At first glance, yes, but it is important to keep in mind that, as a rule, special functions like the Bessel functions are defined as the solutions of hard problems---canonical solutions of ordinary differential equations, integrals that cannot be expressed in elementary terms, non-summable power series---and when it comes to evaluating them in practice, they generally require at least some degree of numerical calculation.
In this regard, then, what is to stop us from packaging up one of the numerical problems that face us, be it a full TDSE simulation or one of its modular components, calling it a special function, and declaring that methods that use it are `analytical'?

\speak{Numerio}
That sounds rather absurd.

\speak{Analycia}
Indeed it does, at face value, but it is not all that far from how special functions are actually defined, i.e., as the solution of a tough differential equation, or to dodge the fact that a given integral cannot be evaluated in elementary terms by re-christening it as an integral representation.
More importantly, it encodes a serious question---what happens when the `back end' of analytical methods involves more numerical calculations than the TDSE simulations they were intended to replace?

\speak{Numerio}
They should give the job to me, of course!
\end{dialogue}

\noindent
\paragraph{The analytical toolset}
These issues aside, the analytical methods of strong-field physics, as traditionally understood in the field, form a fairly well-defined set.
This set can be further subdivided into three main classes:
\begin{itemize}
\item 
\textbf{Fully quantum models}, 
which retain the full coherence of the quantum-mechanical framework.
These frameworks date back to key conceptual leaps in the early days of laser-matter interaction~\cite{keldysh1965, Faisal1973, Reiss1980,perelomov1966ionization, ammosov1986tunnel}, 
but they also include applications of more standard per\-tur\-ba\-tion-theory tools.

The central method in this category is known as the Strong-Field Approximation (SFA) \cite{keldysh1965, Faisal1973, Reiss1980} (see Ref.~\cite{Amini2019} for a recent review), which builds on the solvability of the field-driven free-particle problem, used to great effect for HHG~\cite{Lewenstein1994Mar}.
The SFA is more properly a family of related methods~\cite{galstyan2016} with the key commonality of taking the driving laser field as the dominant factor after the electron has been released; in its fully-quantum version, it produces observables in the form of highly-oscillatory time integrals.

\item 
\textbf{Semiclassical models},
which bridge the gap between the full quantum description and the classical realm by incorporating recognizable trajectory language but still keeping the quantum coherence of the different pathways involved.
The paradigmatic example is the quantum-orbit version of the SFA~\cite{Popruzhenko2014}, obtained by applying saddle-point approximations to the SFA's time integrals, which results in trajectory-centred models analogous to Feynman path integrals~\cite{salieres2001feynman} where the particles' positions are generally evaluated over com\-plex-\allowbreak{}valued times~\cite{Popruzhenko2014, ivanov2014} and are often complex-valued themselves~\cite{torlina2012, Pisanty2016, Pisanty2017, torlina2017}.

As a general structure, the relationship between semiclassical methods and the full TDSE is the same as between ray optics and wave optics for light, a correspondence that can be made rigorous as an eikonal limit~\cite{Smirnova2008}.
This has the caveat that optical tunnelling requires evanescent waves in the classically-forbidden region, with an optical counterpart in the use of complex rays for evanescent light~\cite{einziger1982}.
The presence of these complex values complicates the analysis, but it also presents its own opportunities for insight~\cite{Pisanty2020}.

The recent development of analytical methods has centred on correcting the SFA to account in various ways for the interaction of the photoelectron with its parent ion, 
from straightforward rescattering~\cite{milosevic2007intensity} 
through fuller Coulomb corrections~\cite{Popruzhenko2008,torlina2012}
and explicit path-integral formulations~\cite{Maxwell2017},
which now span a wide family of approaches~\cite{figueira2020}.
On the other hand, it is important to keep in mind that there are also multiple techniques, such as semiclassical propagators~\cite{Zagoya2014}, which are independent of the SFA.

\item 
\textbf{Fully classical models},
which can retain a small core of quantum features (most often, the tunnelling probability obtained from tunnelling theories~\cite{ammosov1986tunnel}) but which generally treat all the particle dynamics using classical trajectories.
This includes the paradigmatic Simple Man's Model~\cite{Corkum1993, Kulander1993} for HHG, 
but it also covers much more elaborate methods, often of a statistical kind, that look at classical trajectory ensembles to understand the dynamics~\cite{Panfili2001}, 
and in particular the specific formulation as Classical Trajectory Monte Carlo~\cite{dimitriou2004origin} `shotgun' approach to predicting photoelectron momentum and energy spectra.

\end{itemize}
These methods are summarised in Table~\ref{tab:analytical-methods}.

\begin{table*}[t]
\centering
\definecolor{headercellbackground}{rgb}{0.92,0.95,1}
{\renewcommand{\arraystretch}{1.2}%
\begin{tabular}{lll}
\multicolumn{3}{l}{
  \cellcolor{headercellbackground}%
  \rule{0pt}{1.2em}%
  Fully quantum models%
  }
\\
Lowest-order perturbation theory & 
  semianalytic & 
  Matrix elements must be taken from quantum chemistry
\\
Strong field approximation~\cite{Lewenstein1994Mar} & 
  semianalytic & 
  Intensive numerical integration
\\
\multicolumn{3}{l}{
  \cellcolor{headercellbackground}%
  \rule{0pt}{1.2em}%
  Semiclassical models%
  }
\\
Quantum Orbit models~\cite{Popruzhenko2014,figueira2020phases} & 
  semianalytic & 
  Saddle-point equations must be solved numerically
\\
`Improved' SFA~\cite{milosevic2007intensity} & 
  semianalytic & 
  Saddle-point equations must be solved numerically
\\
Coulomb-Corrected SFA~\cite{Popruzhenko2008} & 
  semianalytic & 
  Significant numerical integration
\\
Path-integral approaches~\cite{Maxwell2017} & 
  not analytic & 
  Numerical solutions for equations of motion
\\
Semiclassical propagators~\cite{Zagoya2014} & 
  not analytic & 
  Numerical solutions for equations of motion
\\
\multicolumn{3}{l}{
  \cellcolor{headercellbackground}%
  \rule{0pt}{1.2em}%
  Fully classical models%
  }
\\
Simple Man's Model~\cite{Corkum1993, Kulander1993} & 
  semianalytic &
  Trajectories are only analytical for a few driving fields
\\
Classical Ensemble Models~\cite{Panfili2001} &
  not analytic &
  Equations of motion must be solved numerically
\\
Classical Trajectory Monte Carlo~\cite{dimitriou2004origin} &
  not analytic &
  Equations of motion must be solved numerically
\\
\end{tabular}
}
\caption{
  Rough survey of analytical methods of strong-field physics and attoscience.
  }
\label{tab:analytical-methods}
\end{table*}

\subsection{Hybrid methods}
In addition to the purely-numerical and purely-analytical approaches discussed above, it is also possible to use hybrid approaches, which involve nontrivial analytical manipulations coupled with numerical approaches that incur significant computational expense.

This class of methods can include relatively simple variations on standard themes, such as multi-channel SFA approaches that include transition amplitudes and dipole transition matrix elements derived from quantum chemistry~\cite{Mairesse2010},
but it also includes long-standing pillars like Molecular Dynamics and other rate-equation approaches---which use \emph{ab initio} potential-energy surfaces and cross-sections, but discard the quantum coherence---that are now being applied within attoscience~\cite{Shi2019}.
Beyond these simpler cases, there is also a wide variety of novel and creative methods, 
such as the classical-ensemble back-\allowbreak{}pro\-pa\-ga\-tion of TDSE results employed recently to analyse tunnelling times~\cite{Ni2016},
which hold significant promise for the future of attoscience.

\subsection{Friction points}
\narrator{%
Now with a complete set of basic definitions, our combatants {\sc Analycia} and {\sc Numerio} turn their discussion to more specific aspects of analytical and \emph{ab initio} methods.
}

\subsubsection{Numerical $\neq$ \emph{ab initio}}
\begin{dialogue}

\speak{Analycia}
It seems to me that the classification of numerical and \emph{ab initio} methods, as presented in the `\emph{ab initio} toolset', is out of step with the definitions as originally stated. 
Are you using the terms `\emph{ab initio}' and `numerical' interchangeably?

\speak{Numerio}
You are right.
It is important to emphasize the difference between numerical methods and \emph{ab initio} methods. 
Both classes share and benefit from the development and application of `computational thinking', but strictly speaking the latter category is a subset of the former.
On the other hand, in the literature, the two terms are often used interchangeably.

\speak{Analycia}
That may be, but then that is a problem with the literature.
There are many methods on the toolset that are very far from \emph{ab initio} as you defined it.

The clearest examples of this are the methods based on the SAE approximation~\cite{Abusamha2010, Kukk2013, Saito2004, Gozem2015, Labeye2018, vandenWildenberg2019}.
This approach neglects, in an extremely crude way, the two-body nature of the Coulomb electrostatic repulsion between the different electrons, which is often called `electron correlation'. 
Should these methods really be called `\emph{ab initio}'?

\speak{Numerio}
Most of these methods try---with varying degrees of success---to correct for the neglect of electron correlations by introducing various parameterisations of effective one-particle Hamiltonians.
However, these constructions are for the most part semi-empirical, and as such they introduce significant physics beyond the fundamental laws, and definitely cannot be called \emph{ab initio} methods.

\speak{Analycia}
It is good to see that laid out clearly. 
In a similar vein, what about DFT and TD-DFT? 
I notice that many of the openly-available DFT packages explicitly market themselves as being `\emph{ab initio}' approaches~\cite{Tancogne2020, Noda2019, Apra2020}.

\speak{Numerio}
DFT is a rigorously \emph{ab initio} method, and it takes its validity from strict theorems (originally for static systems~\cite{Hohenberg1964, Kohn1965} and subsequently extended to time-dependent ones~\cite{runge1984,vanleeuwen1998, vanleeuwen1999}) that show that the complexity of the full multi-electron wavefunction can be reduced to single-electron quantities.
In brief, there exists an `exchange-correlation functional' that allows us to get multi-electron rigour while calculating only single-electron densities.

\speak{Analycia}
That may be the case in the ideal world of mathematicians, but it does not work in the real world.
The formal DFT and TDDFT frameworks only work if one knows what the exchange-correlation functional actually is, as well as the functionals for any observables such as photoelectron spectra.
In practice, however, we can only guess at what those might be.
I have a deep respect for DFT and TDDFT: for large classes of systems it is our only viable tool, and there is a large body of science which validates the functionals it employs.
Nevertheless, the methods for validating the `F' in DFT are semi-empirical, and do not have the full-sense rigour of \emph{ab initio}.

\speak{Numerio}
Yes, those are fair points.
However, it is worth noting that there also exists a rigorous method to construct approximate parameterised functionals. This is based on introducing parameters whose value can be fixed by requiring them to satisfy the known exact properties of the functional. These parameters are of universal nature in the sense that once they have been determined, they are kept fixed for all systems to be calculated.
Having said this, in practice, when the DFT Hamiltonian ends up in the form of a semi-empirical parameterisation~\cite{Tong1997, Stener2005, Toffoli2013}, then this takes it out of the \emph{ab initio} class.%
\footnote{%
It is worth noting that `\emph{ab initio}' can take different meanings in different fields.
There are multiple contexts, such as the study of condensed matter and other extended systems, where wavefunction methods are not feasible, and DFT methods are the closest approach to the \emph{ab initio} ideal as we have defined it here; most descriptions of DFT as \emph{ab initio} appear in such contexts.
}

\speak{Analycia}
So, are there \emph{any} numerical methods which truly satisfy the \emph{ab initio} definition?

\speak{Numerio}
Yes, there are.
Most of the approaches based on quantum chemistry possess potentially full \emph{ab initio} rigour.
In practice, of course, for some applications this full potential is not needed, and the degree of electron correlation in the calculation can be restricted in order to reduce the calculation time.
However, even in those cases, there is still an \emph{ab initio} method underlying the computation.

That said, even within an \emph{ab initio} method, it is common to introduce semi-empirical parametrisations of the Hamiltonian.
This happens most often when we cannot describe (or do not need to) every term in the Hamiltonian to an \emph{ab initio} standard. 

\speak{Analycia}
What kind of interactions would this approach apply to? 

\speak{Numerio}
The introduction of pseudo-potentials can be used, for example, to model the effect of core electrons in an atom or molecule.
Another common case is the effect of spin-orbit interactions in a semi-relativistic regime. 
This can be seen as a non-\emph{ab initio} description of certain degrees of freedom or interactions whose effect is not dominant within a given physical process or regime. 
Sometimes this has a limited scope, but it can also extend out to what we have described as `hybrid' methods (such as Molecular Dynamics simulations), which are not fully \emph{ab initio} but which nevertheless maintain a very strong \emph{ab initio} identity.

\speak{Analycia}
This does not really paint a picture of a `single class' of \emph{ab initio} methods: instead, you have depicted a continuum of methods, which goes smoothly from a full accounting of electron correlation down to restricted numerical simulations which operate under substantial approximations.

\speak{Numerio}
I agree, and if you press me I should be able to organise these methods on a spectrum, between approaches which are fully \emph{ab initio} and techniques which are simply numerical approaches.
\end{dialogue}

\subsubsection{Analytical methods generally involve computation}
\label{sec:analytical-involves-numerical}
\narrator{Having conceded that \emph{ab initio} methods span a rather large continuum, \textsc{Numerio} strikes back at just how `analytical' the analytical approaches really are.}

\begin{dialogue}

\speak{Numerio}
Since you are so keen to hold \emph{ab initio} methods to the `golden standard' of the definition, it is only fair that we do the same for analytical methods.
Many of the methods you have listed look rather heavy on the numerics to me, particularly on the fully-quantum side.
To pick on something, perturbation theory is certainly purely analytical on its own, but those models often require accurate matrix elements for the transitions they describe, and those can only be obtained from quantum chemistry, often at great expense.

\speak{Analycia}
Yes, that is true---

\direct{\textsc{Numerio} interrupts \textsc{Analycia}}

\speak{Numerio}
And is that not also the case even for the `stars' of the show?
The SFA, in its time-integrated version, produces integrals which are highly oscillatory, and this generally implies a significant computational cost.

\speak{Analycia}
I agree, the SFA and related methods often involve a large fraction of numerical effort.
Even for the quantum-orbit version, the key stages in the calculation---the actual solution of the saddle-point equations---rely completely on numerical methods.
On the other hand, of course, this is typically at a much lower computational cost than most TDSE simulations.

\speak{Numerio}
For most methods, that is quite clear. 
However, this lower computational cost is much less clear for some of the more recent approaches that implement Coulomb corrections on the SFA.
The analytical complexity of those methods can get very high---does that not come together with a higher computational cost?

\speak{Analycia}
To be honest, the computational expense in some of the more complex Coulomb-corrected approaches (in particular those that utilise ensembles of quantum trajectories) to the SFA can, in fact, exceed that of some of the simpler single-electron TDSE simulations.

\speak{Numerio}
I also notice that you have classified several classical trajectory methods as `analytical', including statistical ensemble and Monte Carlo approaches that often involve substantial computation expense calculating aggregates of millions of trajectories.
More importantly, this goes beyond the raw numbers---the Newtonian equations of motion are only solvable in closed form for field-driven \textit{free} particles.
As soon as any sort of atomic or molecular potential is included, one must turn to numerical integration.

\speak{Analycia}
Yes, that is also correct.
The character of the method is different to a direct simulation of the TDSE, but the numerical component cannot be denied.

\speak{Numerio}
So, similarly to the continuum of `\emph{ab-initio}-ness' we agreed upon earlier, what you are saying is that for analytical methods there is also a continuous spectrum between fully analytical and exclusively numerical.

\speak{Analycia}
Yes, I suppose I am.
We should then be able to place the theoretical methods of attoscience on a two-dimensional spectrum depending on how much they have an analytical and \emph{ab initio} character.
\end{dialogue}

\begin{figure}
  \centering
  \includegraphics[width=0.99\columnwidth]{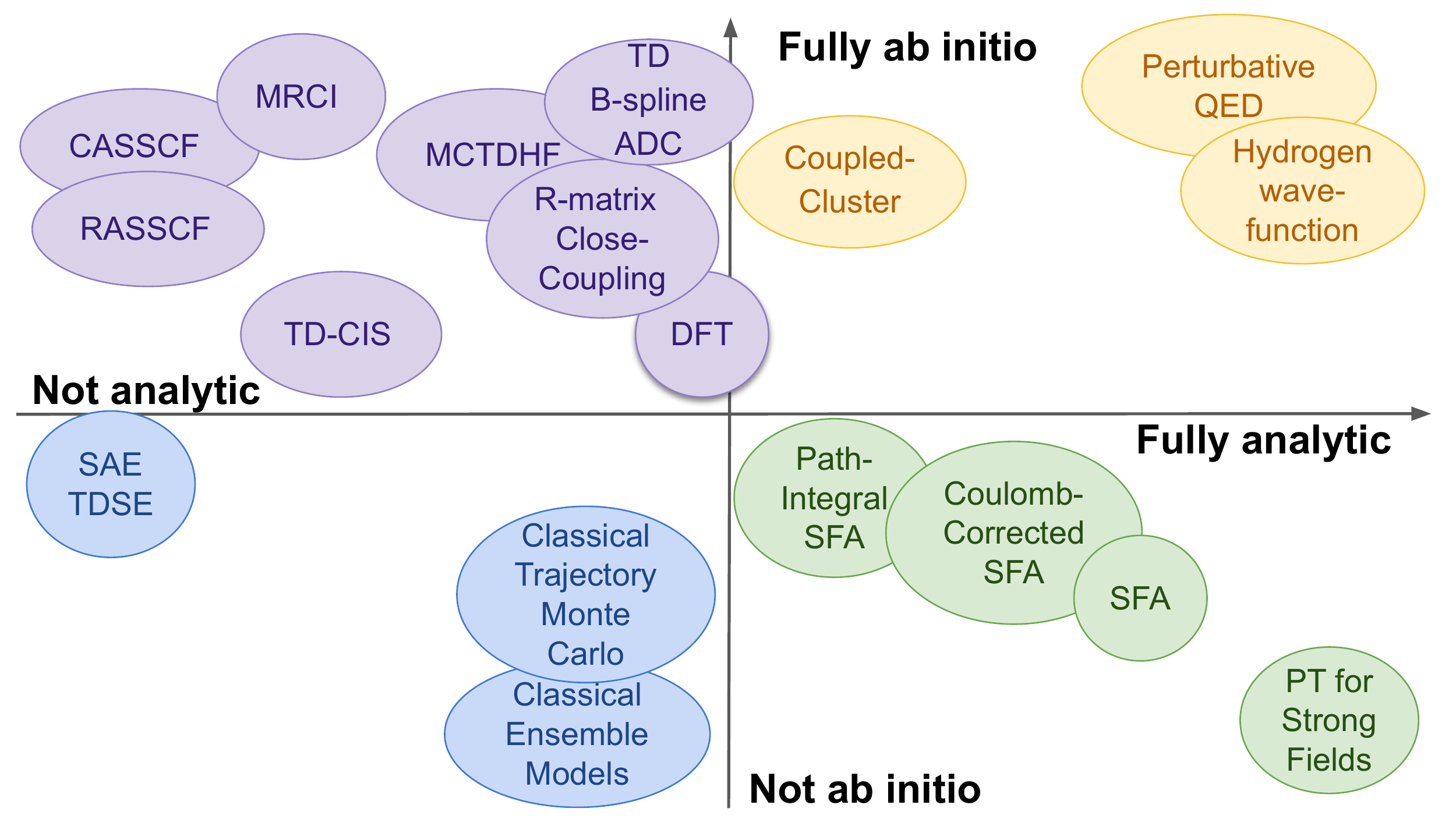}
  \caption{Rough spectrum of the theoretical methods of attoscience, ranked by their analytical (horizontal) and \emph{ab initio} (vertical) character.}
  \label{fig:spectrum}
\end{figure}

\narrator{Our two theorists sit down to chart the methods they have discussed so far, and report their findings in Fig.~\ref{fig:spectrum}.}

\subsubsection{Quantitative vs qualitative insights}
\label{sec:quant-qual}
\begin{dialogue}

\speak{Numerio}
It seems we have completely eliminated the dichotomy that we started with between analytical and \textit{ab initio} methods.

\speak{Analycia}
So it seems, at least on the surface, but there is still a clear difference between the two approaches.
In this regard, I would like to make a somewhat contentious claim: it is more important to distinguish methods according to whether the insights we can obtain from them are of a more quantitative nature or of a more qualitative one.
It seems to me that it is the spectrum between \emph{those} two extremes that carries more value.

\speak{Numerio}
Speaking of `qualitative methods' is certainly unusual in the physical sciences, and to me it feels like it carries some negative connotations.

\speak{Analycia}
Perhaps this is because that phrase has been mistakenly associated too tightly with biological and social sciences, and physical scientists sometimes want to distance themselves from that perception?
If that is so, then it is important to work to de-stigmatise that classification.

In any case, though, I am curious to know whether the attoscience community agrees that this is a more important distinction.

\direct{The audience response to this poll is presented in Table~\ref{tab:polls}, in Section~\ref{sec:discussion} below.}
\end{dialogue}

\subsubsection{Analytical $\neq$ approximate}
\begin{dialogue}

\speak{Numerio}
You mentioned above that some of the `analytical' methods of attoscience can involve substantial computational expense.
What is the point of performing such computations, for approaches that can only ever be approximate.

\speak{Analycia}
This is a common misconception. 
`Analytical' does not necessarily mean `approximate', and there \textit{are} problems where analytical approaches can be fully exact.
Some of this list is limited to the canonical examples (the particle in a box, the harmonic oscillator, the hydrogen atom, the free particle driven by an electromagnetic field),
but it is important to emphasise that it also covers perturbation theory, which is exact in the regimes where it holds.
And, in that sense, it includes the exact solutions for single- and few-photon ionisation and excitation processes, which are crucial to large sections of attoscience, particularly when it comes to matter interacting with XUV light.

\speak{Numerio}
With the caveat we discussed above, surely? 
The perturbation-theory calculations are exact in their own right, but their domain of applicability without numerical calculations is extremely limited.
\end{dialogue}

\subsubsection{\emph{Ab initio} $\neq$ exact}
\begin{dialogue}

\speak{Analycia}
One striking aspect which is implicit in your description of \emph{ab initio} methods, and in how they are handled in the broader literature, is the implication that any \emph{ab initio} method is automatically exact.

\speak{Numerio}
No, that is inaccurate. 
The two descriptors are distinct and they should not be considered as synonyms.

\speak{Analycia}
Perhaps the term is used to somehow overvalue the results of a numerical simulation?
It is easy to fall into the trap of thinking that a result obtained in an \emph{ab initio} fashion is automatically quantitatively accurate, but that is a misconception. 
The clearest examples of this difference are simulations that work in reduced dimensionality, e.g.\ 1D or 2D, but, more generally, plenty of \emph{ab initio} approaches make full use of approximations when they are necessary.

\speak{Numerio}
That is true, and if a method's approximations cannot be lifted then it does not really fit the definition of \emph{ab initio}.
However, it is common to use `lighter', more flexible numerical methods---which use approximations to reduce the cost---for more intensive investigations, while still benchmarking them against an orthodox, fully \emph{ab initio} approach, and then we can be confident in the accuracy of the more flexible methods.

\speak{Analycia}
But that is no different to how we benchmark analytical methods.
What is it about \emph{ab initio} approaches that singles them out as the `gold standard', then?

\speak{Numerio}
I would say that the key feature is the existence of a systematic way to improve the accuracy which does not rely on any empirical fittings or parametrisations, as the central part of the numerical convergence of the method.
When this is present, 
we can expect to get a description at the same level of accuracy for the same physical observables, even if we change the system in question, 
and we can also estimate the error we make in a systematic way.
Under these conditions, then, the \emph{ab initio} methods can achieve fidelities that are so high that they can be considered to be fully exact.%
\footnote{Having said this, it is important to notice a difference between wavefunction-based and functional-based \emph{ab intio} methods. Although non-semiempirical functionals can, in principle, be systematically improved, the nature of this improvement will generally be highly system-specific and, in contrast to wavefunction methods, a better description is not always guaranteed for a particular system of interest.}

\end{dialogue}

\subsubsection{The choice of basis set}
\label{sec:ChoiceOfBasisSet}
\begin{dialogue}

\speak{Analycia}
You mentioned that a key part of `the \emph{ab initio} toolset' is the development of suitable basis sets.
This sounds odd to me: any two basis sets should be equivalent, so long as they are both complete---which, on the \emph{ab initio} side, corresponds to numerical convergence.

\speak{Numerio}
That is formally true, but it is not very useful in practice.
The basis set used to implement an \emph{ab initio} method---to formulate and solve the Schrödinger equation---is a crucial factor in the numerical aspects, 
and it determines 
the level of accuracy of the calculations as well as 
the computational cost required to reach convergence
to a stable solution that captures the full physics of your problem.

More broadly, this is a source of approximation (and thus an entry point for errors), as well as a powerful ally in the search for new physics.
In short, the choice of basis set largely determines the subspace of the solutions that we can reasonably explore, and this in turn influences the physics that can be investigated with the method.

\speak{Analycia}
You mentioned that many of the \emph{ab initio} approaches in attoscience have their roots in quantum chemistry, 
and I understand that quantum chemists have worked very hard at optimising basis sets for their work.
Why can't those sets be used in attoscience?

\speak{Numerio}
The basis sets most commonly used in traditional quantum chemistry, particularly Gaussian-type orbitals (GTOs), have indeed been highly optimised by tailored fitting procedures over many years and, as a result, they have enabled the flourishing of \emph{ab initio} quantum chemical methods~\cite{szabo1996}.
There, the driving goal is to have accurate and fast numerical convergence for the physical quantities that interest quantum chemists, such as ground-state energies and electric polarizabilities.

\speak{Analycia}
Ah---and these goals do not align well with attoscience?

\speak{Numerio}
Exactly.
These basis sets are generally poorly suited to describe free electrons in the continuum. 
As such, traditional basis sets struggle when describing molecular ionisation over a wide range of photoelectron kinetic energies \cite{Ruberti2013, Ruberti2014}.
By extension, this limits our ability to describe general attosecond and strong-field physics. 

\speak{Analycia}
So this is where the attoscience-specific development of basis sets comes in, then.

\speak{Numerio}
Yes. For attoscience the key requirement is an accurate description of wavefunctions with oscillatory behaviour far away from the parent molecular region, and this drives the development when existing basis sets are insufficient.

\speak{Analycia}
So what determines the choice of basis set in any given situation?

\speak{Numerio}
This depends on a number of factors---%
some down to numerical convenience in the specific implementation,
but also, often, determined by the physics that the method seeks to describe.
Within any particular \emph{ab initio} framework, the use of new basis sets allows us to explore different parts of the Hilbert space of the system under investigation, and to look for new and interesting solutions there.

\speak{Analycia}
This sounds reasonable enough, but it also speaks against the strict definition of `\emph{ab initio}' as you formulated it, which requires us not to input any physics beyond the fundamental interactions.
To the extent that the basis-set choice determines the subspace where solutions will play out, that represents an additional input about the physics which is built directly into the code.
This can then limit the reach of the method; one clear example of this is the elimination of double-ionisation effects if a continuum with doubly-ionised states is not included in the basis set.
Given these limitations, can we ever truly reach the \emph{ab initio} ideal?

\speak{Numerio}
When phrased in those terms, then I agree that it is an ideal, but there is also no denying the practical reality that has been achieved in describing the full complexity of quantum mechanics as regards attoscience.
And, I would argue, the methods we have available do offer systematic ways to ensure convergence in a controlled fashion, and we can very well say that we are approaching physics in an \emph{ab initio} way.
\end{dialogue}

\section{Advantages and disadvantages of analytical and numerical methods}
\label{sec:Advantages_Disadvantage}

In the previous section we developed, through our combatants \textsc{Numerio} and \textsc{Analycia}, a framework that allows us to place the theoretical methods of attoscience in a continuous spectrum: from analytical to numerical, and from \emph{ab initio} to approximate, as well as from methods that offer qualitative insights to ones whose output is most valuable in its quantitative aspects.
In this section we move on, to focus on the strengths and weaknesses of methods across the theoretical spectrum established in Fig.~\ref{fig:spectrum}.
This analysis is crucial, as it enables an impartial evaluation of different methods, which in turn allows attoscientists to use the most suitable tools for the job at hand. 
Understanding the advantages and disadvantages of different methods, as well as their successes and shortcomings, allows us to highlight the most efficient one---or the most effective combination---for the chosen application, and it is an important guide in the development of hybrid methods.

\subsection{Fundamental strengths and weaknesses}
\narrator{Continuing the conversation \textsc{Analycia} and \textsc{Numerio} each make a case for their respective methods, and attempt to scrutinise the shortcomings of each other's favoured methods.
}
\begin{dialogue}

\speak{Numerio}
The main advantage that has struck me in recent years is the impressive progress in the application of numerical methods to problems of increasing complexity. A number of problems which were once well beyond our reach are now possible. This has been achieved both through the development and refinement of efficient computational methods, as well as the increasing availability of high-performance computing (HPC) platforms. 
Such methods can also act as benchmarks against which to test the validity of simpler, smaller-scale, or more approximate methods. 
Their other clear advantage is their generality, which enables their application to a variety of physical problems.

\speak{Analycia}
True, but despite these advantages, you must admit there can be a heavy price to pay. As you mention, application of these methods can require large-scale HPC resources, and such calculations can be extremely time-consuming, even if optimised codes and efficient numerical methods are used. It may not be possible to perform a large number of such calculations, which then makes it infeasible to perform scans over laser parameters that are often crucial to understand the physics. Additionally, an inherent difficulty in many methods is the rapid increase in required numerical effort with the number of degrees of freedom of the target system. This often restricts methods, for instance, to the treatment of one active electron, or to linearly-polarised laser fields. Releasing these restrictions, and others, then incurs a significant computational cost. 

Analytical methods, however, are not encumbered with many of the difficulties encountered by numerical methods. Their inherent approximations afford them a large speed advantage as well as a high degree of modularity. These qualities allow them to provide an intuitive physical picture of the complex dynamics. 
They may also avoid the unfavourable scaling properties with which numerical methods can be saddled, allowing them to explore a more expansive parameter space. This, coupled with the understanding they provide, can be used to direct more resource-expensive numerical or experimental approaches. 

\speak{Numerio}
Yes, I'm aware that analytical methods provide a number of advantages, but the price tag is the required level of approximation that enables analyticity. Approximation is a double-edged sword: ideally we would only discard unnecessary details in order to highlight the important processes, but, most commonly, we also end up discarding important details, and this may imply that some physical processes are not accurately captured. Approximations also often carry more restrictive regimes of validity, and this makes them less general than \emph{ab initio} approaches.  So, despite the advantage they may have with regard to scaling properties, they can also be rather restricted in some respects. This can often come in the form of rather unrealistic assumptions, such as the assumption of monochromatic laser pulses. 
\end{dialogue}

\narrator{%
Unable to find common ground on their favoured methods, \textsc{Numerio} and \textsc{Analycia} decide to look at the specific example of NSDI.
}

\subsection{In context: non-sequential double ionisation}
\narrator{This case study on NSDI will explore the impact that the characteristics of various methods can have in understanding a physical process. However, before we rejoin our debating combatants, we will present a few of the key concepts of NSDI.}
\vspace{0.5ex}

\begin{figure}
  \centering
  \includegraphics[width=0.7\linewidth]{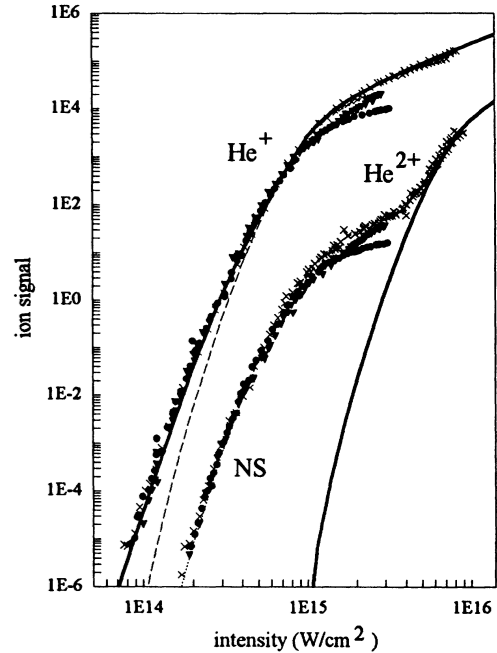}
  \caption{
    Singly- and doubly-charged helium ion yields as a function of laser intensity, at a wavelength of $\lambda = \SI{780}{nm}$, showing the `knee' structure associated with the transition between NSDI and sequential double ionisation~\cite{Walker1994}.%
    \setcounter{footnote}{2}%
    \protect\footnotemark
  }
  \label{Fig:Knee}
\end{figure}
\footnotetext{Reprinted with permission from Ref.~\cite{Walker1994}. {\textcopyright} 1994 by the American Physical Society.}
 
\begin{figure}
  \centering
  \includegraphics[width=\linewidth]{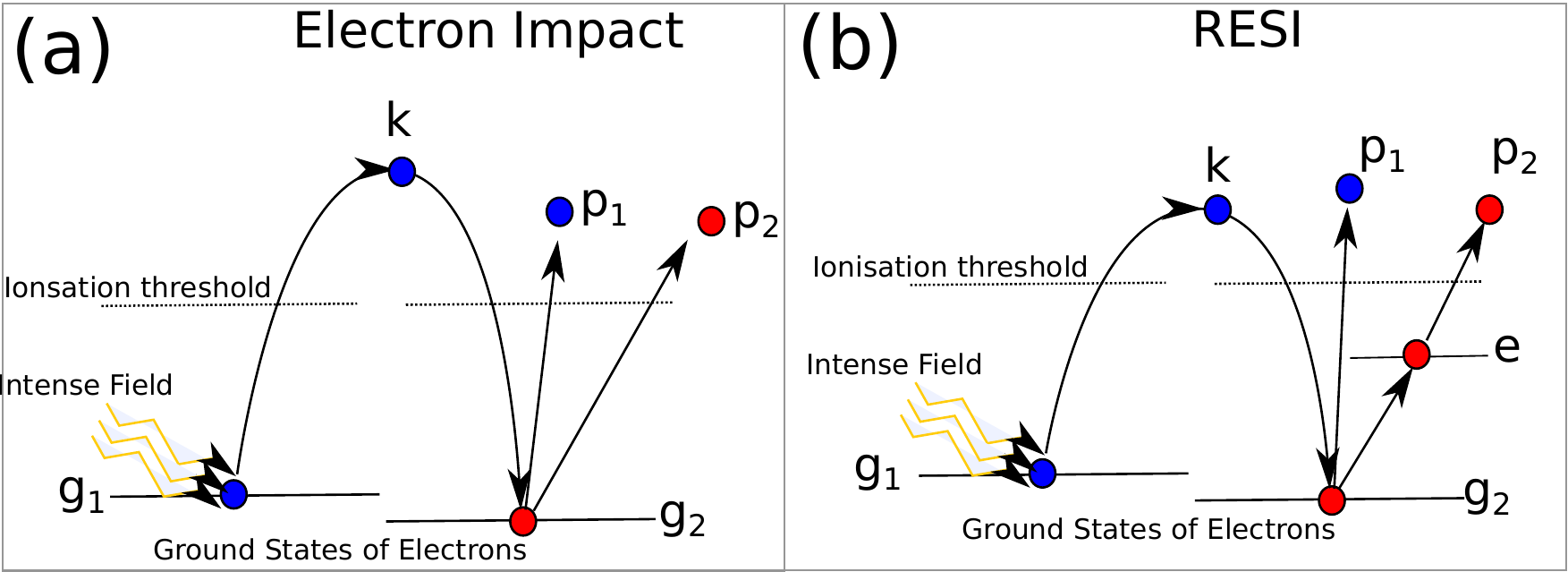}
  \caption{%
    Schematic of the two main mechanisms in NSDI~\cite{MaxwellThesis2019}. 
    Panel (a) shows electron impact ionisation (EI), where the recolliding electron mediates direct double ionisation. 
    Panel (b) shows the recollision excitation with subsequent ionisation (RESI) mechanism, where the recolliding electron excites the bound electron, which is subsequently released by the field.
  }
  \label{Fig:Mech}
\end{figure}

\noindent
NSDI has been studied using a wide variety of analytical and numerical methods. These include both classical and quantum approaches, solving the Newtonian equations of motions and the TDSE. This range of methods is a testament to the difficulty in modelling this process, and thus makes it an ideal case study.

\paragraph{What is NSDI?}
Put simply, NSDI is a correlated double ionisation process, where the recollision of a photoelectron with its parent ion leads to the ionisation of a second electron. Historically, NSDI was discovered as an anomaly, where the experimental ionisation rate did not agree with analytical computations for sequential double ionisation for lower laser intensities, giving rise to the famous `knee' structure (see Fig.~\ref{Fig:Knee})~\cite{LHuillier1983, Agostini1984, Agostini1985, Fittinghoff1992, Kondo1993, Walker1994}. 
Originally, there was contention over the precise mechanism, but over time the three-step model~\cite{Kuchiev1987, Corkum1993, Schafer1993} involving the laser-driven recollision was accepted. The three steps of this model are (i) strong-field ionisation of one electron, (ii) propagation of this electron in the continuum, and (iii) laser-driven recollision and the release of two electrons.
This classical description is based on strong approximations, and it is generally considered to be an analytical method (although, as we discussed in Section~\ref{sec:analytical-involves-numerical}, it generally relies on some numerical computations).
In particular, the exploitation of classical trajectories gives it the intuitive descriptive power of an analytical method.

Within the three-step model, two main mechanisms have been identified for NSDI. The first is electron impact (EI) ionisation, where the returning electron has enough energy to release the second electron, leading to simultaneous emission of both electrons, as depicted in Fig.~\ref{Fig:Mech}(a). The alternative mechanism is recollision with subsequent ionisation (RESI), which occurs when the returning electron only has enough energy to excite the second electron (but not remove it directly), and this second electron is subsequently released by the strong field, leading to a delay between the ionisation of the first and second electron, as shown in Fig.~\ref{Fig:Mech}(b). The separation of these mechanisms is best expressed by semi-analytic models based on the SFA~\cite{Becker1994, Becker1996}, where the mechanisms can be represented as Feynman diagrams and linked to rescattering events~\cite{Becker1999}.

\paragraph{The NSDI Toolset}
Here we summarise some of the methods that are available to model NSDI. For detailed reviews on these methods see Refs.~\cite{becker2011,figueira2011}.
\begin{itemize}

\item 
\textbf{Three-step model:} This simple and intuitive classical description neglects the Coulomb potential and quantum effects~\cite{Corkum1993, Schafer1993, Kuchiev1987}. Nonetheless, this formulation has  become the accepted mechanism of NSDI.
    
\item
\textbf{Classical Models:} These can be split into those with some quantum ingredients like a tunnelling rate~\cite{Ye2008, Emmanouilidou2008, Emmanouilidou2009, Brabec1996, Chen2000} and those that are fully classical, so that ionisation only occurs by overcoming a potential barrier~\cite{Panfili2001, Haan2008, Haan2008PRL, Ho2005, Haan2002, Panfili2002, Goreslavskii2001, Popruzhenko2002}. The electron dynamics are approximated by classical trajectories, which permits a clear and intuitive description. The contributions of classes of trajectory can be analysed, which is crucial in tracing the origin of certain physical processes.
However, the model neglects quantum phenomena such as interference~\cite{maxwell2016, Hao2014}.
    
\item 
\textbf{Semi-classical SFA:} The Coulomb potential is neglected but the dynamics can be understood via intuitive quantum orbits, and the different mechanisms can easily be separated~\cite{Becker1994,Becker1996, Becker1999, Becker2000, Goreslavskii2001, Popruzhenko2002, Quan2009, Shaaran2010, Shaaran2012, figueira2012, Maxwell2015, maxwell2016}. This also allows quantum effects such as tunnelling and interference to be included, with interference effects in NSDI being predicted~\cite{Maxwell2015, maxwell2016,Hao2014} and measured~\cite{Liao2017} fairly recently.
    
\item 
\textbf{Reduced-dimensionality TDSE simulations:} Solution of the TDSE assuming that a particular aspect of the motion can be restricted to the laser polarisation axis. 
One-dimensional treatments restrict the entire electron motion to this axis~\cite{lein2000}, and two-di\-men\-sional treatments restrict the centre of mass~\cite{ruiz2006}, while treating electron correlation in full dimensionality. 
Similar approximations are made in other methods, such as the multi-configurational time-dependent Hartree method \cite{sukiasyan2009}, which treats NSDI with the assumption of planar electron motion.
        
\item 
\textbf{\emph{Ab initio} full dimensional TDSE simulation:} Full quantum mechanical treatment of a two-electron atom through direct solution of the time-dependent close-coupling equations~\cite{smyth1998, parker2001, pindzola1998, pindzola2007, colgan2012, parker2006}. Such methods are computationally intensive, although efficiency improvements have been made in recent years. To date, these methods have not been extended to treat molecules or atoms other than helium.
\end{itemize}

\narrator{We rejoin our two debating attoscientists, whose discussion has now moved on to the specifics of different analytic and \emph{ab initio} methods in NSDI. The discussion begins with a debate on the positive and negative aspects of a direct \emph{ab initio} approach.}

\subsubsection{Full-dimensional numerical solution of the TDSE}
\begin{dialogue}

\speak{Numerio}
As a numericist, I often feel that there is no substitute for solving the TDSE in its full dimensionality. In the context of NSDI, this is a daunting computational task, involving solution of many coupled radial equations---often thousands---on a two-dimensional grid. The first code development to do this began in the late 1990s~\cite{smyth1998,parker2001,pindzola1998} and, by 2006, calculations could be carried out for double ionisation of helium at \SI{390}{nm}~\cite{parker2006}. 
However, these calculations typically required enormous computational resources---an entire supercomputer, in fact---using all 16,000 cores available at that time on the UK's national high-end computing platform (HECToR).

Following the literature over the next few years, I noted the development of a number of similar approaches~\cite{feist2008,feist2009,hu2010,nepstad2010,hu2013,djiokap2012,djiokap2015,donsa2019,donsa2019prl}. In NSDI applications in particular, I was struck by the significant progress made in reducing the scale of such calculations by the {\sc tsurff} method~\cite{scrinzi2012,zielinski2016,trecx,zhu2020}. 
This approach allowed calculations for double ionisation of helium at \SI{800}{nm} to be carried out using only 128 CPUs for around 10 days~\cite{zielinski2016}. Fig.~\ref{he780fig} shows a recent highlight of this work, the two-electron momentum distribution for helium at \SI{780}{nm}~\cite{zielinski2016}. The calculation successfully displayed the expected minimum in the distribution when both electrons attain equal momenta greater than $2U_p$. Watching these developments unfold over the past 15 years, it has become clear to me that even a daunting problem such as this is well within our grasp, and should be attempted.

\begin{figure}[t]
  \centering
  \includegraphics{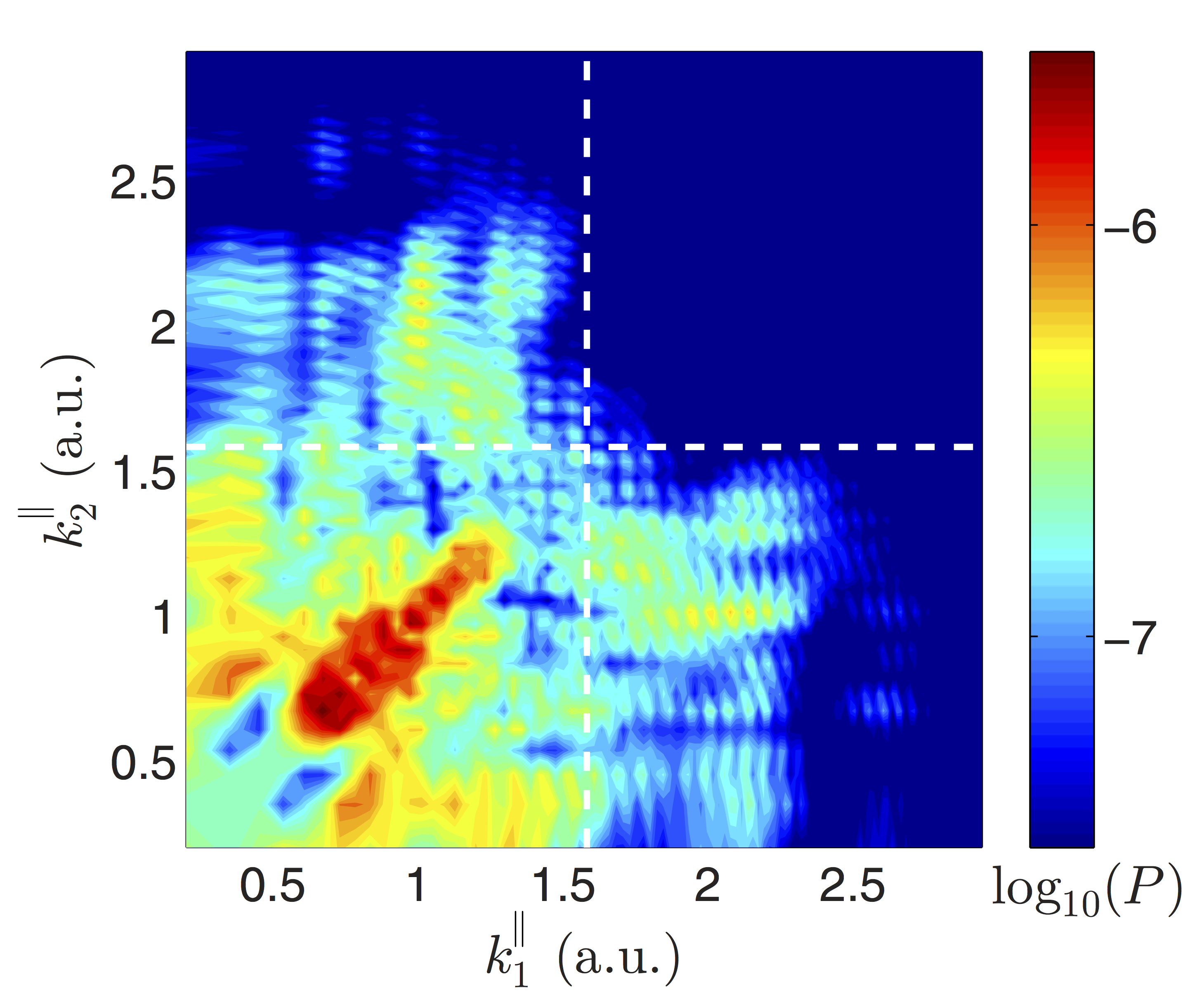}
  \caption{%
    Two-electron momentum distribution for double ionisation of helium at \SI{780}{nm}, calculated using the \textsc{tsurff} method~\cite{zielinski2016}.%
    \setcounter{footnote}{3}%
    \protect\footnotemark
  }
  \label{he780fig}
\end{figure}
\footnotetext{Reprinted with permission from Ref.~\cite{zielinski2016}. {\textcopyright} 2016 by the American Physical Society.}

\speak{Analycia} These are quite intensive calculations, so my first question would be: is it always worth it?
Calculations should not only be feasible---they should also be justifiable. 
The large scale of each single calculation can be a very limiting factor, since you may need further computations, perhaps to perform intensity averaging, or to scan over a particular laser parameter. Here you may encounter additional hurdles, since it is well known that the computational cost can scale very unfavourably with certain laser parameters, particularly wavelength. 
Even with the efficiency savings that you mention, the method may struggle to perform calculations at longer wavelengths, or in sufficient quantity to scan over experimental uncertainties.

Secondly, it's true that significant progress has made these large-scale calculations more tractable. However, this does not necessarily mean that the results will be easy to analyse. Disentangling the complex web of physical processes included in such calculations can be very difficult. This requires tools and switches within the method, for example to evaluate the role of certain interactions, and thereby aid your understanding. Even with such analysis tools at hand, gaining strong physical insight may be an arduous procedure, involving further large-scale calculations, and these may not even be guaranteed to provide the insights you desire.

\speak{Numerio}
Absolutely, you have highlighted the main difficulties with \emph{ab initio} methods that I have encountered. The scale of the calculations can impose a limit on their scope, and their complexity can obscure interpretation. On the other hand, simpler methods avoid these difficulties, but they rely on approximations which need to be justified. For me, the ideal tool would be a method with qualities representing the best of both worlds --- a method where many small-scale but accurate TDSE calculations could be carried out to provide detailed interpretation. Although this is feasible in some fields, in the context of NSDI currently it is not. However, equipped with an arsenal of \emph{ab initio} methods, there is an opportunity to benchmark simpler methods which fall short of a full \emph{ab initio} treatment. If their approximations can be validated by such comparisons, then their interpretive power will be valuable. 

\speak{Analycia} I think now we are beginning to agree.
\end{dialogue}

\narrator{The debate above highlights that both calculation and interpretation are important. Often, an \emph{ab initio} approach can provide a calculation, but detailed interpretation may require analytical techniques. To discuss these techniques further, the debate now moves to focus on the merits of analytical methods used to study NSDI.}

\subsubsection{Analytical approaches}

\begin{figure}
  \centering
  \includegraphics[width=\linewidth]{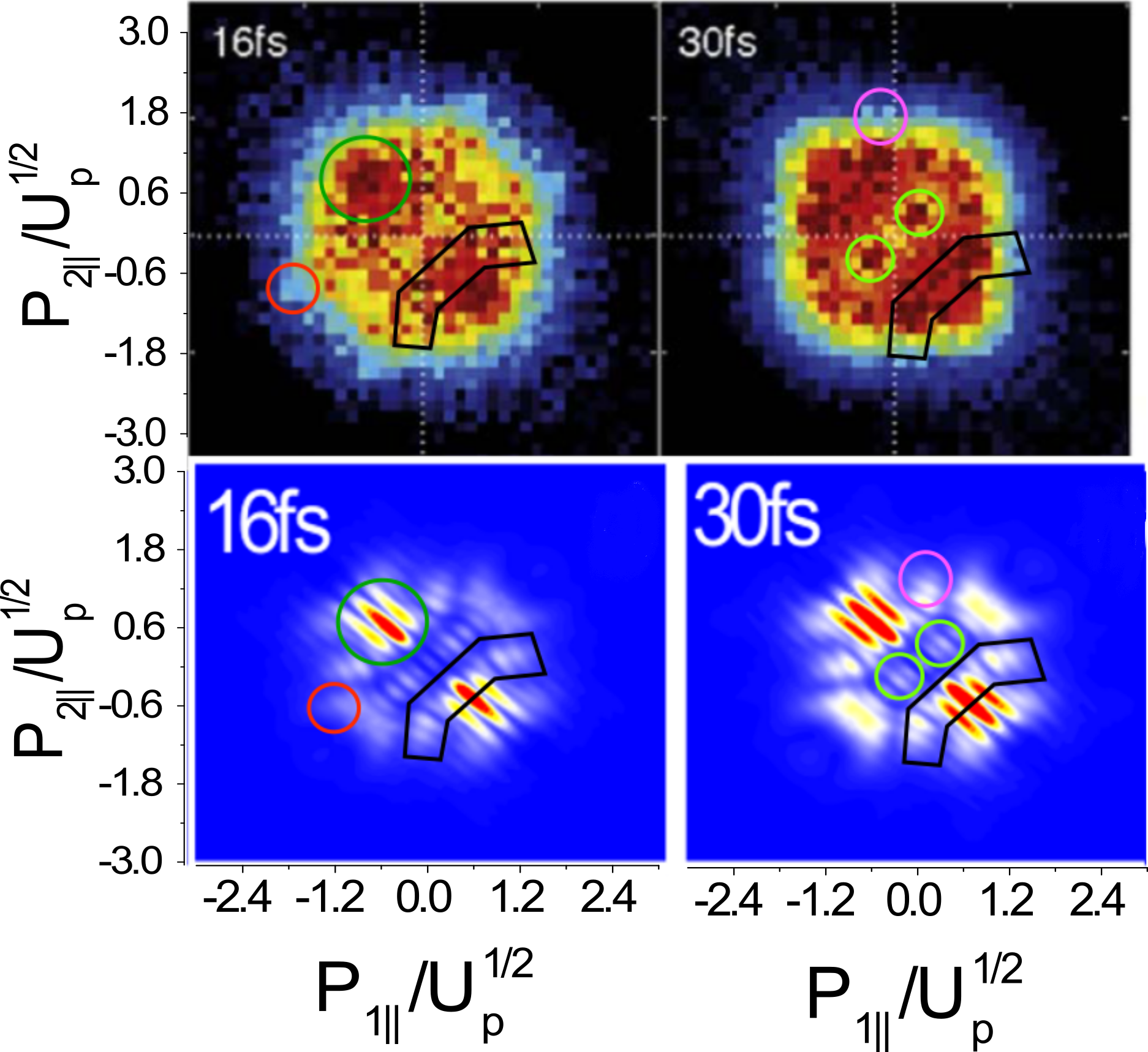}
  \caption{%
    \setcounter{footnote}{4}%
    Comparison of experimental data (upper row)~\cite{Kubel2014}%
    \protect\footnotemark \ %
    with theoretical focal-averaged distributions (lower row) selected from~\cite{maxwell2016}.%
    \protect\footnotemark \ 
    The left and right columns present \SI{16}{fs} and \SI{30}{fs} laser pulse lengths, respectively, with $\lambda=\SI{800}{nm}$ ($\omega=\SI{0.057}{\au}$) and $I = \SI{e14}{W/cm^2}$ ($U_p = \SI{0.22}{\au}$). 
    Specific features associated with quantum interference are marked by polygons in both upper and lower panels. 
    It was necessary to account for interference effects in the theoretical results to get this agreement.%
    }
\label{fig:NSDI_Interference}
\end{figure}
\footnotetext[\numexpr\thefootnote-1]{Reproduced from Ref.~\cite{Kubel2014} under a \href{https://creativecommons.org/licenses/by/3.0/}{CC BY} license.}
\footnotetext{Reprinted with permission from Ref.~\cite{maxwell2016}. {\textcopyright} 2016 by the American Physical Society.}

\begin{dialogue}

\speak{Analycia} 
You see in my experience working on NSDI, descriptive power is often enabled by the high degree of modularity that analytical methods possess. 
This modularity may be harnessed to determine the physical origin of an effect by switching certain interactions on and off.
Like intermediate-rigour numerical methods, the light computational demand means that large sets of individual calculations may be carried out where necessary. 

A good example of the power of modularity in analytical models is the use of interference in SFA models for NSDI to match experimental results~\cite{maxwell2016, Hao2014}. In Fig.~\ref{fig:NSDI_Interference} we see experimental results~\cite{Kubel2014} for two pulse lengths. The lower panels show the results of the SFA model~\cite{maxwell2016} that uses a superposition of different excited states in the RESI mechanism of NSDI. Including interference leads to a good match, which provides strong evidence for interference effects in NSDI. This was only possible because interference effects could be switched on and off,%
\footnote{%
The consequences of this difficulty were discussed in detail in Battle 2 of the \textit{Quantum Battles in Attoscience} conference~\cite{battle2}, reported in a companion paper in this Special Issue~\cite{Amini2020}.
}
thereby allowing analysis of the different shapes and structures within the distribution. Each of these shapes could then be directly attributed to different excited states, which demonstrates the power of the modularity of analytical methods in providing an intuitive understanding of the physics.

\begin{figure}
  \centering
  \includegraphics[width=\linewidth]{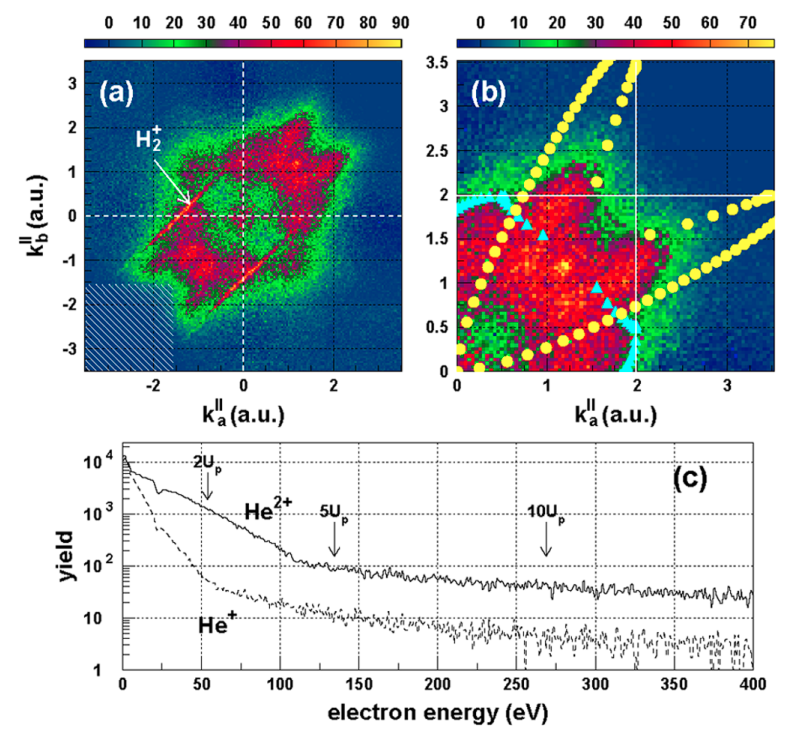}
  \caption{%
  Photoelectron spectra for NSDI in helium driven by a wavelength of \SI{800}{nm} and intensity of \SI{4.5e14}{W/cm^2}~\cite{Staudte2007}, showing
  (a) the correlated momentum distribution,
  with a detail (b) shown with superimposed results from a classical electron scattering model,
  as well as (c) the electron energy spectrum of He$^{2+}$ and He$^{+}$.%
    \setcounter{footnote}{7}%
    \protect\footnotemark
  }
\label{fig:Staudte}
\end{figure}
\footnotetext{Reprinted with permission from Ref.~\cite{Staudte2007}. {\textcopyright} 2016 by the American Physical Society.}

\speak{Numerio} 
The interpretive power is certainly valuable, and the availability of switches such as these is often the key to a good physical understanding.  My main concern, however, is that the approximations may affect the accuracy of the results. In particular, the SFA neglects the Coulomb potential, and it is known that this influences the famous finger-like structure in NSDI seen in Fig.~\ref{fig:Staudte}, causing a suppression of two-electron ejection with equal momenta. Furthermore, we would expect a host of other Coulomb effects just as there are in single electron ionisation \cite{figueira2020phases}.
Thus, care must be taken with the conclusions that you draw from such an analytical model. As I said earlier, many numerical methods may not afford this degree of modularity, but it would strengthen my confidence in the conclusions if an \emph{ab initio} method also observed these effects. In this way, a numerical method could be guided by analytical predictions to assess the accuracy of certain approximations.

\speak{Analycia} 
This is a fair point, but the considerable speed advantage means that you can often do additional checks and analysis to get around this problem.
The SFA model presented  could be solved in five minutes on a desktop computer, whereas, as you mentioned, \emph{ab initio} models will take days on hundreds of cores. The fast SFA calculations can then account for additional factors such as focal volume averaging, even though it increases the overall runtime by a factor of ten or more. It can also perform scans through intensity and frequency in a timely manner. Such scans can provide important insights, for example in Fig.~\ref{fig:Contour} where the contributions of various excited states is monitored as a function of laser intensity and frequency. Their relative contributions then explain the shapes appearing in various regions of the momentum distribution. The extra analysis can increase the overall runtime by factors of 100--1000, which is still perfectly manageable for the SFA, but would be out of the question for most \emph{ab initio} methods.

Furthermore, there is always a place for analytical methods in performing computationally inexpensive initial investigations, which then provide the evidence needed to commit to using more expensive \emph{ab initio} or experimental efforts. In recent work on interference in NSDI, motivated by predictions of SFA models, experimental work was done to investigate interference effects~\cite{quan2017}.

\begin{figure}
  \centering
  \includegraphics[width=\linewidth]{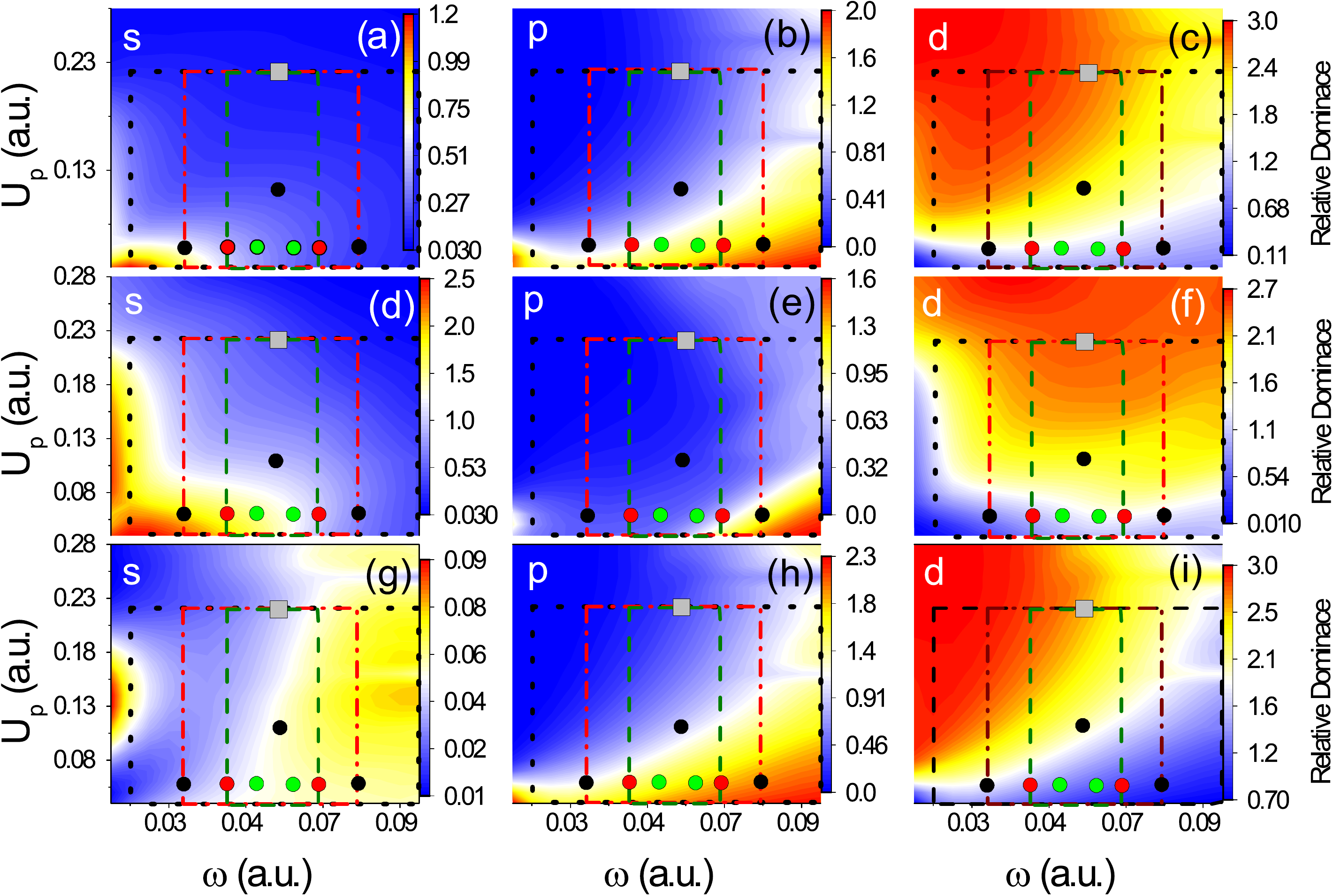}
  \caption{%
    Scan over intensity ($U_p$) and frequency ($\omega$), that attributed preferential excitation to states with different orbital angular momenta $l$ (s-, p- and d-states) in the RESI process for different pulse lengths~\cite{maxwell2016} to the shapes found in~\cite{Kubel2014}.
    The contributions of s-states are displayed in (a), (d) and (g), those of p-states in (b), (e) and (h) and those of d-states in (c), (f) and (i).%
    \setcounter{footnote}{8}%
    \protect\footnotemark
  }
  \label{fig:Contour}
\end{figure}
\footnotetext{Reprinted with permission from Ref.~\cite{maxwell2016}. {\textcopyright} 2016 by the American Physical Society.}

\speak{Numerio} 
Yes, I agree in \emph{some} cases the extra analysis is beneficial. However, \emph{ab initio} methods are still much more generalised than their analytical counterparts. Take Fig.~\ref{he780fig}, where many different processes contribute, including both the RESI and EI mechanisms, together with sequential double ionisation. The presented SFA model includes only the RESI mechanism. 

\speak{Analycia} 
There are two sides to this: it is nice to be able to clearly separate EI and RESI in the SFA, but it is true that it introduces a lack of flexibility. 

With the goal of reaching some kind of agreement, I would posit that the benefits of both types of models outweigh the negatives. In the case of classical and semi-classical models, they have clearly led to huge leaps in understanding for the mechanisms of NSDI. Furthermore, I would add that NSDI in particular is a good candidate for hybrid models. Strongly correlated dynamics and multi-electron effects are well-suited to an \emph{ab initio} approach, while the main ionisation dynamics are well-described by semi-classical models.

That said, I would also like to know how the broader community feels about this.

\direct{The audience response to this poll is presented as Poll 2 in Table~\ref{tab:polls}, in Section~\ref{sec:discussion} below.}
\end{dialogue}

\narrator{Within the context of NSDI, our combatants have discussed the merits and drawbacks of their respective approaches, and have begun to appreciate the computational and interpretational qualities that analytical and \emph{ab initio} contribute. In the following section, we focus on how progress in scientific discovery can be aided by both types of method.}

\section{Scientific discovery}
\label{sec:discovery}

The seed of a scientific discovery can be planted in the form of a bump or a dip on a smooth curve of experimental data, as a whimsical term in the denominator of some equation, or as a quirky splash in numerical results. In other words, a scientific discovery can be triggered by experimental results or theoretical ones, either analytical or numerical. As soon as something new has been spotted, to become a real full-grown discovery it has to be examined and explained by each of these aforementioned components, each branch of research, and in the end there has to be an agreement among all of them. 

In some cases, the initiating site is analytical and the others come next, as in the case of optical tunnelling ionisation, predicted in 1965~\cite{keldysh1965} before its much later experimental observation in 1989~\cite{Corkum1989Mar}.%
\footnote{%
  Optical tunnelling ionisation also gives rise to the question of the tunnelling time, which was discussed in depth during Battle 1 of the \textit{Quantum Battles in Attoscience} conference~\cite{battle1}, reported in a companion paper in this Special Issue~\cite{Hofmann2021}.%
  }
Sometimes, the role of a trigger is played by numerical calculations, as for coherent multi-channel strong-field ionisation~\cite{Rohringer2009May}, which was shortly followed by its experimental validation~\cite{Goulielmakis2010Aug}. 
It can even be a little bit of both, 
as in the first description of the RABBITT scheme in 1990~\cite{veniard_phase_1990}, 
or in single-photon laser-enabled Auger decay (sp-LEAD), 
which was predicted in 2013~\cite{Cooper2013Aug}, first observed in 2017~\cite{Iablonskyi2017Aug}, and further characterised in~\cite{You2019}.
There are also theoretical predictions---%
both analytical, like molecular Auger interferometry~\cite{Khokhlova2019Jun},
and numerical, like HHG in topological solids~\cite{Bauer2018, Silva2019Dec, Chacon2020}---%
which have already sparked experimental efforts to confirm them, but which are still waiting for their observations to come.
On the other side, we have discoveries which arise from experimental observations and are then explained theoretically, such as NSDI, which was discussed in detail in Section~\ref{sec:Advantages_Disadvantage}.

In this section we tell another story of scientific discovery in attoscience, through the case study of resonant HHG, which also starts from recorded experimental data.

\subsection{Experimental kick-off}
By the year 2000, HHG was already a full-grown discovery. It had been observed~\cite{mcpherson_studies_1987, ferray_mutliple_1988} and theoretically modelled~\cite{Kuchiev1987, Corkum1993, krause_hihg-order_1992, Lewenstein1994Mar} a decade previously. After this breakthrough, many features of HHG were under active investigation, both experimentally and theoretically. In particular, resonances in the HHG spectrum had been extensively studied since the 1990s. Some structures in the HHG spectra of atomic gases had been very early on attributed to single-atom resonances~\cite{lhuillier_coherence_1992, balcou_phase-matching_1993}. Then more recent measurement~\cite{toma_resonance-enhanced_1999} and theoretical works~\cite{figueira2002Jan, Taieb2003Sep} explained these structures with multiphoton resonances with bound excited states linked to enhancement of specific electron trajectories that were recolliding multiple times with the ionic core.

\begin{figure*}[t]
    \centering
    \begin{overpic}[width=0.3\linewidth]{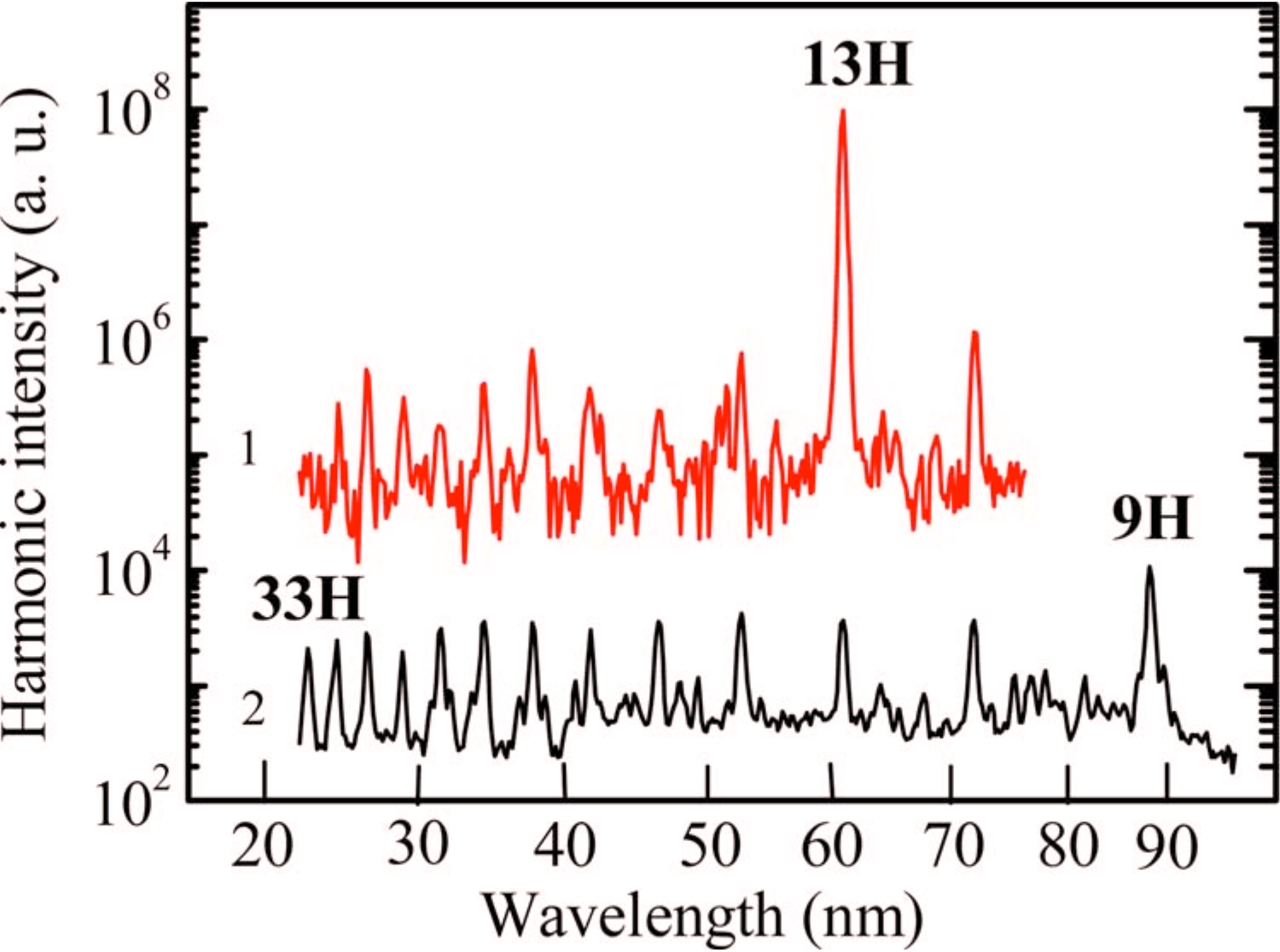}
    \put (18,65) {\bf (a)}
    \end{overpic}
    \begin{overpic}[width=0.32\linewidth]{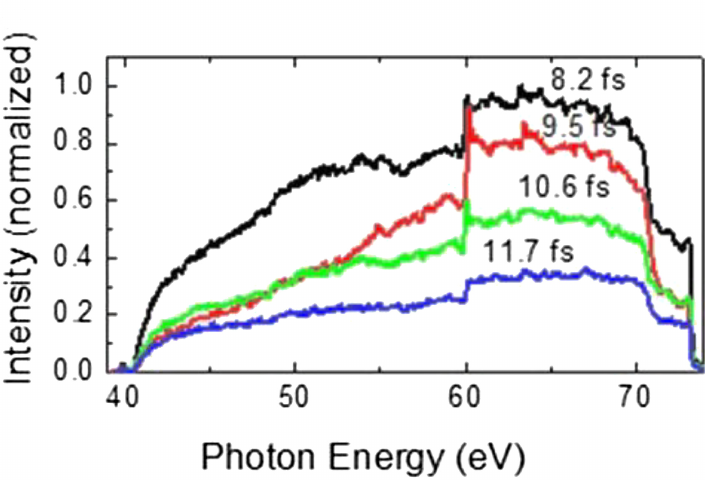}
    \put (18,65) {\bf (b)}
    \end{overpic}
    \begin{overpic}[width=0.36\linewidth]{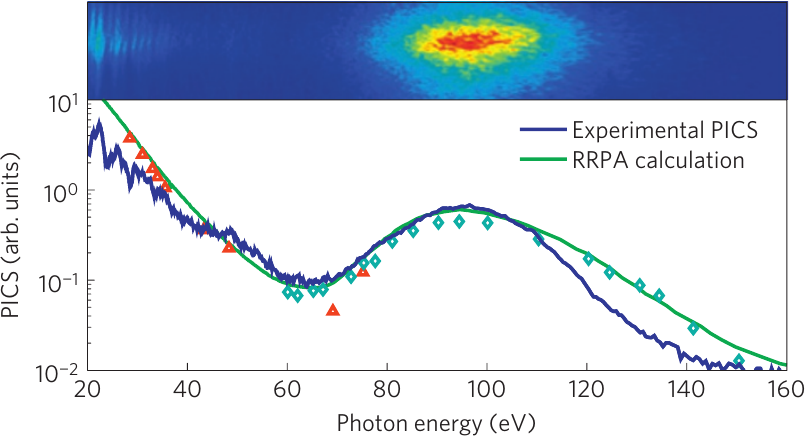}
    \put (15,58) {\bf (c)}
    \end{overpic}
    \caption{
    First observations of resonant HHG. 
    (a) Figure taken from~\cite{Ganeev2006Jun}: High-order harmonic spectra from (1) indium and (2) silver plumes.%
    \setcounter{footnote}{10}%
    \protect\footnotemark{} 
    (b) Figure taken from~\cite{Gilbertson2008Sep}: Spectra of the harmonic supercontinuum generated with double optical gating from the different values of input pulse duration for a helium target gas.%
    \protect\footnotemark{} 
    (c) Figure taken from~\cite{Shiner2011Jun}: Top, the raw HHG spectrum from xenon at an intensity of \SI{1.9e14}{W/cm^{2}}. Bottom, experimental HHG spectrum divided by the krypton wave packet (blue) and the relativistic random-phase approximation (RRPA) calculation of the xenon photoionisation cross-section from~\cite{kutzner_extended_1989} (green). The red and green symbols are PICS measurements from~\cite{fahlman_xe_1984} and \cite{becker_subshell_1989} respectively, each weighted using the anisotropy parameter calculated in~\cite{kutzner_extended_1989}.
    }
    \label{fig:resonant HHG observation}
\end{figure*}
\footnotetext[\numexpr\thefootnote-1]{Reprinted with permission from Ref.~\cite{Ganeev2006Jun} \textcopyright The Optical Society.} 
\footnotetext{Reprinted from Ref.~\cite{Gilbertson2008Sep}, with the permission of AIP Publishing} 
%

In this context, Ganeev et al.~\cite{Ganeev2006Jun} first measured, in 2006, a strong enhancement of a single harmonic, by two orders of magnitude, in the HHG spectrum of plasma plumes. Their result is shown in Fig.~\ref{fig:resonant HHG observation}(a). At that time, they attributed this resonance to the multiple recolliding electron trajectories that had previously been observed and modelled in atoms~\cite{toma_resonance-enhanced_1999, Taieb2003Sep}, and they related these trajectories to multiphoton resonances with excited states.

Then, in 2008, when studying the spectra of single attosecond pulses generated in noble gases, Gilbertson et al. measured for the first time a strong enhancement in the HHG spectrum of helium~\cite{Gilbertson2008Sep}, as shown on Fig.~\ref{fig:resonant HHG observation}(b). Since they employed single attosecond pulses, they recorded continuous spectra which allowed them to see the enhancement perfectly, as it would otherwise fall between two harmonics if observed in an attosecond pulse train. They did not give any explanation for this enhancement, as it was not their main focus, but they observed that it appears at the energy of the $2s2p$ autoionising state (AIS) of helium.

Then in 2011, Shiner et al.\ measured a strong enhancement at \SI{100}{eV} in the HHG spectrum of xenon gas~\cite{Shiner2011Jun,Schmidt2012Mar}, shown in Fig.~\ref{fig:resonant HHG observation}(c). The experimental HHG spectrum is displayed on the upper panel. From that spectrum the authors extracted the photoionisation cross-section (PICS) by first dividing by the spectrum of krypton (obtained at the same conditions), and then multiplying by the photoionisation cross-section of krypton from Ref.~\cite{huang_theoretical_1981}. The obtained experimental PICS is shown as a blue line in the lower panel. The green curve is the photoionisation cross-section of xenon from Ref.~\cite{kutzner_extended_1989}. The very good agreement between the two curves, combined with the qualitative agreement with a toy model including only the $4d$ and $5p$ states of xenon, allowed the authors to relate the enhancement at \SI{100}{eV} to the giant resonance of xenon.

Thus, by the late 2000s, there were observations of resonant enhancement features in HHG from plasma plumes as well as few- and many-electron rare-gas atoms---but no solid theoretical explanation.

\narrator{We now hand again the stage to our theoretical acquaintances, who have begun discussing the ingredients required for a theoretical model for resonant HHG and will  guide us through the rest of the story.}

\subsection{Building the model}
\begin{dialogue}

\speak{Numerio}
An explanation of the observed process demands the creation of a model, and this requires a thorough analysis of the experimental data revealing the same phenomenon and distinguishing its essential features. The essential feature of resonant HHG, which is common for all observations independently from the medium---gaseous or plasma---, is the enhancement of one or of a group of high harmonics. This does not sound like a lot to start with. However, this already gives a hint that the desired explanation has no connection to propagation effects. This fact restricts the model to an account of the single-particle response only. 

\speak{Analycia}
There have been a number of attempts to create the model describing resonant HHG. One group of theories is based on bound-bound transitions~\cite{Gaarde2001Jun, figueira2002Jan, Taieb2003Sep, Ishikawa2003Jul}, but it cannot be applied for plateau harmonics due to the crucial role~\cite{Ganeev2006Jun} played by the free-electron motion. Another group of theories mentions a connection of the multi-electron excited states to the enhanced yield of harmonics~\cite{Milosevic2007Aug, Frolov2009Jun, Frolov2010Aug}. In particular, the enhancement of high harmonics generated in xenon~\cite{Shiner2011Jun} was associated~\cite{Frolov2009Jun} with the region of the well-known ‘giant’ dipole resonance in the photoionisation (photorecombination) cross-section of xenon atoms. 

\speak{Numerio}
This sounds closer to the ingredients that are likely required to explain the phenomenon. Does this not get us closer to resolving the puzzle?

\speak{Analycia}
Indeed it does! After revealing a similar correspondence between experimental HHG enhancements~\cite{Gilbertson2008Sep,Ganeev2006Jun} and transitions with high oscillator strengths between the ground state and AIS of the generating ions~\cite{Chan1991Jul,Duffy2001Mar}, the model of resonant HHG was forged in the form of the ‘four-step model’~\cite{Strelkov2010Mar}. 

\speak{Numerio}
It seems like there should be a vivid similarity with the common three-step model for HHG, should there not?

\speak{Analycia}
Of course! The four-step model~\cite{Strelkov2010Mar} extends the three-step model~\cite{Corkum1993,Schafer1993,Kuchiev1987} to include the resonant harmonic emission along with the `classic', nonresonant one. The first two steps of the four-step model---(i) tunnelling ionisation, and (ii) free-electron motion---repeat those of its forerunner. Then, if the energy of the electron returning back to the parent ion is close to the one of the ground--AIS transitions, the third step of the three-step model turns into two: (iii) electron capture into the AIS, and (iv) relaxation from the AIS down to the ground state, accompanied by the XUV emission. 

\speak{Numerio}
Sure, that seems like a possible chain of events, but how can it lead to a higher emission probability if it requires an extra step?

\speak{Analycia} 
You are right that a substitution of one step by two of them should intuitively cause a decrease in probability, but the combination of higher probability for electron capture into the AIS (corresponding to the looser localisation of the AIS), together with the high oscillator strength of the transition between the AIS and the ground state, results in an increase of the resonant harmonic yield by several orders of magnitude.
Perhaps you could argue this is similar to how NSDI may dominate over sequential double ionisation despite having more steps, as discussed in Section~\ref{sec:Advantages_Disadvantage}.
\end{dialogue}

\narrator{By this point a convincing model has been suggested; however, it is far from an end of the story of ‘scientific discovery of resonant HHG', and a series of hurdles still has to be surmounted.}

\subsection{Challenging the model: numerical calculations}
\begin{figure}
    \centering
    \begin{overpic}[
      scale=1.0,
      unit=1mm,
      ]{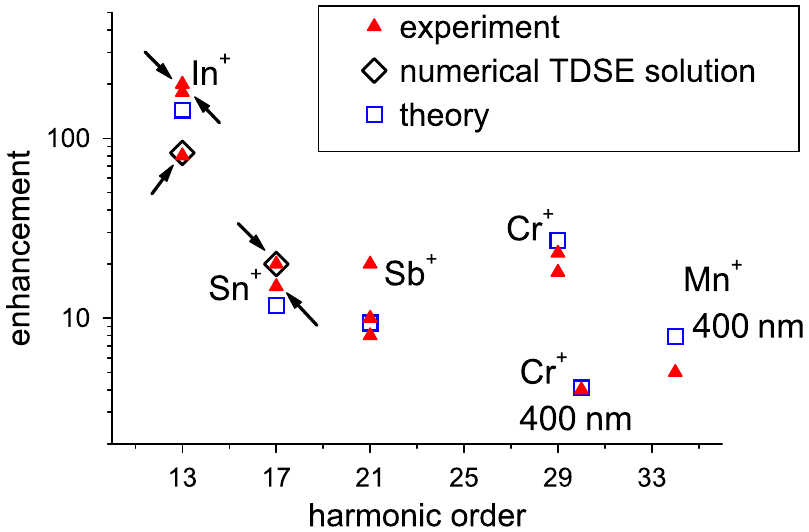}
    \put(15,38){\small\cite{Ganeev2006Jun}}
    \put(15,60){\small\cite{ganeev_harmonic_2006}}
    \put(47,25){\small\cite{ganeev_harmonic_2006}}
    \put(27,39){\small\cite{suzuki_anomalous_2006}}
    \put(43,35){\small\cite{suzuki_intense_2007}}
    \put(71,35){\small\cite{ganeev_strong_2007}}
    \put(71,30){\small\cite{ganeev_systematic_2007}}
    \put(85,18){\small\cite{ganeev_systematic_2007}}
    \put(71,20){\small\cite{ganeev_systematic_2007}}
    \put(43,20){\small\cite{ganeev_systematic_2007}}
    \put(34,22){\small\cite{ganeev_systematic_2007}}
    \put(27,46.5){\small\cite{ganeev_systematic_2007}}
    \end{overpic}
    \caption{
    Comparison of experimental measurements~\cite{Ganeev2006Jun, ganeev_harmonic_2006, suzuki_anomalous_2006, suzuki_intense_2007, ganeev_strong_2007, ganeev_systematic_2007} with analytical theory and single-electron TDSE simulations~\cite{Strelkov2010Mar} for the enhancement factor in resonant HHG in plasma-plume ions, as reported in Ref.~\cite{Strelkov2010Mar}.%
    \setcounter{footnote}{12}%
    \protect\footnotemark
    }
    \label{fig:strelkov-TDSE}
\end{figure}
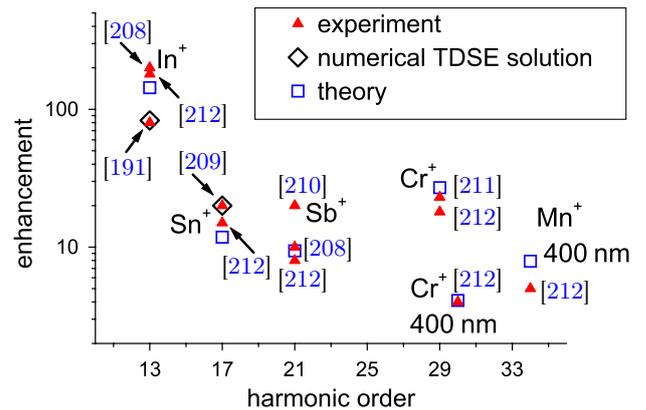

\begin{dialogue}

\speak{Numerio}
Alright, this model sounds physically reasonable enough, but we still need some actual proof that it's describing the experiment properly. Small-scale numerical simulations were of great help in that matter. When building the four-step model, Strelkov also performed some TDSE simulations at the SAE level, and compared it with several experimental results for singly-ionised indium and tin, as shown in Fig.~\ref{fig:strelkov-TDSE}. The very good agreement shows that a single active electron is able to accurately model the process. 

\speak{Analycia} 
Sure, that is an important result, but in the paper, Strelkov also made an analytical estimate of the enhancement using the oscillator strength and lifetime of the resonant transition. The result of this estimate is shown in Fig.~\ref{fig:strelkov-TDSE} as blue squares for several singly-ionised atoms. The good agreement both with experiment and with the TDSE calculations marks another step in the confirmation of the four-step model.

\speak{Numerio}
Indeed, that was quite convincing already, but all these considerations were time independent. When building a model in attosecond science, it is often useful to have a dynamical point of view on the process under study. Tudoroskaya and Lein investigated resonant HHG and the four-step model using time-frequency analysis~\cite{tudorovskaya_high-order_2011}. They solved the SAE TDSE for 1D model potentials with a shape resonance that models an AIS. They were able to reproduce an enhancement of more than two orders of magnitude at the harmonic order corresponding to the shape resonance. Their time-frequency analysis confirmed that the harmonic emission at resonance starts when the electron returns to the ionic core. More interestingly, it shows that the duration of the emission at resonance is much longer than the emission duration at the other harmonic orders. More precisely, the emission duration at resonance corresponds to the shape resonance lifetime, indicating that the electron gets trapped in the resonance and emits from there, thus validating the four-step model.
\end{dialogue}

\narrator{After this convincing achievement, the model seems to be validated, especially from \textsc{Numerio}'s point of view. But for \textsc{Analycia} the story is not finished yet.}

\footnotetext{Adapted with permission from Ref.~\cite{Strelkov2010Mar}. {\textcopyright} 2010 by the American Physical Society.}

\subsection{Generalisation: analytical theory}
\begin{figure}
    \centering
    \includegraphics[width=\linewidth]{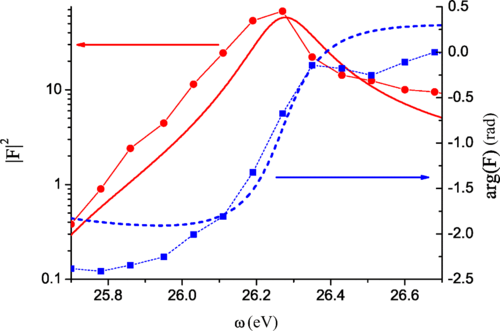}
    \caption{
    Squared absolute value (red) and phase of the Fano-like factor calculated analytically (solid) and numerically within SAE TDSE (with symbols) for HHG in tin plasma plume~\cite{Strelkov2014May}.%
    \setcounter{footnote}{13}%
    \protect\footnotemark
    }
    \label{fig:Fano-like-factor}
\end{figure}

\begin{dialogue}

\speak{Numerio}
Perfect, we now have the model of resonant HHG in our arsenal, which allows us to conduct a qualitative analysis and to make qualitative predictions. Moreover, we also possess quantitative answers based on SAE TDSE solutions for a number of generating particles in given laser fields. 
So I believe we have all we wanted then?

\speak{Analycia}
Not so fast!
Even though there is a tool providing us with a quantitative answer, it cannot be easily re-applied for a different generating system or slightly different field parameters, in other words, there is a lack of generality. This creates a strong demand for a computationally cheap and more flexible tool. 

\speak{Numerio}
Do you have some concrete solutions in mind?

\speak{Analycia}
This theoretical demand has been satisfied within the introduction of the analytical theory of resonant HHG~\cite{Strelkov2014May}. The analytical theory is built on two pillars: Lewenstein's SFA-based theory~\cite{Lewenstein1994Mar} (conventional for HHG), and Fano's theory~\cite{Fano1961Dec}, which guides the treatment of AISs originated from the configuration interaction. 

\speak{Numerio}
I understand, each of these theories is indeed very successful in describing the two physical processes at hand. But how do you combine them to reproduce the experimental observations?

\speak{Analycia}
The resonant HHG theory delivers the answer---the spectrum of the dipole moment of the system---as a product of the spectrum of the nonresonant dipole moment and a Fano-like factor. The nonresonant dipole moment is the same as in the well-known Lewenstein theory, which captures the field configuration and the major characteristics of the ground state of the generating particle. On the other hand, the Fano-like factor encodes the resonance, and depends on the AIS’s features: its energy and its energy width, as well as the dipole matrix element for the transition between the AIS and the ground state.

As a result, the harmonic spectrum in the resonant case is identical to the one in the nonresonant case far from the resonance, while in the vicinity of the resonance it acquires a Lorentzian-like-shape profile due to the Fano-like factor (see Fig.~\ref{fig:Fano-like-factor}). This Fano-like profile around the resonance carries the information about two major properties of the resonant harmonics---their behaviour in amplitude and phase---which result in an enhancement and an emission time delay of resonant harmonics, respectively. 

\speak{Numerio}
Ok, I agree this analytical theory provides a much more general picture of the process. But is it really that useful? I mean, what could we do with resonant HHG?

\speak{Analycia}
The two features of resonant harmonics, amplitude and phase, provide us with extra handles for improving the generation of attosecond pulses, an intensity boost and an elongation of duration, and they also provide an opportunity to study the structure of the AIS using the harmonic spectrum.
\end{dialogue}

\narrator{With a robust framework in place, our scientists discuss the final obstacles faced by the theory.}

\footnotetext{Reprinted with permission from Ref.~\cite{Strelkov2014May}. {\textcopyright} 2014 by the American Physical Society.}

\subsection{Closure: \emph{ab initio} calculations}
\begin{dialogue}

\speak{Analycia}
Although the model and the analytical theory of resonant HHG coincide with the results of numerical TDSE calculations in the SAE approximation, this theory encountered significant resistance, both in conferences and in peer review, insofar as the model potential used in these calculations is artificial and does not reflect the fully multi-electron nature of AISs. 
What is your opinion regarding this issue?

\speak{Numerio}
I would say that, on the one hand, this is an instance of a broader discussion regarding the role, advantages and disadvantages of the use of model potentials in numerical calculations.
On the other hand, however, this limitation can be addressed using fully-\emph{ab initio} calculations, eliminating this final uncertainty. 

Recent first-principle calculations for resonant HHG by manganese atoms and ions~\cite{Wahyutama2019Jun} show the characteristic enhancement observed earlier in the energy region around a group of AISs. 

\speak{Analycia}
Finally! These results close the remaining questions in the theoretical understanding and description of resonant HHG, and open a wide front of study into the applications of this process, equipped with a full toolset: analytical theory as well as numerical (SAE and \emph{ab initio}) calculations.

\speak{Numerio}
I agree, we are not always on great terms, but we really made a nice team on this one!

\direct{The audience opinion on the necessity of combining different approaches is presented as Poll 3 in Table~\ref{tab:polls} in Section~\ref{sec:discussion} below.}
\end{dialogue}

\narrator{After this constructive exchange, the two agree to work more tightly together from now on. 
}

$\ $

\noindent
As we discuss below in Section~\ref{sec:audience-questions} as a response to Audience Question~1,
we are not always necessarily after discoveries in our field, but also after finding and solving interesting problems. 
Nonetheless, any scientific production, or creative activity, before it can considered as scientific, requires confrontation of different points of view. We argue here that this confrontation is all the more efficient and constructive when it involves all the different aspects of scientific work: experimental, analytical, numerical, and \emph{ab initio}. 
As we have seen at the start of this section, the initial trigger can be pushed by any of them, but the actual scientific progress generally happens afterwards, when they collaborate together.

\section{Discussions}
\label{sec:discussion}

The dialogue between proponents of analytical and \emph{ab initio} approaches, as we have followed it so far, opens a number of additional questions for deeper examination.
We now turn to these more specific points, as well as our (combatants') responses to the questions raised by audience members during the talk.

During the online conference~\cite{battle}, in addition to the talk, several questions were directed to the audience in the form of polls, both over the Zoom platform as well as to a wider public over Twitter.
We present in Table~\ref{tab:polls} a summary of the results of these polls.

\vspace{1ex}

\narrator{Our combatants return to the stage to resolve several still-itching questions that remain from their conversation.}

\subsection{Is approximation a strength or a weakness?}

\narrator{The degree of approximation made by a particular method is a typical source of contention between numericists and analyticists. Here, {\sc Numerio} and {\sc Analycia} discuss how they feel approximation should be characterised.}

\begin{dialogue}
\speak{Analycia} It has been said that approximation is a downside of analytical methods. 
However, I would like to argue---perhaps somewhat provocatively---that approximation is more of a strength. 
Approximation is what drives the interpretation---the qualitative picture---of a physical process as constructed by an analytical model. 
If you can remove all that is unnecessary, and still achieve reasonable agreement with \emph{ab initio} simulations or with experimental results, then this is when you actually start to gain some real understanding and interpretation of physical processes. 

In other words, I do not think any method, analytical or numerical, is scientifically useful by itself. Science stems from the comparison and interplay of different methods, and particularly of different levels of approximations.

\speak{Numerio}
I tend to agree that, as an ideal, this is where approximations can really bring clarity to the table.
However, in practice, most of the time when we approximate we end up dropping some of the things that we would like to retain.
In that sense, approximation is both a blessing and a curse: it simplifies the picture so we can better understand it, but we generally lose out on some of the physics we want to describe.
There is rarely a `happy medium' where approximation is purely a strength.

Having said that, though, I should also point out that these benefits and disadvantages of approximation are equally applicable to numerical methods.
If an approximate calculation matches experiment or an \emph{ab initio} simulation in full rigour, then we can be confident that we have captured the physics.

\speak{Analycia}
Wait! \direct{eagerly} I think I see what you mean---there is no demand that this approximate calculation needs to be analytical?

\speak{Numerio}
Yes.
Moreover, approximation also mitigates some of the problems in numerical methods regarding the complexity of interpretation. 
An overly-complex method may yield information that is simply too fine-grained to be analysed easily, %
but an approximate numerical method can strip away much of that complexity by focusing on a suitable subspace of solutions, and if it correctly matches the rigorous outcome then we can be confident that we understand the physics.

\end{dialogue}

\narrator{%
\textsc{Numerio} and \textsc{Analycia} agree to treat approximation as both a strength and a weakness, and as a vital way to obtain new perspectives on physics,
and move on to the question of modularity in \emph{ab initio} methods.
}

\subsection{Modularity in {\em ab initio} methods}
\label{subsec:modularity}

\begin{dialogue}

\speak{Numerio}
It is often thought that {\em ab initio} methods do not provide the level of modularity that analytical methods can. However, even if the solutions provided by {\em ab initio} methods are numerical, the Hamiltonian is typically comprised of a set of analytical terms. By switching these terms on and off, we can gain insights into their role in a particular aspect of the physics in question. 

To give one example of this, in Fig.~\ref{fig:adc1} I show an intensity scan of the degree of coherence of the remaining ion in strong-field ionisation of CO$_2$. To gain physical insight, we can deactivate a number of interactions, and then compare the result to the `true' coherence.
This comparison shows how the interplay of different mechanisms contributes, in a non-trivial way, to the total coherence.

\begin{figure}
    \centering
    \includegraphics[width=\columnwidth, trim={0 0 4cm 0}]{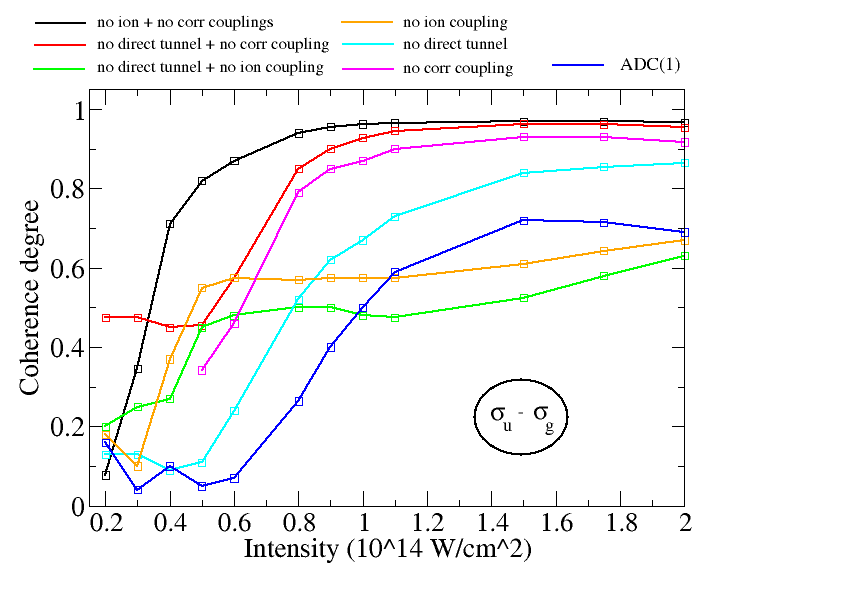}
    \caption{
    Modularity in \textit{ab initio} calculations: the quantum coherence between the $\sigma_{g}$ and 
    $\sigma_{u}$ ionic states of CO$_2$, as a function of laser intensity, can be examined by switching different couplings on and off, providing a valuable window into which effects are most essential.
    Plotted from unpublished data obtained during the initial phase of the research reported in~\cite{ruberti2018}.
    }
    \label{fig:adc1}
\end{figure}

\speak{Analycia}
This is certainly a good demonstration of modularity within an {\em ab initio} method! 
However, would you say this is a typical example? 
I imagine that many \emph{ab initio} methods would struggle to match this level of modularity. 
In this case, each individual calculation should come at a reasonably small cost in terms of computational time and resource, so that many calculations can be carried out. 
For some methods, the massive scale of individual calculations means that this level of modularity cannot be afforded.

More broadly, in numerical methods you cannot always do this type of switching procedure, particularly when it comes to spatial or momentum interference patterns~\cite{battle2, Amini2020}. 
It is extremely rare to come across numerical methods that are able to split between these, and provide clear assignments to the different channels that are interfering. 
So sometimes, yes, you can switch interactions on and off and assign things in a modular way, but \emph{ab initio} methods are often limited in the degree to which they can do this.

\speak{Numerio}
Yes, the degree of modularity often depends on the problem at hand. 
Activating and deactivating Hamiltonian terms provides insight in certain problems, but others will not be aided by this procedure. 
I suppose the ideal situation would be to have a method which can solve a given problem using reasonable computational resources, while keeping enough modularity to provide the required physical understanding. 

In that regard, one of the strongest tools is the use of approximations specifically tailored to the situation---which is one clear instance of approximation being a strength, as we have just agreed.

\end{dialogue}

\subsection{Are \emph{both} analytical and numerical methods required in scientific discovery?}

\narrator{\textsc{Numerio} and \textsc{Analycia}, agreeing that analytical and \emph{ab initio} methods are not always used in equal measure, turn to discuss the impact that this has on knowledge and discovery.}

\begin{dialogue}
\speak{Numerio} 
We have presented a case for analytical and numerical methods working best for scientific discovery when they are used in equal measure. However, is this always necessary? 
Take Fig.~\ref{fig:strelkov-TDSE}, where the analytical model matches experiment just as well as the TDSE model.
My natural inclination in a case like this would be to carry out a large, multi-electron calculation for this problem---but since the analytical model has described the experiment so well, would it be worthwhile?
In some cases, like this one, analytical methods can stand on their own, while in other cases it will not get you very far and you need to really crank the handle of big codes.

\speak{Analycia}
I agree---often one method will dominate, and it may be because it works better or it may be due to historical reasons.
However, while you can prove wrong a model when its results do not match experiments, it does not work the other way around: you can never prove that a physical model is right. Therefore the agreement of different theoretical approaches is all the more precious in that regards. 
In addition to what you said, analytical and \emph{ab initio} methods are two different powerful tools, which lead to differences in our understanding and interpretations. 
In different situations one method is more useful than the other for advancing knowledge. 

\speak{Numerio}
Yes, the methods we use will affect our understanding, but maybe we should not be too hung up on this.
We should mostly be driven by moving between the discoveries of new knowledge.
For instance, when we explore different systems such as high-harmonic generation in liquids
(e.g. \cite{Flettner2003, Luu2018})
or in solids 
(e.g. \cite{Ghimire2011, Vampa2017})%
, we start from the knowledge we had in the gas phase and push its limits and extend this knowledge.
Whether this knowledge originates from analytical, \textit{ab initio}, or experimental studies is ultimately not so important.

\speak{Analycia}
I can see what you are getting at, but in practice we cannot ignore biases different methods imbue in our knowledge. 
Let us see what our community thinks about this?

\direct{
The audience responses to this question are presented as poll 3 in Table~\ref{tab:polls}. 
}

\speak{Numerio}
It seems it is a mixed bag, with most hedging their bets in how often each method should be used.

\speak{Analycia}
We should take this result with a pinch of salt but perhaps we can agree this means it is very contextual. Physical processes we study should be attacked by exploring all of the approaches we have at hand. 

\speak{Numerio}
Yes, but also we should prioritise these methods by their range of applicability as well as the level of insight they elucidate.

\end{dialogue}

\narrator{%
Our combatants decide that they will each include more methods in their arsenal, as well as working together, to aid the process of scientific discovery. 
However, \textsc{Numerio} still has one final bone to pick.
}

\begin{table*}[t]
\centering%
\definecolor{headercellbackground}{rgb}{0.92,0.95,1}%
\definecolor{lg}{gray}{0.97}%
\definecolor{dg}{gray}{0.65}%
{\renewcommand{\arraystretch}{1.2}%
\setlength{\tabcolsep}{4.5pt}%
\begin{tabular}{llllllllllr}
\cellcolor{headercellbackground}
1 &
\multicolumn{10}{p{0.9\textwidth}}{
  \cellcolor{headercellbackground}%
  \rule{0pt}{1.2em}%
  The distinction of methods that provide quantitative vs qualitative insights is more useful in practice than the analytical and \textit{ab initio} distinction?%
   \protect
  \hfill
  (\href{https://twitter.com/quantumbattles/status/1279011339760001027}{Twitter~link})%
  }
\\ 
\cellcolor{lg}
& \cellcolor{lg} Agree & \cellcolor{lg} 84\%
& Disagree & 16\%
& \cellcolor{lg} & \cellcolor{lg} 
& &
& \cellcolor{lg} Sample size & \cellcolor{lg} 38
\\
\cellcolor{lg}
& \cellcolor{lg} Agree & \cellcolor{lg} 100\%
& Disagree & 0\%
& \cellcolor{lg} & \cellcolor{lg} 
& &
& \cellcolor{lg} Sample size & \cellcolor{lg} 10
\\
\cellcolor{headercellbackground}
2 &
\multicolumn{10}{p{0.9\textwidth}}{
  \cellcolor{headercellbackground}%
  \rule{0pt}{1.2em}%
  Would you feel more confident in planning an experiment based on guidance from analytical theory, \textit{ab initio} simulations or a hybrid model?%
  \hfill
  (\href{https://twitter.com/quantumbattles/status/1279014734004719616}{Twitter~link})%
  }
\\ 
\cellcolor{lg}
& \cellcolor{lg} Analytical & \cellcolor{lg} 19\%
& \textit{ab initio} & 38\%
& \cellcolor{lg} hybrid & \cellcolor{lg} 43\%
& &
& \cellcolor{lg} Sample size & \cellcolor{lg} 42
\\
\cellcolor{lg}
& \cellcolor{lg} Analytical & \cellcolor{lg} 29\%
& \textit{ab initio} & 57\%
& \cellcolor{lg} hybrid & \cellcolor{lg} 14\%
& &
& \cellcolor{lg} Sample size & \cellcolor{lg} 7
\\
\cellcolor{headercellbackground}
3 &
\multicolumn{10}{p{0.9\textwidth}}{
  \cellcolor{headercellbackground}%
  \rule{0pt}{1.2em}%
  Combining analytical, numerical, \textit{ab initio} and experimental methods in equal measure is the best route to discovery?%
  \hfill
  (\href{https://twitter.com/quantumbattles/status/1279017380488806402}{Twitter~link})%
  }
\\
\cellcolor{lg}
& \cellcolor{lg} Always & \cellcolor{lg} 15\%
& Mostly & 41\%
& \cellcolor{lg} Sometimes & \cellcolor{lg} 44\%
& Never & 0\%
& \cellcolor{lg} Sample size & \cellcolor{lg} 46
\\
\cellcolor{lg}
& \cellcolor{lg} Always & \cellcolor{lg} 18\%
& Mostly & 36\%
& \cellcolor{lg} Sometimes & \cellcolor{lg} 36\%
& Never & 9\%
& \cellcolor{lg} Sample size & \cellcolor{lg} 11
\\
\cellcolor{headercellbackground}
4 &
\multicolumn{10}{p{0.9\textwidth}}{
  \cellcolor{headercellbackground}%
  \rule{0pt}{1.2em}%
  What do you agree with more: \newline
  (1) Computing (and quantum computing power) will advance so much we will not need analytical methods anymore. \newline
  (2) We do not actually really need such computational power to understand physics, as our understanding is more closely represented by analytical models.%
  \hfill
  (\href{https://twitter.com/quantumbattles/status/1279022555265937408}{Twitter~link})%
  }
\\
\cellcolor{lg}
& \cellcolor{lg} I agree with (1) & \cellcolor{lg} 11\%
& I agree with (2) & 34\%
& \cellcolor{lg} I agree with neither & \cellcolor{lg} 55\%
& & 
& \cellcolor{lg} Sample size & \cellcolor{lg} 43
\\
\cellcolor{lg}
& \cellcolor{lg} I agree with (1) & \cellcolor{lg} 50\%
& I agree with (2) & 50\%
& \cellcolor{lg} I agree with neither & \cellcolor{lg} 0\%
& & 
& \cellcolor{lg} Sample size & \cellcolor{lg} 2
\end{tabular}
}
\caption{
  Audience polls taken over the Zoom (upper rows) and Twitter (lower rows) platform during the presentation.%
  \protect\footnotemark
  }
\label{tab:polls}
\end{table*}

\footnotetext{%
As part of the audience response to this poll, J.~Tennyson argued that this is a rhetorical question with the wording placing the velvet on the answer `yes', which should be considered when interpreting the poll results.
}


%

\subsection{The role of increasing computational power}


\narrator{%
The consequences of increasing computational power are a common theme in the development of modern physics mentioned time and time again.
\textsc{Numerio} turns to its role in attoscience.
}

\begin{dialogue}

\speak{Numerio}
One aspect that was very apparent as we looked at the evolution of numerical and \emph{ab initio} methods in attoscience is that, even given the considerable challenges initially faced by the field, these methods have achieved many tasks that would have seemed completely impossible even a scant few years ago.

Going out on a limb, I would even claim that these improvements will continue and accelerate, particularly once quantum computers become available, and that these advancements will drastically reduce the need for analytical methods, or even---des\-pite their advantages, which we discussed earlier--- eliminate it altogether.
And, I wonder, does our community agree with this?

\direct{
The response to the poll is shown in poll 4, on Table~\ref{tab:polls}.
}

\speak{Analycia}
I find it quite interesting that you should use a phrasing of the form `computing power will make analytical theory obsolete'---because of how \emph{old} that idea is.
That concept dates back a full six decades~\cite{berry2007}, to when electronic computers were first being developed in the 1960s (to replace \emph{human} computers).
Within that context, it is understandable that people got the impression that analytical theory---with its emphasis on special functions, integral transforms and asymptotic methods---would be displaced by raw computation.

However, over the past sixty years, time and time again the facts have demonstrated the opposite: we now place a \emph{higher} value on special functions and asymptotic methods than we did back then.
Of course, it is possible that at least some of the analytical methods of attoscience will be displaced by raw simulations, at least of the single-electron TDSE, but whenever this narrative starts to look appealing, it is important to take the long view and keep this historical context in mind.

\end{dialogue}

\subsection{Audience Questions and Comments}
\label{sec:audience-questions}

Over the course of the panel discussion~\cite{battle}, questions and comments were raised by the audience which helped challenge and develop the arguments being fielded by the combatants.  
We present them here, voicing our answers through \textsc{Numerio} and \textsc{Analycia}, and referencing answers already given above.

\begin{dialogue}
\speak{\bfseries Reinhard D\"orner} 
\textbf{Is our field really after discoveries? Is it not more about finding and solving interesting puzzles?}
    
\direct{This question was motivated by the distinction made by Thomas Kuhn in his famous book \emph{The structure of scientific revolutions} {\rm \cite{Kuhn2012}}. Therein, he argued that the times where the most progress is steadily made are times of `normal science', where what scientists do is best described as solving riddles with the tools of the paradigm they are working in~\cite{Doerner2020}.}
    
\speak{Numerio}
I agree that we are not looking for new fundamental laws. 
Here it is instructive to connect back to the definition of `\emph{ab initio}', particularly to remind ourselves that we have fixed the fundamental, theoretical `reference frame'. 
In attosecond physics, we are not yet looking for new fundamental laws: we already have established fundamental laws, the `rules of the game', and we are looking for new solutions to the fundamental quantum-mechanical equations of motion. 
The space of solutions is potentially infinite, as is the amount of new physical phenomena yet to be described.
In our case, we are interested in understanding the physics of atoms and molecules, driven by light-matter interactions, in new and unexplored regimes---%
and I would agree that this can be described as finding and solving interesting puzzles.

\speak{Analycia} 
I would disagree: the fact that something is not `fundamental' does not stop it from being a discovery.
If nothing else, that viewpoint completely disregards discoveries made in other sciences which are not `fundamental'.
I would say that it is still discovery if it is new knowledge.

\speak{Numerio} 
Perhaps this is a matter of terminology: to me, speaking of `discovery' entails finding new laws or entirely novel particles or dimensions, which do not occur in attoscience.
In our domain, the basic rules are already set, and we are solving a puzzle which is as interesting as it is difficult. 
There are many different ways of arranging the pieces of this puzzle, with each one representing, in principle, a different physical scenario that we can tackle with our theoretical methods, be they \emph{ab initio}, analytical, or hybrid.
That said, we know only a limited set of such scenarios and I agree that, when we find a new one, it can also be seen as a discovery.

\speak{Analycia} 
Yes, I see what you mean---but there is not always such a clear split between rules and scenarios, i.e., between laws of physics and their solutions.
There is a level where we only have the fundamental laws, but there are also higher levels of understanding and abstraction where the behaviour of a set of solutions can become a `rule', a law of physics, in itself.
And, I would argue, our role in attoscience is to discover these laws.
However, I do agree that our work mostly takes place within the fixed paradigm of a single set of fundamental laws.

\end{dialogue}

\narrator{The conclusion that \textsc{Analycia} and \textsc{Numerio} take away from this is that a solution to a problem may still be interesting and useful, irrespective of whether it is called a discovery.}


\begin{dialogue}
\speak{\bfseries Tom Meltzer} 
\textbf{Do you think the range of applicability of a model is not more important than whether it's \emph{ab initio}, numerical, analytical, semiclassical and so on?}

\speak{Numerio}
I agree that this is an important aspect of any method.
That said, I would also say that an even more important lens is whether the model gives us insights of a qualitative or quantitative nature, as we argued in more detail in Section~\ref{sec:quant-qual}.

\speak{Analycia}
I have a similar view on this.
Here it is important to remark that one of the big reasons why, I would argue, we should move away from the `analytical-versus-\emph{ab initio}' view is that, ultimately, it is impossible to have a method which is truly \emph{ab initio}.
We discussed some of this in detail in Section~\ref{sec:ChoiceOfBasisSet}, regarding the impact of the choice of basis set has on a method: to the extent that we must supply physics insights into that choice, it takes the method away from the \emph{ab initio} ideal.

However, I would go beyond that, since there are many other ways in which the base assumptions of how we phrase the problem---which often go unquestioned---can affect the physics.
The most obvious example in attoscience is macroscopic effects coming from the propagation of light inside our sample, 
but there are also other, more esoteric aspects---%
say, the appearance of collective quantum effects such as superfluorescence~\cite{Mercadier2019, Rohringer2020qbattles}, 
or effects coming from field quantisation~\cite{Lewenstein2020quantum}---%
which are ruled out by the basic framing, and this takes us away from ever reaching the \emph{ab initio} ideal.

\speak{Numerio}
Those are fair points, but there is also a danger of throwing out the baby with the bathwater here, in discarding the valuable work done in pursuit of that ideal.
In that regard, I would argue that a better definition for `\emph{ab initio}' could be 
`models with approximations that have well-defined error bounds for an explicit parameter range, such that any neglected physics will lie within these bounds'.
We know what physics gets neglected, and we should be able to quantify those well enough to know they are not relevant (as well as the types of questions that become inaccessible); for any additional sources of error, like the choice of basis set, the error must be quantifiable.

Here is, I would say, where the range of applicability of the model is most important, as it dictates whether those sources of error are quantifiable and negligible---what one could call `allowed' approximations---or into a regime with unquantified approximations.
This is then a major component that determines where we can place our method on the qualitative-versus-quantitative spectrum.

\end{dialogue}

\narrator{In summary, our combatants agree the range of applicability is a central aspect to consider, which is essential for reshaping ideas on what is `\emph{ab initio}'.
However, they also assert that the characterisation of whether a model provides qualitative or quantitative insights is the most important feature to consider. }


\begin{dialogue}

\speak{\bfseries (Anonymous)} \textbf{Can the single-configuration time-dependent Hartree-Fock method be used effectively to study multi-electron effects on atomic and molecular systems?}

\speak{Numerio} 
The method you mention (TDHF) can definitely be used to describe multi-electron effects. It is equivalent, in its linear-response reformulation, to the Random Phase Approximation with exchange (RPAX) method, which has been widely used (mainly by the condensed-matter community) and which can provide accurate molecular excitation energies and transition moments. Using TDHF in its full time-dependent character to study non-perturbative dynamics (beyond a perturbative approach) is certainly possible. 

\speak{Analycia}
That sounds quite complicated for limited gain. Is it really worth it?

\speak{Numerio}
This is a good method, but we also have available several multi-configuration versions, including MCTDHF and TD CAS- and RAS-SCF, which are generally more effective.
This makes TDHF, in my opinion, only a computationally-cheaper alternative which should be considered for large systems that are not amenable to the full multi-configurational treatment. 

\end{dialogue}


\begin{dialogue}

\speak{\bfseries Jens Biegert} \textbf{An experiment is like doing an \emph{ab initio} simulation in the sense that one can change the boundary conditions, but it does not necessarily allow you to disentangle what happens. However, analytic, semi-analytical and hybrid methods do allow insight.}

\speak{Numerio} 
It is true that \emph{ab initio} calculations are often characterised as the theoretical analogue of experiments, and that analytical methods drill down on the insightful details. 
However, in my experience, I feel that this is a mischaracterisation, as I am aware of many instances where \emph{ab initio} methods were able to disentangle a variety of physical interactions, by virtue of their modular properties.

\speak{Analycia} Oh, really? I would be interested to hear more, as this is an area where I always felt we analyticists held an advantage.

\end{dialogue}

\narrator{Given the level of interest in this topic, the combatants broadened its scope into the discussion given in Sec.~\ref{subsec:modularity}.}

\section*{Author contributions}
All the authors were involved in the preparation of the manuscript.
All the authors have read and approved the final manuscript.
\section*{Author ORCID iDs}

\begin{enumerate}[
label={\smash{\includegraphics[width=8pt]{figures/ORCID-icon.png}}}
]
\item Gregory S.\ J.\ Armstrong: \linkedorcid{0000-0001-5949-2626}
\item Margarita A.\ Khokhlova: \linkedorcid{0000-0002-5687-487X}
\item Andrew S.\ Maxwell: \linkedorcid{0000-0002-6503-4661}
\item Emilio Pisanty: \linkedorcid{0000-0003-0598-8524}
\item Marco Ruberti: \linkedorcid{0000-0003-0424-3643}
\end{enumerate}

\section*{Acknowledgements}
We (the authors, and our combatants \textsc{Analycia} and \textsc{Numerio}) are deeply grateful to the organisers of the \textit{Quantum Battles in Attoscience} conference as well as to the audience members for their active participation, and we are especially grateful to the Battle's `referee', Stefanie Gr\"afe, for her insightful and even-handed moderation during the panel discussion.
We also thank S.D.~Bartlett, T.~Rudolph and R.W.~Spekkens~\cite{Bartlett2006} as well as G.~Galilei~\cite{Galilei1632} for inspiration for the format of this paper.

GSJA acknowledges funding from the UK Engineering and Physical Sciences Research Council (EPSRC) under grant EP/T019530/1.
MAK acknowledges funding from the Alexander von Humboldt Foundation.
ASM acknowledges grant EP/P510270/1 funded by the UK EPSRC.
ASM and EP acknowledge support from ERC AdG NOQIA, Spanish Ministry of Economy and Competitiveness (``Severo Ochoa'' program for Centres of Excellence in R\&D (CEX\allowbreak{}2019-000910-S), Plan National FIDEUA PID2019-106901\allowbreak{}GB-I00/10.13039/501100011\allowbreak{}033, FPI), Fundació Privada Cellex, Fundació Mir-Puig, and from Generalitat de Catalunya (AGAUR Grant No.\ 2017 SGR 1341, CERCA program, QuantumCAT \_U16-011424, co-funded by the ERDF Operational Program of Catalonia 2014-2020), MINECO-EU QUANTERA MAQS (funded by State Research Agency (AEI) PCI2019-111828-2/10.\allowbreak{}13039/\allowbreak{}501100011033), EU Horizon 2020 FET-OPEN OPTOLogic (Grant No 899794), Marie Sklodowska-Curie grant STRETCH No.\ 101029393, and the National Science Centre, Poland-Symfonia Grant No.\ 2016/20/W/ST4/00314.
MR acknowledges funding from the EPSRC/DSTL MURI grant EP/N018680/1.

%
\bibliographystyle{arthur}
%
\interlinepenalty=10000
\let\oldbibitem\bibitem
\def\bibitem{\vfill\oldbibitem}
\bibliography{bibliographyqb}

\begin{thebibliography}{100}
\providecommand{\url}[1]{\texttt{#1}}
\providecommand{\urlprefix}{URL }
\expandafter\ifx\csname urlstyle\endcsname\relax
  \providecommand{\doi}[1]{doi:\discretionary{}{}{}#1}\else
  \providecommand{\doi}{doi:\discretionary{}{}{}\begingroup
  \urlstyle{rm}\Url}\fi
\providecommand{\selectlanguage}[1]{\relax}
\providecommand{\bibAnnoteFile}[1]{%
  \IfFileExists{#1}{\begin{quotation}\noindent\textsc{Key:} #1\\
  \textsc{Annotation:}\ \input{#1}\end{quotation}}{}}
\providecommand{\bibAnnote}[2]{%
  \begin{quotation}\noindent\textsc{Key:} #1\\
  \textsc{Annotation:}\ #2\end{quotation}}
\providecommand{\eprint}[2][]{\url{#2}}

\bibitem{brabec2000}
{T.~Brabec and F.~Krausz}.
\newblock Intense few-cycle laser fields: Frontiers of nonlinear
  optics\href{http://doi.org/10.1103/RevModPhys.72.545}{.
\newblock \emph{Rev. Mod. Phys.} \textbf{72} no.~2, pp. 545--591 (2000)}.
\bibAnnoteFile{brabec2000}

\bibitem{krausz2009}
{F.~Krausz and M.~Ivanov}.
\newblock Attosecond physics\href{http://doi.org/10.1103/RevModPhys.81.163}{.
\newblock \emph{Rev. Mod. Phys.} \textbf{81} no.~1, pp. 163--234 (2009)}.
\newblock
  \href{https://nrc-publications.canada.ca/eng/view/ft/?id=1245a958-9c93-4116-bfdb-f447e8a53c48}{NRC
  eprint}.
\bibAnnoteFile{krausz2009}

\bibitem{calegari2016}
{F.~Calegari, G.~Sansone, S.~Stagira, C.~Vozzi and M.~Nisoli}.
\newblock Advances in attosecond
  science\href{http://doi.org/10.1088/0953-4075/49/6/062001}{.
\newblock \emph{J. Phys. B: At. Mol. Opt. Phys.} \textbf{49} no.~6, p. 062001
  (2016)}.
\bibAnnoteFile{calegari2016}

\bibitem{Goulielmakis2010Aug}
{E.~Goulielmakis, Z.-H. Loh, A.~Wirth} et~al.
\newblock Real-time observation of valence electron
  motion\href{http://doi.org/10.1038/nature09212}{.
\newblock \emph{Nature} \textbf{466}, pp. 739--743 (2010)}.
\newblock
  \href{http://www.mpq-ksu-lmu.org/fileadmin/user_upload/04_Publications/paper_Nature_Y2010_M08_D05_09212.pdf}{Author
  eprint}.
\bibAnnoteFile{Goulielmakis2010Aug}

\bibitem{Calegari2014}
{F.~Calegari, D.~Ayuso, A.~Trabattoni} et~al.
\newblock Ultrafast electron dynamics in phenylalanine initiated by attosecond
  pulses\href{http://doi.org/10.1126/science.1254061}{.
\newblock \emph{Science} \textbf{346} no. 6207, pp. 336--339 (2014)}.
\newblock \href{http://hdl.handle.net/10486/679967}{UAM eprint}.
\bibAnnoteFile{Calegari2014}

\bibitem{Goulielmakis2004direct}
{E.~Goulielmakis, M.~Uiberacker, R.~Kienberger} et~al.
\newblock Direct measurement of light
  waves\href{http://doi.org/10.1126/science.1100866}{.
\newblock \emph{Science} \textbf{305} no. 5688, pp. 1267--1269 (2004)}.
\newblock
  \href{http://www.attoworld.de/kienberger-group/Documents/papers_kienberger/Science305p1267_2004.pdf}{Author
  eprint}.
\bibAnnoteFile{Goulielmakis2004direct}

\bibitem{battle}
{G.~S.~J. Armstrong, M.~A. Khokhlova, M.~Labeye} et~al.
\newblock {Quantum Battle 3 -- Numerical vs Analytical Methods}.
\newblock \emph{\href{https://www.quantumbattles.com/}{Quantum Battles in
  Attoscience}} online conference (1-3 July 2020).
\newblock
  \href{https://youtu.be/VJnFfHVDym4}{youtu.be/\allowbreak{}VJnFf\allowbreak{}HVD\allowbreak{}ym4}.
\bibAnnoteFile{battle}

\bibitem{Cohen-Tannoudji1997}
{C.~Cohen-Tannoudji, J.~Dupont-Roc and G.~Grynberg}.
\newblock \emph{Photons and Atoms: Introduction to Quantum Electrodynamics}
  (Wiley, 1997).
\bibAnnoteFile{Cohen-Tannoudji1997}

\bibitem{dirac1929}
{P.~A.~M. Dirac}.
\newblock Quantum mechanics of many-electron
  systems\href{http://doi.org/10.1098/rspa.1929.0094}{.
\newblock \emph{Proc. Roy. Soc. Lond., Ser. A} \textbf{123} no. 792, pp.
  714--733 (1929)}.
\newblock \href{http://www.jstor.org/stable/95222}{JSTOR:95222}.
\bibAnnoteFile{dirac1929}

\bibitem{Lewenstein2020quantum}
{M.~Lewenstein, M.~F. Ciappina, E.~Pisanty} et~al.
\newblock The quantum nature of light in high harmonic generation.
\newblock
  \href{https://arxiv.org/abs/2008.10221}{arXiv:\allowbreak{}2008.\allowbreak{}10221},
  (2020).
\bibAnnoteFile{Lewenstein2020quantum}

\bibitem{scrinzi2014}
{A.~Scrinzi}.
\newblock Time-dependent {Schrödinger} equation{.
\newblock In {T.~Schultz and M.~Vrak\-king} (eds.), \emph{Attosecond and {XUV}
  Physics: Ultrafast Dynamics and Spectroscopy}, pp. 257--292 (Wiley-{VCH},
  Weinheim, 2014)}.
\bibAnnoteFile{scrinzi2014}

\bibitem{Bauer2006}
{D.~Bauer and P.~Koval}.
\newblock \textsc{Qprop}: A {Schr\"{o}dinger}-solver for intense laser--atom
  interaction\href{http://doi.org/10.1016/j.cpc.2005.11.001}{.
\newblock \emph{Comput. Phys. Commun.} \textbf{174} no.~5, pp. 396--421
  (2006)}.
\newblock \href{https://arxiv.org/abs/physics/0507089}{arXiv:physics/0507089}.
\bibAnnoteFile{Bauer2006}

\bibitem{Tulsky2020}
{V.~Tulsky and D.~Bauer}.
\newblock \textsc{Qprop} with faster calculation of photoelectron
  spectra\href{http://doi.org/10.1016/j.cpc.2019.107098}{.
\newblock \emph{Comput. Phys. Commun.} \textbf{251}, p. 107098 (2020)}.
\newblock \href{https://arxiv.org/abs/1907.08595}{arXiv:1907.08595}.
\bibAnnoteFile{Tulsky2020}

\bibitem{Patchkovskii2016}
{S.~Patchkovskii and H.~G. Muller}.
\newblock Simple, accurate, and efficient implementation of 1-electron atomic
  time-dependent {Schr\"{o}dinger} equation in spherical
  coordinates\href{http://doi.org/10.1016/j.cpc.2015.10.014}{.
\newblock \emph{Comput. Phys. Commun.} \textbf{199}, pp. 153--169 (2016)}.
\newblock \href{https://core.ac.uk/display/82248458}{CORE eprint}.
\bibAnnoteFile{Patchkovskii2016}

\bibitem{Fritzsche2019}
{S.~Fritzsche}.
\newblock A fresh computational approach to atomic structures, processes and
  cascades\href{http://doi.org/10.1016/j.cpc.2019.01.012}{.
\newblock \emph{Comput. Phys. Commun.} \textbf{240}, pp. 1--14 (2019)}.
\newblock \href{https://repository.gsi.de/record/219896}{GSI eprint}.
\bibAnnoteFile{Fritzsche2019}

\bibitem{Ullrich2012}
{C.~Ullrich}.
\newblock \emph{Time-dependent density functional theory: concepts and
  applications} (Oxford University Press, 2012).
\bibAnnoteFile{Ullrich2012}

\bibitem{Kohn1965}
{W.~Kohn and L.~J. Sham}.
\newblock Self-consistent equations including exchange and correlation
  effects\href{http://doi.org/10.1103/PhysRev.140.A1133}{.
\newblock \emph{Phys. Rev.} \textbf{140}, pp. A1133--A1138 (1965)}.
\bibAnnoteFile{Kohn1965}

\bibitem{DeGiovannini2013}
{U.~De~Giovannini, G.~Brunetto, A.~Castro, J.~Walkenhorst and A.~Rubio}.
\newblock Simulating pump–probe photoelectron and absorption spectroscopy on
  the attosecond timescale with time‐dependent density functional
  theory\href{http://doi.org/10.1002/cphc.201201007}{.
\newblock \emph{ChemPhysChem} \textbf{14}, pp. 1363--1376 (2013)}.
\newblock \href{https://arxiv.org/abs/1301.1958}{arXiv:1301.1958}.
\bibAnnoteFile{DeGiovannini2013}

\bibitem{Wopperer2017}
{P.~Wopperer, U.~De~Giovannini and A.~Rubio}.
\newblock Efficient and accurate modeling of electron photoemission in
  nanostructures with tddft\href{http://doi.org/10.1140/epjb/e2017-70548-3}{.
\newblock \emph{Eur. Phys. J. B} \textbf{90}, p. 1307 (2017)}.
\newblock \href{https://arxiv.org/abs/1608.02818}{arXiv:1608.02818}.
\bibAnnoteFile{Wopperer2017}

\bibitem{DeGiovannini2018}
{U.~De~Giovannini and A.~Castro}.
\newblock Real-time and real-space time-dependent density-functional theory
  approach to attosecond
  dynamics\href{http://doi.org/10.1039/9781788012669-00424}{.
\newblock In {M.~Vrakking and F.~Lepine} (eds.), \emph{Attosecond Molecular
  Dynamics}, vol.~13 of \emph{Theoretical and Computational Chemistry series},
  pp. 424--461 (Royal Society Chemistry, Cambridge, 2018)}.
\bibAnnoteFile{DeGiovannini2018}

\bibitem{Lara-Astasio2018}
{M.~M.~Lara-Astiaso, M.~Galli, A.~Trabattoni} et~al.
\newblock Attosecond pump–\allowbreak{}probe spectroscopy of charge dynamics
  in tryptophan\href{http://doi.org/10.1021/acs.jpclett.8b01786}{.
\newblock \emph{J. Phys. Chem. Lett.} \textbf{9}, pp. 4570--4577 (2018)}.
\newblock \href{http://dx.doi.org/10.3204/PUBDB-2018-03687}{DESY eprint}.
\bibAnnoteFile{Lara-Astasio2018}

\bibitem{Tancogne2020}
{N.~Tancogne-Dejean, M.~J.~T. Oliveira, X.~Andrade} et~al.
\newblock Octopus, a computational framework for exploring light-driven
  phenomena and quantum dynamics in extended and finite
  systems\href{http://doi.org/10.1063/1.5142502}{.
\newblock \emph{J. Chem. Phys.} \textbf{152} no.~12, p. 124119 (2020)}.
\bibAnnoteFile{Tancogne2020}

\bibitem{Noda2019}
{M.~Noda, S.~A. Sato, Y.~Hirokawa} et~al.
\newblock {SALMON: Scalable Ab-initio Light--\allowbreak{}Matter simulator for
  Optics and Nanoscience}\href{http://doi.org/10.1016/j.cpc.2018.09.018}{.
\newblock \emph{Comput. Phys. Commun.} \textbf{235}, pp. 356--365 (2019)}.
\newblock \href{https://arxiv.org/abs/1804.01404}{arXiv:1804.01404}.
\bibAnnoteFile{Noda2019}

\bibitem{Apra2020}
{E.~Apr{\`{a}}, E.~J. Bylaska, W.~A. de~Jong} et~al.
\newblock {NWChem}: Past, present, and
  future\href{http://doi.org/10.1063/5.0004997}{.
\newblock \emph{J. Chem. Phys.} \textbf{152} no.~18, p. 184102 (2020)}.
\bibAnnoteFile{Apra2020}

\bibitem{Garcia2020Siesta}
{A.~Garc{\'{\i}}a, N.~Papior, A.~Akhtar} et~al.
\newblock \textsc{Siesta}: Recent developments and
  applications\href{http://doi.org/10.1063/5.0005077}{.
\newblock \emph{J. Chem. Phys.} \textbf{152} no.~20, p. 204108 (2020)}.
\newblock \href{https://arxiv.org/abs/2006.01270}{arXiv:2006.01270}.
\bibAnnoteFile{Garcia2020Siesta}

\bibitem{BIRS-CMO-Workshop}
{E.~Lorin, A.~Bandrauk and B.~{Sanders {\rm (eds.)}}}.
\newblock {BIRS-CMO Workshop `Mathematical and Numerical Methods for
  Time-Dependent Quantum Mechanics -- from Dynamics to Quantum Information'}.
\newblock Oaxaca de Ju\'arez, M\'exico, 13-18 August 2017. Video proceedings
  available at \url{https://www.birs.ca/events/2017/5-day-workshops/17w5010}.
\bibAnnoteFile{BIRS-CMO-Workshop}

\bibitem{Perfetto2018}
{E.~Perfetto and G.~Stefanucci}.
\newblock {CHEERS}: a tool for correlated hole-electron evolution from
  real-time simulations\href{http://doi.org/10.1088/1361-648X/aae675}{.
\newblock \emph{J. Phys: Condens. Matter} \textbf{30}, p. 465901 (2018)}.
\newblock \href{https://arxiv.org/abs/1810.03322}{arXiv:1810.03322}.
\bibAnnoteFile{Perfetto2018}

\bibitem{Perfetto2018JPCL}
{E.~Perfetto, D.~Sangalli, A.~Marini and G.~Stefanucci}.
\newblock Ultrafast charge migration in {XUV} photoexcited phenylalanine: A
  first-principles study based on real-time nonequilibrium {Green}'s
  functions\href{http://doi.org/10.1021/acs.jpclett.8b00025}{.
\newblock \emph{J. Phys. Chem. Lett.} \textbf{9}, pp. 1353--1358 (2018)}.
\newblock \href{https://arxiv.org/abs/1803.05681}{arXiv:1803.05681}.
\bibAnnoteFile{Perfetto2018JPCL}

\bibitem{szabo1996}
{A.~Szabo and N.~S. Ostlund}.
\newblock \emph{Modern quantum chemistry: Introduction to advanced electronic
  structure theory} (Dover Publications, Mineola, N.Y., 1996).
\bibAnnoteFile{szabo1996}

\bibitem{Schirmer2018}
{J.~Schirmer}.
\newblock \emph{Many-Body Methods for Atoms, Molecules and Clusters}, vol.~94
  of \emph{Lecture Notes in Chemistry} (Springer, 2018).
\bibAnnoteFile{Schirmer2018}

\bibitem{Evarestov2012}
{R.~Evarestov}.
\newblock \emph{Quantum Chemistry of Solids: LCAO Treatment of Crystals and
  Nanostructures}, vol. 153 of \emph{Springer Series in Solid-State Sciences}
  (Springer, 2012).
\bibAnnoteFile{Evarestov2012}

\bibitem{Ruberti2014JCP}
{M.~Ruberti, V.~Averbukh and P.~Decleva}.
\newblock {B}-spline algebraic diagrammatic construction: Application to
  photoionization cross-sections and high-order harmonic
  generation\href{http://doi.org/10.1063/1.4900444}{.
\newblock \emph{J. Chem. Phys.} \textbf{141}, p. 164126 (2014)}.
\newblock \href{http://hdl.handle.net/10044/1/21876}{ICL eprint}.
\bibAnnoteFile{Ruberti2014JCP}

\bibitem{Simpson2016}
{E.~Simpson, A.~Sanchez-Gonzalez, D.~Austin} et~al.
\newblock Polarisation response of delay dependent absorption modulation in
  strong field dressed helium atoms probed near
  threshold\href{http://doi.org/10.1088/1367-2630/18/8/083032}{.
\newblock \emph{New J. Phys.} \textbf{18}, p. 083032 (2016)}.
\bibAnnoteFile{Simpson2016}

\bibitem{Averbukh2018}
{V.~Averbukh and M.~Ruberti}.
\newblock First-principles many-electron dynamics using the {B}-spline
  algebraic diagrammatic construction
  approach\href{http://doi.org/10.1039/9781788012669-00068}{.
\newblock In {M.~Vrakking and F.~Lepine} (eds.), \emph{Attosecond Molecular
  Dynamics}, vol.~13 of \emph{Theoretical and Computational Chemistry series},
  pp. 68--102 (Royal Society Chemistry, Cambridge, 2018)}.
\bibAnnoteFile{Averbukh2018}

\bibitem{Ruberti2018PCCP}
{M.~Ruberti, P.~Decleva and V.~Averbukh}.
\newblock Multi-channel dynamics in high harmonic generation of aligned
  {CO}$_2$: \emph{ab initio} analysis with time-dependent {B}-spline algebraic
  diagrammatic construction\href{http://doi.org/10.1039/C7CP07849H}{.
\newblock \emph{Phys. Chem. Chem. Phys.} \textbf{20}, pp. 8311--8325 (2018)}.
\newblock \href{http://hdl.handle.net/10044/1/58557}{ICL eprint}.
\bibAnnoteFile{Ruberti2018PCCP}

\bibitem{ruberti2018}
{M.~Ruberti, P.~Decleva and V.~Averbukh}.
\newblock Full ab initio many-electron simulation of attosecond molecular
  pump--probe spectroscopy\href{http://doi.org/10.1021/acs.jctc.8b00479}{.
\newblock \emph{J. Chem. Theory Comput.} \textbf{14} no.~10, pp. 4991--5000
  (2018)}.
\bibAnnoteFile{ruberti2018}

\bibitem{Ruberti2019}
{M.~Ruberti}.
\newblock Restricted correlation space {B}-spline {ADC} approach to molecular
  ionization: Theory and applications to total photoionization
  cross-sections\href{http://doi.org/10.1021/acs.jctc.9b00288}{.
\newblock \emph{J. Chem. Theory Comput.} \textbf{15} no.~6, pp. 3635--3653
  (2019)}.
\bibAnnoteFile{Ruberti2019}

\bibitem{Ruberti2019PCCP}
{M.~Ruberti}.
\newblock Onset of ionic coherence and charge dynamics in attosecond molecular
  ionization\href{http://doi.org/10.1039/C9CP03074C}{.
\newblock \emph{Phys. Chem. Chem. Phys.} \textbf{21}, pp. 17584--17604 (2019)}.
\bibAnnoteFile{Ruberti2019PCCP}

\bibitem{Ruberti2021}
{M.~Ruberti}.
\newblock Quantum electronic coherences by attosecond transient absorption
  spectroscopy: \emph{ab initio} {B}-spline {RCS}-{ADC}
  study\href{http://doi.org/10.1039/D0FD00104J}{.
\newblock \emph{Faraday Discuss.} \textbf{228} no.~0, pp. 286--311 (2021)}.
\bibAnnoteFile{Ruberti2021}

\bibitem{schwickert2020}
{D.~Schwickert, M.~Ruberti, P.~Koloren{\v{c}}} et~al.
\newblock Electronic quantum coherence in glycine probed with femtosecond
  {X}-rays.
\newblock \href{https://arxiv.org/abs/2012.04852}{arXiv:2012.04852}, (2020).
\bibAnnoteFile{schwickert2020}

\bibitem{majety2015hacc}
{V.~P. Majety, A.~Zielinski and A.~Scrinzi}.
\newblock Photoionization of few electron systems: a hybrid coupled channels
  approach\href{http://doi.org/10.1088/1367-2630/17/6/063002}{.
\newblock \emph{New J. Phys.} \textbf{17} no.~6, p. 063002 (2015)}.
\bibAnnoteFile{majety2015hacc}

\bibitem{Majety2015PRL}
{V.~P. Majety and A.~Scrinzi}.
\newblock Dynamic exchange in the strong field ionization of
  molecules\href{http://doi.org/10.1103/PhysRevLett.115.103002}{.
\newblock \emph{Phys. Rev. Lett.} \textbf{115}, p. 103002 (2015)}.
\newblock \href{https://arxiv.org/abs/1505.03349}{arXiv:1505.03349}.
\bibAnnoteFile{Majety2015PRL}

\bibitem{Beck2000}
{M.~H. Beck, A.~J{\"{a}}ckle, G.~A. Worth and H.-D. Meyer}.
\newblock The multiconfiguration time-dependent {Hartree} {(MCTDH)} method: a
  highly efficient algorithm for propagating
  wavepackets\href{http://doi.org/10.1016/S0370-1573(99)00047-2}{.
\newblock \emph{Phys. Rep.} \textbf{324} no.~1, pp. 1--105 (2000)}.
\newblock \href{https://www.pci.uni-heidelberg.de/tc/usr/mctdh/lit/rev.pdf}{UH
  eprint}.
\bibAnnoteFile{Beck2000}

\bibitem{sukiasyan2009}
{S.~Sukiasyan, C.~McDonald, C.~Van~Vlack} et~al.
\newblock Correlated few-electron dynamics in intense laser
  fields\href{http://doi.org/10.1016/j.chemphys.2009.10.001}{.
\newblock \emph{Chem. Phys.} \textbf{366} no.~1, pp. 37--45 (2009)}.
\bibAnnoteFile{sukiasyan2009}

\bibitem{Marante2017}
{C.~Marante, M.~Klinker, I.~Corral} et~al.
\newblock Hybrid-basis close-coupling interface to quantum chemistry packages
  for the treatment of ionization
  problems\href{http://doi.org/10.1021/acs.jctc.6b00907}{.
\newblock \emph{J. Chem. Theory Comput.} \textbf{13} no.~2, p. 499 (2017)}.
\bibAnnoteFile{Marante2017}

\bibitem{Marante2017PRA}
{C.~Marante, M.~Klinker, T.~Kjellsson} et~al.
\newblock Photoionization using the xchem approach: Total and partial cross
  sections of ne and resonance parameters above the 2s2 2p5
  threshold\href{http://doi.org/10.1103/PhysRevA.96.022507}{.
\newblock \emph{Phys. Rev. A} \textbf{96}, p. 022507 (2017)}.
\bibAnnoteFile{Marante2017PRA}

\bibitem{Klinker2018}
{M.~Klinker, C.~Marante, L.~Argenti, J.~Gonzalez-Vazquez and F.~Martin}.
\newblock Electron correlation in the ionization continuum of molecules:
  Photoionization of {N}2 in the vicinity of the hopfield series of
  autoionizing states\href{http://doi.org/10.1021/acs.jpclett.7b03220}{.
\newblock \emph{J. Phys. Chem. Lett.} \textbf{9}, p. 756 (2018)}.
\newblock \href{http://hdl.handle.net/10486/685202}{UAM eprint}.
\bibAnnoteFile{Klinker2018}

\bibitem{Toffoli2016}
{D.~Toffoli and P.~Decleva}.
\newblock A multichannel least-squares {B}-spline approach to molecular
  photoionization: Theory, implementation, and applications within the
  configuration-interaction singles
  approximation\href{http://doi.org/10.1021/acs.jctc.6b00627}{.
\newblock \emph{J. Chem. Theory Comput.} \textbf{12}, p. 4996 (2016)}.
\bibAnnoteFile{Toffoli2016}

\bibitem{clarke2018}
{D.~D.~A. Clarke, G.~S.~J. Armstrong, A.~C. Brown and H.~W. van~der Hart}.
\newblock {$R$}-matrix-with-time-de\-pen\-den\-ce theory for ultrafast atomic
  processes in arbitrary light
  fields\href{http://doi.org/10.1103/PhysRevA.98.053442}{.
\newblock \emph{Phys. Rev. A} \textbf{98} no.~5, p. 053442 (2018)}.
\bibAnnoteFile{clarke2018}

\bibitem{brown2020}
{A.~C. Brown, G.~S.~J. Armstrong, J.~Benda} et~al.
\newblock {RMT}: {R}-matrix with time-\allowbreak{}de\-pen\-den\-ce. {Solving}
  the semi-relativistic, time-dependent {Schr\"odinger} equation for general,
  multielectron atoms and molecules in intense, ultrashort, arbitrarily
  polarized laser pulses\href{http://doi.org/10.1016/j.cpc.2019.107062}{.
\newblock \emph{Comput. Phys. Commun.} \textbf{250}, p. 107062 (2020)}.
\newblock
  \href{https://pureadmin.qub.ac.uk/ws/files/190837429/main_paper.pdf}{QUB
  eprint}.
\bibAnnoteFile{brown2020}

\bibitem{Benda2020}
{J.~Benda, J.~Gorfinkiel, Z.~Mašín} et~al.
\newblock Perturbative and nonperturbative photoionization of {H}$_2$ and
  {H$_2$O} using the molecular {$R$}-matrix-with-time
  method\href{http://doi.org/10.1103/PhysRevA.102.052826}{.
\newblock \emph{Phys. Rev. A} \textbf{102}, p. 052826 (2020)}.
\newblock
  \href{https://pure.qub.ac.uk/en/publications/perturbative-and-nonperturbative-photoionization-of-h2-and-h2o-us}{QUB
  eprint}.
\bibAnnoteFile{Benda2020}

\bibitem{Miyagi2013}
{H.~Miyagi and L.~B. Madsen}.
\newblock Time-dependent restricted-active-space self-consistent-field theory
  for laser-driven many-electron
  dynamics\href{http://doi.org/10.1103/PhysRevA.87.062511}{.
\newblock \emph{Phys. Rev. A} \textbf{87}, p. 062511 (2013)}.
\newblock \href{https://arxiv.org/abs/1304.5904}{arXiv:1304.5904}.
\bibAnnoteFile{Miyagi2013}

\bibitem{Miyagi2014}
{H.~Miyagi and L.~B. Madsen}.
\newblock Time-dependent restricted-active-space self-consistent-field theory
  for laser-driven many-electron dynamics. {II}. {E}xtended formulation and
  numerical analysis\href{http://doi.org/10.1103/PhysRevA.89.063416}{.
\newblock \emph{Phys. Rev. A} \textbf{89}, p. 063416 (2014)}.
\newblock \href{https://arxiv.org/abs/1405.5380}{arXiv:1405.5380}.
\bibAnnoteFile{Miyagi2014}

\bibitem{sato2013}
{T.~Sato and K.~L. Ishikawa}.
\newblock Time-dependent com\-plete-\allowbreak{}ac\-tive-space
  self-con\-sis\-tent-field method for multi\-electron dynamics in intense
  laser fields\href{http://doi.org/10.1103/PhysRevA.88.023402}{.
\newblock \emph{Phys. Rev. A} \textbf{88} no.~2, p. 023402 (2013)}.
\newblock \href{https://arxiv.org/abs/1304.5835}{arXiv:1304.5835}.
\bibAnnoteFile{sato2013}

\bibitem{Greenman2010}
{L.~Greenman, P.~J. Ho, S.~Pabst} et~al.
\newblock Implementation of the time-dependent configuration-interaction
  singles method for atomic strong-field
  processes\href{http://doi.org/10.1103/PhysRevA.82.023406}{.
\newblock \emph{Phys. Rev. A} \textbf{82}, p. 023406 (2010)}.
\newblock \href{http://dx.doi.org/10.3204/PHPPUBDB-22196}{DESY eprint}.
\bibAnnoteFile{Greenman2010}

\bibitem{NguyenDang2014}
{T.-T. Nguyen-Dang, E.~Couture-Bienvenue, J.~Viau-Trudel and A.~Sainjon}.
\newblock Time-dependent quantum chemistry of laser driven many-electron
  molecules\href{http://doi.org/10.1063/1.4904102}{.
\newblock \emph{J. Chem. Phys.} \textbf{141}, p. 244116 (2014)}.
\bibAnnoteFile{NguyenDang2014}

\bibitem{Vrakking2018}
{M.~Vrakking and F.~Lepine} (eds.).
\newblock \emph{Attosecond Molecular Dynamics}, vol.~13 of \emph{Theoretical
  and Computational Chemistry series} (Royal Society Chemistry, Cambridge,
  2018).
\bibAnnoteFile{Vrakking2018}

\bibitem{trecx}
{A.~Scrinzi}.
\newblock \urlprefix\url{https://trecx.physik.lmu.de}.
\bibAnnoteFile{trecx}

\bibitem{Carette2013}
{T.~Carette, J.~Dahlström, L.~Argenti and E.~Lindroth}.
\newblock Multiconfigurational {Hartree-Fock} close-coupling ansatz:
  Application to the argon photoionization cross section and
  delays\href{http://doi.org/10.1103/PhysRevA.87.023420}{.
\newblock \emph{Phys. Rev. A} \textbf{87}, p. 023420 (2013)}.
\newblock \href{http://hdl.handle.net/10486/668841}{UAM eprint}.
\bibAnnoteFile{Carette2013}

\bibitem{Sawada2016}
{R.~Sawada, T.~Sato and K.~L. Ishikawa}.
\newblock Implementation of the multiconfiguration time-dependent
  {Hatree}-{Fock} method for general molecules on a multiresolution {Cartesian}
  grid\href{http://doi.org/10.1103/PhysRevA.93.023434}{.
\newblock \emph{Phys. Rev. A} \textbf{93}, p. 023434 (2016)}.
\bibAnnoteFile{Sawada2016}

\bibitem{keldysh1965}
{L.~Keldysh}.
\newblock Ionization in the field of a strong electromagnetic wave{.
\newblock \emph{Sov. Phys. {JETP}} \textbf{20} no.~5, p. 1307 (1965)}.
\newblock
  [\href{http://www.jetp.ac.ru/cgi-bin/e/index/r/47/5/p1945?a=list}{\textit{Zh.
  Eksp. Teor. Fiz.} \textbf{47} no.~5, p.~1945 (1965)}].
\bibAnnoteFile{keldysh1965}

\bibitem{berry2007}
{M.~Berry}.
\newblock Why are special functions
  special?\href{http://doi.org/10.1063/1.1372098}{.
\newblock \emph{Phys. Today} \textbf{54} no.~4, p.~11 (2007)}.
\newblock
  \href{https://michaelberryphysics.files.wordpress.com/2013/07/berry326.pdf}{Author
  eprint}.
\bibAnnoteFile{berry2007}

\bibitem{Borwein2013closed}
{J.~Borwein and R.~Crandall}.
\newblock Closed forms: what they are and why we
  care\href{http://doi.org/10.1090/NOTI936}{.
\newblock \emph{Not. Am. Math. Soc.} \textbf{60}, pp. 50--65 (2013)}.
\bibAnnoteFile{Borwein2013closed}

\bibitem{Braak2011}
{D.~Braak}.
\newblock Integrability of the {Rabi}
  model\href{http://doi.org/10.1103/PhysRevLett.107.100401}{.
\newblock \emph{Phys. Rev. Lett.} \textbf{107} no.~10, p. 100401 (2011)}.
\newblock \href{https://arxiv.org/abs/1103.2461}{arXiv:1103.2461}.
\bibAnnoteFile{Braak2011}

\bibitem{andrews1999special}
{G.~Andrews, R.~Askey and R.~Roy}.
\newblock \emph{Special Functions}, vol.~71 of \emph{Encyclopedia of
  Mathematics and its Applications} (Cambridge University Press, Cambridge,
  1999).
\bibAnnoteFile{andrews1999special}

\bibitem{Faisal1973}
{F.~H.~M. Faisal}.
\newblock Multiple absorption of laser photons by
  atoms\href{http://doi.org/10.1088/0022-3700/6/4/011}{.
\newblock \emph{J. Phys. B: At. Mol. Phys.} \textbf{6} no.~4, p. L89 (1973)}.
\bibAnnoteFile{Faisal1973}

\bibitem{Reiss1980}
{H.~Reiss}.
\newblock Effect of an intense electromagnetic field on a weakly bound
  system\href{http://doi.org/10.1103/PhysRevA.22.1786ssss}{.
\newblock \emph{Phys. Rev. A} \textbf{22} no.~5, pp. 1786--1813 (1980)}.
\bibAnnoteFile{Reiss1980}

\bibitem{perelomov1966ionization}
{A.~Perelomov, V.~Popov and M.~Terent'ev}.
\newblock Ionization of atoms in an alternating electric field{.
\newblock \emph{Sov. Phys. JETP} \textbf{23} no.~5, p. 924 (1966)}.
\newblock
  [\href{http://www.jetp.ac.ru/cgi-bin/e/index/e/23/5/p924?a=list}{\textit{Zh.
  Eksp. Teor. Fiz.} \textbf{50} no.~5, p.~1393 (1966)}].
\bibAnnoteFile{perelomov1966ionization}

\bibitem{ammosov1986tunnel}
{M.~Ammosov, N.~Delone and V.~Krainov}.
\newblock Tunnel ionization of complex atoms and of atomic ions in an
  alternating electromagnetic field{.
\newblock \emph{Sov. Phys. JETP} \textbf{64} no.~6, p. 1191 (1986)}.
\newblock
  [\href{http://www.jetp.ac.ru/cgi-bin/e/index/r/91/6/p2008?a=list}{\textit{Zh.
  Eksp. Teor. Fiz.} \textbf{91} no.~6, p.~2008 (1986)}].
\bibAnnoteFile{ammosov1986tunnel}

\bibitem{Amini2019}
{K.~Amini, J.~Biegert, F.~Calegari} et~al.
\newblock Symphony on strong field
  approximation\href{http://doi.org/10.1088/1361-6633/ab2bb1}{.
\newblock \emph{Rep. Prog. Phys.} \textbf{82} no.~11, p. 116001 (2019)}.
\newblock \href{https://arxiv.org/abs/1812.11447}{arXiv:1812.11447}.
\bibAnnoteFile{Amini2019}

\bibitem{Lewenstein1994Mar}
{M.~Lewenstein, {\relax Ph}.~Balcou, M.~{\relax Yu}. Ivanov, A.~L{'}Huillier
  and P.~B. Corkum}.
\newblock Theory of high-harmonic generation by low-frequency laser
  fields\href{http://doi.org/10.1103/PhysRevA.49.2117}{.
\newblock \emph{Phys. Rev. A} \textbf{49} no.~3, pp. 2117--2132 (1994)}.
\bibAnnoteFile{Lewenstein1994Mar}

\bibitem{galstyan2016}
{A.~Galstyan, O.~Chuluunbaatar, A.~Hamido} et~al.
\newblock Reformulation of the strong-field approximation for light-matter
  interactions\href{http://doi.org/10.1103/PhysRevA.93.023422}{.
\newblock \emph{Phys. Rev. A} \textbf{93} no.~2, p. 023422 (2016)}.
\newblock
  \href{https://arxiv.org/abs/1512.00681}{arXiv:\allowbreak{}1512.\allowbreak{}00681}.
\bibAnnoteFile{galstyan2016}

\bibitem{Popruzhenko2014}
{S.~V. Popruzhenko}.
\newblock Keldysh theory of strong field ionization: history, applications,
  difficulties and
  perspectives\href{http://doi.org/10.1088/0953-4075/47/20/204001}{.
\newblock \emph{J. Phys. B: At. Mol. Opt. Phys.} \textbf{47} no.~20, p. 204001
  (2014)}.
\bibAnnoteFile{Popruzhenko2014}

\bibitem{salieres2001feynman}
{P.~Sali{\`e}res, B.~Carr{\'e}, L.~Le~D{\'e}roff} et~al.
\newblock Feynman's path-integral approach for intense-laser-atom
  interactions\href{http://doi.org/10.1126/science.108836}{.
\newblock \emph{Science} \textbf{292} no. 5518, pp. 902--905 (2001)}.
\bibAnnoteFile{salieres2001feynman}

\bibitem{ivanov2014}
{M.~Ivanov and O.~Smirnova}.
\newblock Multielectron high harmonic generation: simple man on a complex
  plane{.
\newblock In {T.~Schultz and M.~Vrakking} (eds.), \emph{Attosecond and {XUV}
  Physics: Ultrafast Dynamics and Spectroscopy}, pp. 201--256 (Wiley-{VCH},
  Weinheim, 2014)}.
\newblock \href{http://arxiv.org/abs/1304.2413}{arXiv:\allowbreak{}1304.2413}.
\bibAnnoteFile{ivanov2014}

\bibitem{torlina2012}
{L.~Torlina and O.~Smirnova}.
\newblock Time-dependent analytical {$R$}-matrix approach for strong-field
  dynamics. {I}. {One}-electron
  systems\href{http://doi.org/10.1103/PhysRevA.86.043408}{.
\newblock \emph{Phys. Rev. A} \textbf{86} no.~4, p. 043408 (2012)}.
\bibAnnoteFile{torlina2012}

\bibitem{Pisanty2016}
{E.~Pisanty~Alatorre}.
\newblock \emph{Electron dynamics in complex time and complex
  space}\href{http://hdl.handle.net/10044/1/43538}{.
\newblock {PhD} thesis, Imperial College London (2016)}.
\bibAnnoteFile{Pisanty2016}

\bibitem{Pisanty2017}
{E.~Pisanty and A.~Jim\'{e}nez-Gal\'{a}n}.
\newblock Strong-field approximation in a rotating frame: High-order harmonic
  emission from $p$ states in bicircular
  fields\href{http://doi.org/10.1103/PhysRevA.96.063401}{.
\newblock \emph{Phys. Rev. A} \textbf{96} no.~6, p. 063401 (2017)}.
\newblock \href{https://arxiv.org/abs/1709.00397}{arXiv:1709.00397}.
\bibAnnoteFile{Pisanty2017}

\bibitem{torlina2017}
{L.~Torlina and O.~Smirnova}.
\newblock Coulomb time delays in high harmonic
  generation\href{http://doi.org/10.1088/1367-2630/aa55ea}{.
\newblock \emph{New J. Phys.} \textbf{19} no.~2, p. 023012 (2017)}.
\bibAnnoteFile{torlina2017}

\bibitem{Smirnova2008}
{O.~Smirnova, M.~Spanner and M.~Ivanov}.
\newblock Analytical solutions for strong field-driven atomic and molecular
  one- and two-electron continua and applications to strong-field
  problems\href{http://doi.org/10.1103/PhysRevA.77.033407}{.
\newblock \emph{Phys. Rev. A} \textbf{77} no.~3, p. 033407 (2008)}.
\newblock
  \href{https://nrc-publications.canada.ca/eng/view/object/?id=4b63e08b-df87-4ff4-b4dd-41f1b411fdba}{NRC
  eprint}.
\bibAnnoteFile{Smirnova2008}

\bibitem{einziger1982}
{P.~Einziger and L.~Felsen}.
\newblock Evanescent waves and complex
  rays\href{http://doi.org/10.1109/TAP.1982.1142865}{.
\newblock \emph{IEEE Trans. Antennas Propag.} \textbf{30} no.~4, pp. 594--605
  (1982)}.
\bibAnnoteFile{einziger1982}

\bibitem{Pisanty2020}
{E.~Pisanty, M.~F. Ciappina and M.~Lewenstein}.
\newblock The imaginary part of the high-harmonic
  cutoff\href{http://doi.org/10.1088/2515-7647/ab8f1e}{.
\newblock \emph{J. Phys: Photon.} \textbf{2} no.~3, p. 034013 (2020)}.
\newblock \href{https://arxiv.org/abs/2003.00277}{arXiv:2003.00277}.
\bibAnnoteFile{Pisanty2020}

\bibitem{milosevic2007intensity}
{D.~B. Milo\v{s}evi\'{c}, E.~Hasovi\'{c}, M.~Busulad\v{z}i\'{c},
  A.~Gazibegovi\'{c}-Busulad\v{z}i\'{c} and W.~Becker}.
\newblock Intensity-dependent enhancements in high-order above-thresh\-old
  ionization\href{http://doi.org/10.1103/PhysRevA.76.053410}{.
\newblock \emph{Phys. Rev. A} \textbf{76} no.~5, p. 053410 (2007)}.
\bibAnnoteFile{milosevic2007intensity}

\bibitem{Popruzhenko2008}
{S.~V. Popruzhenko, G.~G. Paulus and D.~Bauer}.
\newblock Coulomb-corrected quantum trajectories in strong-field
  ionization\href{http://doi.org/10.1103/PhysRevA.77.053409}{.
\newblock \emph{Phys. Rev. A} \textbf{77} no.~5, p. 053409 (2008)}.
\newblock \href{https://hdl.handle.net/1969.1/126617}{OAKTrust eprint}.
\bibAnnoteFile{Popruzhenko2008}

\bibitem{Maxwell2017}
{A.~S. Maxwell, A.~Al-Jawahiry, T.~Das and C.~Figueira~de Morisson~Faria}.
\newblock Coulomb-corrected quantum interference in above-threshold ionization:
  Working towards multitrajectory electron
  holography\href{http://doi.org/10.1103/PhysRevA.96.023420}{.
\newblock \emph{Phys. Rev. A} \textbf{96} no.~2, p. 023420 (2017)}.
\newblock \href{https://arxiv.org/abs/1705.01518}{arXiv:1705.01518}.
\bibAnnoteFile{Maxwell2017}

\bibitem{figueira2020}
{C.~Figueira~de Morisson~Faria and A.~S. Maxwell}.
\newblock It is all about phases: ultrafast holographic photoelectron
  imaging\href{http://doi.org/10.1088/1361-6633/ab5c91}{.
\newblock \emph{Rep. Progr. Phys.} \textbf{83} no.~3, p. 034401 (2020)}.
\bibAnnoteFile{figueira2020}

\bibitem{Zagoya2014}
{C.~Zagoya, J.~Wu, M.~Ronto, D.~V. Shalashilin and C.~Figueira~de
  Morisson~Faria}.
\newblock Quantum and semiclassical phase-space dynamics of a wave packet in
  strong fields using initial-value
  representations\href{http://doi.org/10.1088/1367-2630/16/10/103040}{.
\newblock \emph{New J. Phys.} \textbf{16} no.~10, p. 103040 (2014)}.
\bibAnnoteFile{Zagoya2014}

\bibitem{Corkum1993}
{P.~B. Corkum}.
\newblock Plasma perspective on strong field multiphoton
  ionization\href{http://doi.org/10.1103/PhysRevLett.71.1994}{.
\newblock \emph{Phys. Rev. Lett.} \textbf{71} no.~13, pp. 1994--1997 (1993)}.
\newblock
  \href{https://nrc-publications.canada.ca/eng/view/object/?id=c16dde3a-0d05-4437-bca6-3292fdd9d9ff}{NRC
  eprint}.
\bibAnnoteFile{Corkum1993}

\bibitem{Kulander1993}
{K.~C. Kulander, K.~J. Schafer and J.~L. Krause}.
\newblock Dynamics of short-pulse excitation, ionization and harmonic
  conversion{.
\newblock In {B.~Piraux, A.~L'Huillier and K.~Rz\k{a}\.zewski} (eds.),
  \emph{\href{https://www.springer.com/gp/book/9780306445873}{Super-Intense
  Laser Atom Physics}}, vol. 316 of \emph{NATO Advanced Studies Institute
  Series B: Physics}, pp. 95--110 (Plenum, New York, 1993)}.
\bibAnnoteFile{Kulander1993}

\bibitem{Panfili2001}
{R.~Panfili, J.~H. Eberly and S.~L. Haan}.
\newblock Comparing classical and quantum dynamics of strong-field double
  ionization\href{http://doi.org/10.1364/OE.8.000431}{.
\newblock \emph{Opt. Express} \textbf{8} no.~7, pp. 431--435 (2001)}.
\bibAnnoteFile{Panfili2001}

\bibitem{dimitriou2004origin}
{K.~I. Dimitriou, D.~G. Arb\'{o}, S.~Yoshida, E.~Persson and
  J.~Burgd\"{o}rfer}.
\newblock Origin of the double-peak structure in the momentum distribution of
  ionization of hydrogen atoms driven by strong laser
  fields\href{http://doi.org/10.1103/PhysRevA.70.061401}{.
\newblock \emph{Phys. Rev. A} \textbf{70} no.~6, p. 061401 (2004)}.
\bibAnnoteFile{dimitriou2004origin}

\bibitem{figueira2020phases}
{C.~Figueira~de Morisson~Faria and A.~S. Maxwell}.
\newblock It is all about phases: ultrafast holographic photoelectron
  imaging\href{http://doi.org/10.1088/1361-6633/ab5c91}{.
\newblock \emph{Rep. Prog. Phys.} \textbf{83} no.~3, p. 034401 (2020)}.
\newblock \href{https://arxiv.org/abs/1906.11781}{arXiv:1906.11781}.
\bibAnnoteFile{figueira2020phases}

\bibitem{Mairesse2010}
{Y.~Mairesse, J.~Higuet, N.~Dudovich} et~al.
\newblock High harmonic spectroscopy of multichannel dynamics in strong-field
  ionization\href{http://doi.org/10.1103/PhysRevLett.104.213601}{.
\newblock \emph{Phys. Rev. Lett.} \textbf{104} no.~21, p. 213601 (2010)}.
\bibAnnoteFile{Mairesse2010}

\bibitem{Shi2019}
{X.~Shi and H.~B. Schlegel}.
\newblock Controlling the strong field fragmentation of {ClCHO$^+$} using two
  laser pulses --an ab initio molecular dynamics
  simulation\href{http://doi.org/10.1002/jcc.25576}{.
\newblock \emph{J. Comput. Chem.} \textbf{40} no.~1, pp. 200--205 (2019)}.
\newblock \href{http://schlegelgroup.wayne.edu/Pub_folder/420.pdf}{Author
  eprint}.
\bibAnnoteFile{Shi2019}

\bibitem{Ni2016}
{H.~Ni, U.~Saalmann and J.-M. Rost}.
\newblock Tunneling ionization time resolved by
  backpropagation\href{http://doi.org/10.1103/PhysRevLett.117.023002}{.
\newblock \emph{Phys. Rev. Lett.} \textbf{117} no.~2, p. 023002 (2016)}.
\newblock
  \href{https://www.pks.mpg.de/fileadmin/user_upload/userpages/us/article/nisa-16.pdf}{Author
  eprint}.
\bibAnnoteFile{Ni2016}

\bibitem{Abusamha2010}
{M.~Abu-samha and L.~B. Madsen}.
\newblock Single-active-electron potentials for molecules in intense laser
  fields\href{http://doi.org/10.1103/PhysRevA.81.033416}{.
\newblock \emph{Phys. Rev. A} \textbf{81}, p. 033416 (2010)}.
\newblock
  \href{https://phys.au.dk/fileadmin/site_files/forskning/ltc/publications/ms_final.pdf}{AU
  eprint}.
\bibAnnoteFile{Abusamha2010}

\bibitem{Kukk2013}
{E.~Kukk, D.~Ayuso, T.~D. Thomas} et~al.
\newblock Effects of molecular potential and geometry on atomic core-level
  photoemission over an extended energy range: The case study of the {CO}
  molecule\href{http://doi.org/10.1103/PhysRevA.88.033412}{.
\newblock \emph{Phys. Rev. A} \textbf{88}, p. 033412 (2013)}.
\newblock \href{http://hdl.handle.net/10486/668670}{UAM eprint}.
\bibAnnoteFile{Kukk2013}

\bibitem{Saito2004}
{N.~Saito, D.~Toffoli, R.~R. Lucchese} et~al.
\newblock Symmetry- and multiplet-resolved n $1s$ photoionization cross
  sections of the {NO}$_2$
  molecule\href{http://doi.org/10.1103/PhysRevA.70.062724}{.
\newblock \emph{Phys. Rev. A} \textbf{70}, p. 062724 (2004)}.
\newblock \href{https://hdl.handle.net/1969.1/182934}{OAKTrust eprint}.
\bibAnnoteFile{Saito2004}

\bibitem{Gozem2015}
{S.~Gozem, A.~O. Gunina, T.~Ichino} et~al.
\newblock Photoelectron wave function in photoionization: Plane wave or
  {Coulomb} wave?\href{http://doi.org/10.1021/acs.jpclett.5b01891}{.
\newblock \emph{J. Phys. Chem. Lett.} \textbf{6}, pp. 4532--4540 (2015)}.
\newblock \href{http://iopenshell.usc.edu/pubs/pdf/jpcl-6-4532.pdf}{USC
  eprint}.
\bibAnnoteFile{Gozem2015}

\bibitem{Labeye2018}
{M.~Labeye, F.~Zapata, E.~Coccia} et~al.
\newblock Optimal basis set for electron dynamics in strong laser fields: The
  case of molecular ion
  {H}$_2^+$\href{http://doi.org/10.1021/acs.jctc.8b00656}{.
\newblock \emph{J. Chem. Theory Comput.} \textbf{14} no.~11, pp. 5846--5858
  (2018)}.
\newblock
  \href{https://www.ncbi.nlm.nih.gov/pmc/articles/PMC6255052/}{PubMedCentral
  eprint}.
\bibAnnoteFile{Labeye2018}

\bibitem{vandenWildenberg2019}
{S.~van~den Wildenberg, B.~Mignolet, R.~Levine and F.~Remacle}.
\newblock Temporal and spatially resolved imaging of the correlated
  nuclear-electronic dynamics and of the ionized photoelectron in a coherently
  electronically highly excited vibrating {LiH}
  molecule\href{http://doi.org/10.1063/1.5116250}{.
\newblock \emph{J. Chem. Phys.} \textbf{151}, p. 134310 (2019)}.
\bibAnnoteFile{vandenWildenberg2019}

\bibitem{Hohenberg1964}
{P.~Hohenberg and W.~Kohn}.
\newblock Inhomogeneous electron
  gas\href{http://doi.org/10.1103/PhysRev.136.B864}{.
\newblock \emph{Phys. Rev.} \textbf{136} no.~3B, pp. B864--B871 (1964)}.
\bibAnnoteFile{Hohenberg1964}

\bibitem{runge1984}
{E.~Runge and E.~K.~U. Gross}.
\newblock Density-functional theory for time-dependent
  systems\href{http://doi.org/10.1103/PhysRevLett.52.997}{.
\newblock \emph{Phys. Rev. Lett.} \textbf{52} no.~12, pp. 997--1000 (1984)}.
\bibAnnoteFile{runge1984}

\bibitem{vanleeuwen1998}
{R.~van Leeuwen}.
\newblock Causality and symmetry in time-dependent density-functional
  theory\href{http://doi.org/10.1103/PhysRevLett.80.1280}{.
\newblock \emph{Phys. Rev. Lett.} \textbf{80}, pp. 1280--1283 (1998)}.
\bibAnnoteFile{vanleeuwen1998}

\bibitem{vanleeuwen1999}
{R.~van Leeuwen}.
\newblock Mapping from densities to potentials in time-dependent
  density-functional theory\href{http://doi.org/10.1103/PhysRevLett.82.3863}{.
\newblock \emph{Phys. Rev. Lett.} \textbf{82}, pp. 3863--3866 (1999)}.
\bibAnnoteFile{vanleeuwen1999}

\bibitem{Tong1997}
{X.-M. Tong and S.-I. Chu}.
\newblock Density-functional theory with optimized effective potential and
  self-interaction correction ground states and autoionizing
  resonances\href{http://doi.org/10.1103/PhysRevA.55.3406}{.
\newblock \emph{Phys. Rev. A} \textbf{55}, pp. 3406--3416 (1997)}.
\newblock \href{https://hdl.handle.net/1808/15974}{KU eprint}.
\bibAnnoteFile{Tong1997}

\bibitem{Stener2005}
{M.~Stener, G.~Fronzoni and P.~Decleva}.
\newblock Time-dependent density-functional theory for molecular
  photoionization with noniterative algorithm and multicenter {B}-spline basis
  set: {CS}$_2$ and {C$_6$H$_6$} case
  studies\href{http://doi.org/10.1063/1.1937367}{.
\newblock \emph{J. Chem. Phys.} \textbf{122}, p. 234301 (2005)}.
\bibAnnoteFile{Stener2005}

\bibitem{Toffoli2013}
{D.~Toffoli and P.~Decleva}.
\newblock Multiphoton core ionization dynamics of polyatomic
  molecules\href{http://doi.org/10.1088/0953-4075/46/14/145101}{.
\newblock \emph{J. Phys. B: At., Mol. Opt. Phys.} \textbf{46}, p. 145101
  (2013)}.
\bibAnnoteFile{Toffoli2013}

\bibitem{Ruberti2013}
{M.~Ruberti, R.~Yun, K.~Gokhberg} et~al.
\newblock Total molecular photoionization cross-sections by algebraic
  diagrammatic construction-{Stieltjes}-{Lanczos} method: Benchmark
  calculations\href{http://doi.org/10.1063/1.4824431}{.
\newblock \emph{J. Chem. Phys.} \textbf{139}, p. 144107 (2013)}.
\bibAnnoteFile{Ruberti2013}

\bibitem{Ruberti2014}
{M.~Ruberti, R.~Yun, K.~Gokhberg} et~al.
\newblock Total photoionization cross-sections of excited electronic states by
  the algebraic diagrammatic construction-stieltjes-lanczos
  method\href{http://doi.org/10.1063/1.4874269}{.
\newblock \emph{J. Chem. Phys.} \textbf{140}, p. 184107 (2014)}.
\bibAnnoteFile{Ruberti2014}

\bibitem{Walker1994}
{B.~Walker, B.~Sheehy, L.~F. DiMauro} et~al.
\newblock Precision measurement of strong field double ionization of
  helium\href{http://doi.org/10.1103/PhysRevLett.73.1227}{.
\newblock \emph{Phys. Rev. Lett.} \textbf{73} no.~9, p. 1227 (1994)}.
\newblock
  \href{https://www.asc.ohio-state.edu/dimauro.6/Publications/DiMauro/1994.Precision.Walker.pdf}{Author
  eprint}.
\bibAnnoteFile{Walker1994}

\bibitem{MaxwellThesis2019}
{A.~S. Maxwell}.
\newblock \emph{Strong-Field Interference of Quantum Trajectories with
  {Coulomb} Distortion and Electron
  Correlation}\href{https://discovery.ucl.ac.uk/id/eprint/10064744/}{.
\newblock {PhD} thesis, University College London (2019)}.
\bibAnnoteFile{MaxwellThesis2019}

\bibitem{LHuillier1983}
{A.~L'Huillier, L.~A. Lompre, G.~Mainfray and C.~Manus}.
\newblock Multiply charged ions induced by multiphoton absorption in rare gases
  at {$\SI{0.53}{\micro m}$}\href{http://doi.org/10.1103/PhysRevA.27.2503}{.
\newblock \emph{Phys. Rev. A} \textbf{27} no.~5, p. 2503 (1983)}.
\bibAnnoteFile{LHuillier1983}

\bibitem{Agostini1984}
{P.~Agostini and G.~Petite}.
\newblock Multiphoton ionisation of calcium with picosecond
  pulses\href{http://doi.org/10.1088/0022-3700/17/23/003}{.
\newblock \emph{J. Phys. B: At. Mol. Opt. Phys.} \textbf{17} no.~23, pp.
  L811--L816 (1984)}.
\bibAnnoteFile{Agostini1984}

\bibitem{Agostini1985}
{P.~Agostini and G.~Petite}.
\newblock Double multiphoton ionisation via above-threshold ionisation in
  strontium atoms\href{http://doi.org/10.1088/0022-3700/18/10/004}{.
\newblock \emph{J. Phys. B: At. Mol. Opt. Phys.} \textbf{18} no.~10, pp.
  L281--L286 (1985)}.
\bibAnnoteFile{Agostini1985}

\bibitem{Fittinghoff1992}
{D.~Fittinghoff, P.~Bolton, B.~Chang and K.~Kulander}.
\newblock Observation of nonsequential double ionization of helium with optical
  tunneling\href{http://doi.org/10.1103/PhysRevLett.69.2642}{.
\newblock \emph{Phys. Rev. Lett.} \textbf{69} no.~18, p. 2642 (1992)}.
\newblock \href{https://zenodo.org/record/1233895}{Zenodo eprint}.
\bibAnnoteFile{Fittinghoff1992}

\bibitem{Kondo1993}
{K.~Kondo, A.~Sagisaka, T.~Tamida, Y.~Nabekawa and S.~Wantanabe}.
\newblock Wavelength dependence of nonsquential double ionization in
  {He}\href{http://doi.org/10.1103/PhysRevA.48.R2531}{.
\newblock \emph{Phys. Rev. A} \textbf{35} no.~6, p. R2531 (1999)}.
\bibAnnoteFile{Kondo1993}

\bibitem{Kuchiev1987}
{M.~Y. Kuchiev}.
\newblock Atomic
  antenna\href{http://www.jetpletters.ac.ru/ps/1241/article_18763.shtml}{.
\newblock \emph{{JETP} Lett.} \textbf{45} no.~7, pp. 404--406 (1987)}.
\newblock
  [\href{http://www.jetpletters.ac.ru/ps/139/article_2404.shtml}{\textit{Pis'ma
  Zh. Eksp. Teor. Fiz.} \textbf{45} no.~7, pp.~319-321 (1987)}].
\bibAnnoteFile{Kuchiev1987}

\bibitem{Schafer1993}
{K.~J. Schafer, B.~Yang, L.~F. DiMauro and K.~C. Kulander}.
\newblock Above threshold ionization beyond the high harmonic
  cutoff\href{http://doi.org/10.1103/PhysRevLett.70.1599}{.
\newblock \emph{Phys. Rev. Lett.} \textbf{70} no.~11, pp. 1599--1602 (1993)}.
\newblock \href{http://dimauro.osu.edu/publications/1993.Above.pdf}{Author
  eprint}.
\bibAnnoteFile{Schafer1993}

\bibitem{Becker1994}
{A.~Becker and F.~H.~M. Faisal}.
\newblock Correlated {Keldysh-Faisal-Reiss} theory of above-threshold double
  ionization of {He} in intense laser
  fields\href{http://doi.org/10.1103/PhysRevA.50.3256}{.
\newblock \emph{Phys. Rev. A} \textbf{50} no.~4, pp. 3256--3264 (1994)}.
\bibAnnoteFile{Becker1994}

\bibitem{Becker1996}
{A.~Becker and F.~H.~M. Faisal}.
\newblock Mechanism of laser-induced double ionization of
  helium\href{http://doi.org/10.1088/0953-4075/29/6/005}{.
\newblock \emph{J. Phys. B: At. Mol. Opt. Phys.} \textbf{29} no.~6, pp.
  L197--L202 (1996)}.
\bibAnnoteFile{Becker1996}

\bibitem{Becker1999}
{A.~Becker and F.~H.~M. Faisal}.
\newblock Interplay of electron correlation and intense field dynamics in the
  double ionization of helium\href{http://doi.org/10.1103/PhysRevA.59.R1742}{.
\newblock \emph{Phys. Rev. A} \textbf{59} no.~3, pp. R1742--R1745 (1999)}.
\bibAnnoteFile{Becker1999}

\bibitem{becker2011}
{W.~Becker, X.~Liu, P.~J. Ho and J.~H. Eberly}.
\newblock Theories of photoelectron correlation in laser-driven multiple atomic
  ionization\href{http://doi.org/10.1103/RevModPhys.84.1011}{.
\newblock \emph{Rev. Mod. Phys.} \textbf{84} no.~3, pp. 1011--1043 (2012)}.
\bibAnnoteFile{becker2011}

\bibitem{figueira2011}
{C.~Figueira~de Morisson~Faria and X.~Liu}.
\newblock Electron–electron correlation in strong laser
  fields\href{http://doi.org/10.1080/09500340.2010.543958}{.
\newblock \emph{J. Mod. Opt.} \textbf{58} no.~13, pp. 1076--1131 (2011)}.
\bibAnnoteFile{figueira2011}

\bibitem{Ye2008}
{D.~F. Ye, X.~Liu and J.~Liu}.
\newblock Classical trajectory diagnosis of a fingerlike pattern in the
  correlated electron momentum distribution in strong field double ionization
  of helium\href{http://doi.org/10.1103/PhysRevLett.101.233003}{.
\newblock \emph{Phys. Rev. Lett.} \textbf{101} no.~23, p. 233003 (2008)}.
\newblock \href{https://arxiv.org/abs/0802.0041}{arXiv:0802.0041}.
\bibAnnoteFile{Ye2008}

\bibitem{Emmanouilidou2008}
{A.~Emmanouilidou}.
\newblock Recoil collisions as a portal to field-assisted ionization at near-uv
  frequencies in the strong-field double ionization of
  helium\href{http://doi.org/10.1103/PhysRevA.78.023411}{.
\newblock \emph{Phys. Rev. A} \textbf{78} no.~2, p. 023411 (2008)}.
\newblock \href{https://arxiv.org/abs/0805.4117}{arXiv:0805.4117}.
\bibAnnoteFile{Emmanouilidou2008}

\bibitem{Emmanouilidou2009}
{A.~Emmanouilidou and A.~Staudte}.
\newblock Intensity dependence of strong-field double-ionization mechanisms:
  From field-assisted recollision ionization to recollision-assisted field
  ionization\href{http://doi.org/10.1103/PhysRevA.80.053415}{.
\newblock \emph{Phys. Rev. A} \textbf{80} no.~5, p. 053415 (2009)}.
\newblock \href{https://arxiv.org/abs/0909.3409}{arXiv:0909.3409}.
\bibAnnoteFile{Emmanouilidou2009}

\bibitem{Brabec1996}
{T.~Brabec, M.~{\relax Yu}. Ivanov and P.~B. Corkum}.
\newblock Coulomb focusing in intense field atomic
  processes\href{http://doi.org/10.1103/PhysRevA.54.R2551}{.
\newblock \emph{Phys. Rev. A} \textbf{54} no.~4, pp. R2551--R2554 (1996)}.
\bibAnnoteFile{Brabec1996}

\bibitem{Chen2000}
{J.~Chen, J.~Liu, L.~B. Fu and W.~M. Zheng}.
\newblock Interpretation of momentum distribution of recoil ions from
  laser-induced nonsequential double ionization by semiclassical rescattering
  model\href{http://doi.org/10.1103/PhysRevA.63.011404}{.
\newblock \emph{Phys. Rev. A} \textbf{63} no.~1, p. 011404 (2000)}.
\bibAnnoteFile{Chen2000}

\bibitem{Haan2008}
{S.~L. Haan, Z.~S. Smith, K.~N. Shomsky and P.~W. Plantinga}.
\newblock Anticorrelated electrons from weak recollisions in nonsequential
  double ionization\href{http://doi.org/10.1088/0953-4075/41/21/211002}{.
\newblock \emph{J. Phys. B: At. Mol. Opt. Phys.} \textbf{41} no.~21, p. 211002
  (2008)}.
\bibAnnoteFile{Haan2008}

\bibitem{Haan2008PRL}
{S.~L. Haan, L.~Breen, A.~Karim and J.~H. Eberly}.
\newblock Variable time lag and backward ejection in full-dimensional analysis
  of strong-field double
  ionization\href{http://doi.org/10.1103/PhysRevLett.97.103008}{.
\newblock \emph{Phys. Rev. Lett.} \textbf{97} no.~10, p. 103008 (2006)}.
\bibAnnoteFile{Haan2008PRL}

\bibitem{Ho2005}
{P.~J. Ho, R.~Panfili, S.~L. Haan and J.~H. Eberly}.
\newblock Nonsequential double ionization as a completely classical
  photoelectric effect\href{http://doi.org/10.1103/PhysRevLett.94.093002}{.
\newblock \emph{Phys. Rev. Lett.} \textbf{94} no.~9, p. 093002 (2005)}.
\newblock \href{https://arxiv.org/abs/physics/0409099}{arXiv:physics/0409099}.
\bibAnnoteFile{Ho2005}

\bibitem{Haan2002}
{S.~L. Haan, P.~S. Wheeler, R.~Panfili and J.~H. Eberly}.
\newblock Origin of correlated electron emission in double ionization of
  atoms\href{http://doi.org/10.1103/PhysRevA.66.061402}{.
\newblock \emph{Phys. Rev. A} \textbf{66} no.~6, p. 061402 (2002)}.
\bibAnnoteFile{Haan2002}

\bibitem{Panfili2002}
{R.~Panfili, S.~L. Haan and J.~H. Eberly}.
\newblock Slow-down collisions and nonsequential double ionization in classical
  simulations\href{http://doi.org/10.1103/PhysRevLett.89.113001}{.
\newblock \emph{Phys. Rev. Lett.} \textbf{89} no.~11, p. 113001 (2002)}.
\bibAnnoteFile{Panfili2002}

\bibitem{Goreslavskii2001}
{S.~P. Goreslavskii, S.~V. Popruzhenko, R.~Kopold and W.~Becker}.
\newblock Electron-electron correlation in laser-induced nonsequential double
  ionization\href{http://doi.org/10.1103/PhysRevA.64.053402}{.
\newblock \emph{Phys. Rev. A} \textbf{64} no.~5, p. 053402 (2001)}.
\bibAnnoteFile{Goreslavskii2001}

\bibitem{Popruzhenko2002}
{S.~V. Popruzhenko, P.~A. Korneev, S.~P. Goreslavski and W.~Becker}.
\newblock Laser-induced recollision phenomena: Interference resonances at
  channel closings\href{http://doi.org/10.1103/PhysRevLett.89.023001}{.
\newblock \emph{Phys. Rev. Lett.} \textbf{89} no.~2, p. 023001 (2002)}.
\bibAnnoteFile{Popruzhenko2002}

\bibitem{maxwell2016}
{A.~S. Maxwell and C.~Figueira~de Morisson~Faria}.
\newblock Controlling below-threshold nonsequential double ionization via
  quantum interference\href{http://doi.org/10.1103/PhysRevLett.116.143001}{.
\newblock \emph{Phys. Rev. Lett.} \textbf{116} no.~14, p. 143001 (2016)}.
\newblock \href{https://discovery.ucl.ac.uk/id/eprint/1479920}{UCL eprint}.
\bibAnnoteFile{maxwell2016}

\bibitem{Hao2014}
{X.~Hao, J.~Chen, W.~Li} et~al.
\newblock Quantum effects in double ionization of argon below the threshold
  intensity\href{http://doi.org/10.1103/PhysRevLett.112.073002}{.
\newblock \emph{Phys. Rev. Lett.} \textbf{112} no.~7, p. 073002 (2014)}.
\bibAnnoteFile{Hao2014}

\bibitem{Becker2000}
{A.~Becker and F.~H.~M. Faisal}.
\newblock Interpretation of momentum distribution of recoil ions from laser
  induced nonsequential double
  ionization\href{http://doi.org/10.1103/PhysRevLett.84.3546}{.
\newblock \emph{Phys. Rev. Lett.} \textbf{84} no.~16, pp. 3546--3549 (2000)}.
\bibAnnoteFile{Becker2000}

\bibitem{Quan2009}
{W.~Quan, X.~Liu and C.~Figueira~de Morisson~Faria}.
\newblock Nonsequential double ionization with polarization-gated
  pulses\href{http://doi.org/10.1088/0953-4075/42/13/134008}{.
\newblock \emph{J. Phys. B: At. Mol. Opt. Phys.} \textbf{42} no.~13, p. 134008
  (2009)}.
\newblock \href{https://arxiv.org/abs/0901.3116}{arXiv:0901.3116}.
\bibAnnoteFile{Quan2009}

\bibitem{Shaaran2010}
{T.~Shaaran, M.~T. Nygren and C.~Figueira~de Morisson~Faria}.
\newblock Laser-induced nonsequential double ionization at and above the
  recollision-excitation-tunneling
  threshold\href{http://doi.org/10.1103/PhysRevA.81.063413}{.
\newblock \emph{Phys. Rev. A} \textbf{81} no.~6, p. 063413 (2010)}.
\newblock \href{https://arxiv.org/abs/1001.5225}{arXiv:1001.5225}.
\bibAnnoteFile{Shaaran2010}

\bibitem{Shaaran2012}
{T.~Shaaran, C.~Figueira~de Morisson~Faria and H.~Schomerus}.
\newblock Causality and quantum interference in time-delayed laser-induced
  nonsequential double
  ionization\href{http://doi.org/10.1103/PhysRevA.85.023423}{.
\newblock \emph{Phys. Rev. A} \textbf{85} no.~2, p. 023423 (2012)}.
\newblock \href{https://eprints.lancs.ac.uk/id/eprint/52796}{LU eprint}.
\bibAnnoteFile{Shaaran2012}

\bibitem{figueira2012}
{C.~Figueira~de Morisson~Faria, T.~Shaaran and M.~T. Nygren}.
\newblock Time-delayed nonsequential double ionization with few-cycle laser
  pulses: Importance of the carrier-envelope
  phase\href{http://doi.org/10.1103/PhysRevA.86.053405}{.
\newblock \emph{Phys. Rev. A} \textbf{86} no.~5, p. 053405 (2012)}.
\newblock \href{https://discovery.ucl.ac.uk/id/eprint/1353962}{UCL eprint}.
\bibAnnoteFile{figueira2012}

\bibitem{Maxwell2015}
{A.~S. Maxwell and C.~Figueira~de Morisson~Faria}.
\newblock Quantum interference in time-delayed nonsequential double
  ionization\href{http://doi.org/10.1103/PhysRevA.92.023421}{.
\newblock \emph{Phys. Rev. A} \textbf{92} no.~2, p. 023421 (2015)}.
\newblock \href{https://arxiv.org/abs/1507.06823}{arXiv:1507.06823}.
\bibAnnoteFile{Maxwell2015}

\bibitem{Liao2017}
{Q.~Liao, Y.~Li, M.~Qin and P.~Lu}.
\newblock Attosecond interference in strong-field nonsequential double
  ionization\href{http://doi.org/10.1103/PhysRevA.96.063408}{.
\newblock \emph{Phys. Rev. A} \textbf{96} no.~6, p. 063408 (2017)}.
\bibAnnoteFile{Liao2017}

\bibitem{lein2000}
{M.~Lein, E.~K.~U. Gross and V.~Engel}.
\newblock Intense-field double ionization of helium: Identifying the
  mechanism\href{http://doi.org/10.1103/PhysRevLett.85.4707}{.
\newblock \emph{Phys. Rev. Lett.} \textbf{85} no.~22, pp. 4707--4710 (2000)}.
\newblock \href{https://lein.itp.uni-hannover.de/papers/LGE_PRL_00.pdf}{Author
  eprint}.
\bibAnnoteFile{lein2000}

\bibitem{ruiz2006}
{C.~Ruiz, L.~Plaja, L.~Roso and A.~Becker}.
\newblock Ab initio calculation of the double ionization of helium in a
  few-cycle laser pulse beyond the one-dimensional
  approximation\href{http://doi.org/10.1103/PhysRevLett.96.053001}{.
\newblock \emph{Phys. Rev. Lett.} \textbf{96} no.~5, p. 053001 (2006)}.
\bibAnnoteFile{ruiz2006}

\bibitem{smyth1998}
{E.~S. Smyth, J.~S. Parker and K.~T. Taylor}.
\newblock Numerical integration of the time-dependent {Schr\"{o}dinger}
  equation for laser-driven
  helium\href{http://doi.org/10.1016/S0010-4655(98)00083-6}{.
\newblock \emph{Comput. Phys. Commun.} \textbf{114} no.~1, pp. 1--14 (1998)}.
\bibAnnoteFile{smyth1998}

\bibitem{parker2001}
{J.~S. Parker, L.~R. Moore, K.~J. Meharg, D.~Dundas and K.~T. Taylor}.
\newblock Double-electron above threshold ionization of
  helium\href{http://doi.org/10.1088/0953-4075/34/3/103}{.
\newblock \emph{J. Phys. B: At. Mol. Opt. Phys.} \textbf{34} no.~3, pp.
  L69--L78 (2001)}.
\newblock \href{https://hdl.handle.net/1811/47978}{OSU eprint}.
\bibAnnoteFile{parker2001}

\bibitem{pindzola1998}
{M.~S. Pindzola and F.~Robicheaux}.
\newblock Time-dependent close-coupling calculations of correlated
  photoionization processes in
  helium\href{http://doi.org/10.1103/PhysRevA.57.318}{.
\newblock \emph{Phys. Rev. A} \textbf{57} no.~1, pp. 318--324 (1998)}.
\bibAnnoteFile{pindzola1998}

\bibitem{pindzola2007}
{M.~S. Pindzola, F.~Robicheaux, S.~D. Loch} et~al.
\newblock The time-dependent close-coupling method for atomic and molecular
  collision processes\href{http://doi.org/10.1088/0953-4075/40/7/r01}{.
\newblock \emph{J. Phys. B: At. Mol. Opt. Phys.} \textbf{40} no.~7, pp.
  R39--R60 (2007)}.
\newblock
  \href{http://users.df.uba.ar/dmitnik/publications/timedependreview.pdf}{UBA
  eprint}.
\bibAnnoteFile{pindzola2007}

\bibitem{colgan2012}
{J.~Colgan and M.~S. Pindzola}.
\newblock Application of the time-dependent close-coupling approach to few-body
  atomic and molecular ionizing
  collisions\href{http://doi.org/10.1140/epjd/e2012-30517-2}{.
\newblock \emph{Eur. Phys. J. D} \textbf{66} no.~11, p. 284 (2012)}.
\bibAnnoteFile{colgan2012}

\bibitem{parker2006}
{J.~S. Parker, B.~J.~S. Doherty, K.~T. Taylor} et~al.
\newblock High-energy cutoff in the spectrum of strong-field nonsequential
  double ionization\href{http://doi.org/10.1103/PhysRevLett.96.133001}{.
\newblock \emph{Phys. Rev. Lett.} \textbf{96} no.~13, p. 133001 (2006)}.
\bibAnnoteFile{parker2006}

\bibitem{feist2008}
{J.~Feist, S.~Nagele, R.~Pazourek} et~al.
\newblock Nonsequential two-photon double ionization of
  helium\href{http://doi.org/10.1103/PhysRevA.77.043420}{.
\newblock \emph{Phys. Rev. A} \textbf{77} no.~4, p. 043420 (2008)}.
\newblock \href{https://arxiv.org/abs/0803.0511}{arXiv:0803.0511}.
\bibAnnoteFile{feist2008}

\bibitem{feist2009}
{J.~Feist, S.~Nagele, R.~Pazourek} et~al.
\newblock Probing electron correlation via attosecond xuv pulses in the
  two-photon double ionization of
  helium\href{http://doi.org/10.1103/PhysRevLett.103.063002}{.
\newblock \emph{Phys. Rev. Lett.} \textbf{103} no.~6, p. 063002 (2009)}.
\newblock \href{https://arxiv.org/abs/0812.0373}{arXiv:0812.0373}.
\bibAnnoteFile{feist2009}

\bibitem{hu2010}
{S.~X. Hu}.
\newblock Optimizing the {FEDVR}-{TDCC} code for exploring the quantum dynamics
  of two-electron systems in intense laser
  pulses\href{http://doi.org/10.1103/PhysRevE.81.056705}{.
\newblock \emph{Phys. Rev. E} \textbf{81} no.~5, p. 056705 (2010)}.
\bibAnnoteFile{hu2010}

\bibitem{nepstad2010}
{R.~Nepstad, T.~Birkeland and M.~F\o{}rre}.
\newblock Numerical study of two-photon ionization of helium using an \emph{ab
  initio} numerical framework\href{http://doi.org/10.1103/PhysRevA.81.063402}{.
\newblock \emph{Phys. Rev. A} \textbf{81} no.~6, p. 063402 (2010)}.
\newblock \href{https://arxiv.org/abs/1004.3216}{arXiv:1004.3216}.
\bibAnnoteFile{nepstad2010}

\bibitem{hu2013}
{S.~X. Hu}.
\newblock Boosting photoabsorption by attosecond control of electron
  correlation\href{http://doi.org/10.1103/PhysRevLett.111.123003}{.
\newblock \emph{Phys. Rev. Lett.} \textbf{111} no.~12, p. 123003 (2013)}.
\bibAnnoteFile{hu2013}

\bibitem{djiokap2012}
{J.~M. Ngoko-Djiokap, S.~X. Hu, W.-C. Jiang, L.-Y. Peng and A.~F. Starace}.
\newblock Enhanced asymmetry in few-cycle attosecond pulse ionization of he in
  the vicinity of autoionizing
  resonances\href{http://doi.org/10.1088/1367-2630/14/9/095010}{.
\newblock \emph{New J. Phys.} \textbf{14} no.~9, p. 095010 (2012)}.
\bibAnnoteFile{djiokap2012}

\bibitem{djiokap2015}
{J.~M. Ngoko-Djiokap, A.~V. Meremianin, N.~L. Ma\-nakov} et~al.
\newblock Multistart spiral electron vortices in ionization by circularly
  polarized {UV} pulses\href{http://doi.org/10.1103/PhysRevA.94.013408}{.
\newblock \emph{Phys. Rev. A} \textbf{94} no.~1, p. 013408 (2016)}.
\newblock \href{https://digitalcommons.unl.edu/physicsstarace/218/}{UNL
  eprint}.
\bibAnnoteFile{djiokap2015}

\bibitem{donsa2019}
{S.~Donsa, I.~B\v{r}ezinov\'a, H.~Ni, J.~Feist and J.~Burgd\"orfer}.
\newblock Polarization tagging of two-photon double ionization by elliptically
  polarized {XUV} pulses\href{http://doi.org/10.1103/PhysRevA.99.023413}{.
\newblock \emph{Phys. Rev. A} \textbf{99} no.~2, p. 023413 (2019)}.
\newblock \href{http://hdl.handle.net/10486/688403}{UAM eprint}.
\bibAnnoteFile{donsa2019}

\bibitem{donsa2019prl}
{S.~Donsa, N.~Douguet, J.~Burgd\"orfer, I.~B\v{r}ezinov\'a and L.~Argenti}.
\newblock Circular holographic ionization-phase
  meter\href{http://doi.org/10.1103/PhysRevLett.123.133203}{.
\newblock \emph{Phys. Rev. Lett.} \textbf{123} no.~13, p. 133203 (2019)}.
\newblock \href{https://arxiv.org/abs/1904.04380}{arXiv:1904.04380}.
\bibAnnoteFile{donsa2019prl}

\bibitem{scrinzi2012}
{A.~Scrinzi}.
\newblock t-{SURFF}: fully differential two-electron photo-emission
  spectra\href{http://doi.org/10.1088/1367-2630/14/8/085008}{.
\newblock \emph{New J. Phys.} \textbf{14} no.~8, p. 085008 (2012)}.
\bibAnnoteFile{scrinzi2012}

\bibitem{zielinski2016}
{A.~Zielinski, V.~P. Majety and A.~Scrinzi}.
\newblock Double photoelectron momentum spectra of helium at infrared
  wavelength\href{http://doi.org/10.1103/PhysRevA.93.023406}{.
\newblock \emph{Phys. Rev. A} \textbf{93} no.~2, p. 023406 (2016)}.
\newblock \href{https://arxiv.org/abs/1511.06655}{arXiv:1511.06655}.
\bibAnnoteFile{zielinski2016}

\bibitem{zhu2020}
{J.~Zhu and A.~Scrinzi}.
\newblock Electron double-emission spectra for helium atoms in intense 400-nm
  laser pulses\href{http://doi.org/10.1103/PhysRevA.101.063407}{.
\newblock \emph{Phys. Rev. A} \textbf{101} no.~6, p. 063407 (2020)}.
\newblock \href{https://arxiv.org/abs/1912.09250}{arXiv:1912.09250}.
\bibAnnoteFile{zhu2020}

\bibitem{Kubel2014}
{M.~K\"{u}bel, K.~J. Betsch, N.~G. Kling} et~al.
\newblock Non-sequential double ionization of {Ar}: from the single- to the
  many-cycle regime\href{http://doi.org/10.1088/1367-2630/16/3/033008}{.
\newblock \emph{New J. Phys.} \textbf{16} no.~3, p. 033008 (2014)}.
\bibAnnoteFile{Kubel2014}

\bibitem{battle2}
{S.~Eckart, M.~K\"ubel, B.~Feti\'c, K.~Amini and A.~Cac\'on}.
\newblock {Quantum Battle 2 -- Quantum interference \& Imaging}.
\newblock \emph{\href{https://www.quantumbattles.com/}{Quantum Battles in
  Attoscience}} online conference (1-3 July 2020).
\newblock
  \href{https://youtu.be/r2erW1u65Jk}{youtu.be/\allowbreak{}r2erW1u65Jk}.
\bibAnnoteFile{battle2}

\bibitem{Amini2020}
{K.~Amini, A.~Chac{\'o}n, S.~Eckart, B.~Feti{\'c} and M.~K{\"u}bel}.
\newblock Quantum interference and imaging using intense laser fields.
\newblock \emph{Eur. Phys. J. D}, under review (2021),
  \href{https://arxiv.org/abs/2103.05686}{arXiv:2103.05686}.
\bibAnnoteFile{Amini2020}

\bibitem{Staudte2007}
{A.~Staudte, C.~Ruiz, M.~Sch\"offler} et~al.
\newblock Binary and recoil collisions in strong field double ionization of
  helium\href{http://doi.org/10.1103/PhysRevLett.99.263002}{.
\newblock \emph{Phys. Rev. Lett.} \textbf{99} no.~26, p. 263002 (2007)}.
\newblock
  \href{https://nrc-publications.canada.ca/eng/view/object/?id=77cdbaab-c33b-4925-881e-5c68e1054226}{NRC
  eprint}.
\bibAnnoteFile{Staudte2007}

\bibitem{quan2017}
{W.~Quan, X.~Hao, Y.~Wang} et~al.
\newblock Quantum interference in laser-induced nonsequential double
  ionization\href{http://doi.org/10.1103/PhysRevA.96.032511}{.
\newblock \emph{Phys. Rev. A} \textbf{96} no.~3, p. 032511 (2017)}.
\bibAnnoteFile{quan2017}

\bibitem{Corkum1989Mar}
{P.~B. Corkum, N.~H. Burnett and F.~Brunel}.
\newblock Above-threshold ionization in the long-wavelength
  limit\href{http://doi.org/10.1103/PhysRevLett.62.1259}{.
\newblock \emph{Phys. Rev. Lett.} \textbf{62} no.~11, pp. 1259--1262 (1989)}.
\bibAnnoteFile{Corkum1989Mar}

\bibitem{battle1}
{N.~Shvetsov-Shilovski, A.~Bray, H.~Ni, C.~Hofmann and W.~Koch}.
\newblock {Quantum Battle 1 -- Tunnelling}.
\newblock \emph{\href{https://www.quantumbattles.com/}{Quantum Battles in
  Attoscience}} online conference (1-3 July 2020).
\newblock
  \href{https://youtu.be/COXjc1GVXhs}{youtu.be/\allowbreak{}COXjc1GVXhs}.
\bibAnnoteFile{battle1}

\bibitem{Hofmann2021}
{C.~Hofmann, A.~Bray, W.~Koch, H.~Ni and N.~I. Shvetsov-Shilovski}.
\newblock Quantum battles in attoscience:
  tunnelling\href{http://doi.org/10.1140/epjd/s10053-021-00224-2}{.
\newblock \emph{Eur. Phys. J. D} \textbf{75} no.~7, pp. 208--13 (2021)}.
\newblock \href{https://arxiv.org/abs/2107.10084}{arXiv:2107.10084}.
\bibAnnoteFile{Hofmann2021}

\bibitem{Rohringer2009May}
{N.~Rohringer and R.~Santra}.
\newblock Multichannel coherence in strong-field
  ionization\href{http://doi.org/10.1103/PhysRevA.79.053402}{.
\newblock \emph{Phys. Rev. A} \textbf{79} no.~5, p. 053402 (2009)}.
\newblock \href{http://dx.doi.org/10.3204/PHPPUBDB-22208}{DESY eprint}.
\bibAnnoteFile{Rohringer2009May}

\bibitem{veniard_phase_1990}
{V.~V\'eniard, R.~Ta\"ieb and A.~Maquet}.
\newblock Phase dependence of ({$N+1$})-color ({$N>1$}) ir-uv photoionization
  of atoms with higher harmonics\href{http://doi.org/10.1103/PhysRevA.54.721}{.
\newblock \emph{Phys. Rev. A} \textbf{54} no.~1, pp. 721--728 (1996)}.
\bibAnnoteFile{veniard_phase_1990}

\bibitem{Cooper2013Aug}
{B.~Cooper and V.~Averbukh}.
\newblock Single-photon laser-enabled {Auger} spectroscopy for measuring
  attosecond electron-hole
  dynamics\href{http://doi.org/10.1103/PhysRevLett.111.083004}{.
\newblock \emph{Phys. Rev. Lett.} \textbf{111} no.~8, p. 083004 (2013)}.
\newblock \href{http://hdl.handle.net/10044/1/10044/1/19480}{ICL eprint}.
\bibAnnoteFile{Cooper2013Aug}

\bibitem{Iablonskyi2017Aug}
{D.~Iablonskyi, K.~Ueda, K.~L. Ishikawa} et~al.
\newblock Observation and control of laser-enabled {Auger}
  decay\href{http://doi.org/10.1103/PhysRevLett.119.073203}{.
\newblock \emph{Phys. Rev. Lett.} \textbf{119} no.~7, p. 073203 (2017)}.
\newblock
  \href{https://researchbank.swinburne.edu.au/items/71e8a0dc-e79e-4e9e-a0d5-4f2d25d3531d/1/}{SUT
  eprint}.
\bibAnnoteFile{Iablonskyi2017Aug}

\bibitem{You2019}
{D.~You, K.~Ueda, M.~Ruberti} et~al.
\newblock A detailed investigation of single-photon laser enabled {Auger} decay
  in neon\href{http://doi.org/10.1088/1367-2630/ab520d}{.
\newblock \emph{New J. Phys.} \textbf{21}, p. 113036 (2019)}.
\bibAnnoteFile{You2019}

\bibitem{Khokhlova2019Jun}
{M.~A. Khokhlova, B.~Cooper, K.~Ueda} et~al.
\newblock Molecular {Auger}
  interferometry\href{http://doi.org/10.1103/PhysRevLett.122.233001}{.
\newblock \emph{Phys. Rev. Lett.} \textbf{122} no.~23, p. 233001 (2019)}.
\newblock \href{https://hdl.handle.net/10044/1/74108}{ICL eprint}.
\bibAnnoteFile{Khokhlova2019Jun}

\bibitem{Bauer2018}
{D.~Bauer and K.~K. Hansen}.
\newblock High-harmonic generation in solids with and without topological edge
  states\href{http://doi.org/10.1103/PhysRevLett.120.177401}{.
\newblock \emph{Phys. Rev. Lett.} \textbf{120} no.~17, p. 177401 (2018)}.
\newblock \href{https://arxiv.org/abs/1711.05783}{arXiv:1711.05783}.
\bibAnnoteFile{Bauer2018}

\bibitem{Silva2019Dec}
{R.~E.~F. Silva, A.~Jim\'{e}nez-Gal\'{a}n, B.~Amorim, O.~Smirnova and
  M.~Ivanov}.
\newblock Topological strong-field physics on sub-laser-cycle
  timescale\href{http://doi.org/10.1038/s41566-019-0516-1}{.
\newblock \emph{Nat. Photon.} \textbf{13}, pp. 849--854 (2019)}.
\bibAnnoteFile{Silva2019Dec}

\bibitem{Chacon2020}
{A.~Chac{\'{o}}n, D.~Kim, W.~Zhu} et~al.
\newblock Circular dichroism in higher-order harmonic generation: Heralding
  topological phases and transitions in {Chern}
  insulators\href{http://doi.org/10.1103/PhysRevB.102.134115}{.
\newblock \emph{Phys. Rev. B} \textbf{102} no.~13, p. 134115 (2020)}.
\newblock \href{https://arxiv.org/abs/1807.01616}{arXiv:1807.01616}.
\bibAnnoteFile{Chacon2020}

\bibitem{mcpherson_studies_1987}
{A.~McPherson, G.~Gibson, H.~Jara} et~al.
\newblock Studies of multiphoton production of vacuum-ultraviolet radiation in
  the rare gases\href{http://doi.org/10.1364/JOSAB.4.000595}{.
\newblock \emph{J. Opt. Soc. Am. B} \textbf{4} no.~4, pp. 595--601 (1987)}.
\bibAnnoteFile{mcpherson_studies_1987}

\bibitem{ferray_mutliple_1988}
{M.~Ferray, A.~L'Huillier, X.~F. Li} et~al.
\newblock Multiple-harmonic conversion of 1064 nm radiation in rare
  gases\href{http://doi.org/10.1088/0953-4075/21/3/001}{.
\newblock \emph{J. Phys. B: At. Mol. Opt. Phys.} \textbf{21} no.~3, p. L31
  (1988)}.
\bibAnnoteFile{ferray_mutliple_1988}

\bibitem{krause_hihg-order_1992}
{J.~L. Krause, K.~J. Schafer and K.~C. Kulander}.
\newblock High-order harmonic generation from atoms and ions in the high
  intensity regime\href{http://doi.org/10.1103/PhysRevLett.68.3535}{.
\newblock \emph{Phys. Rev. Lett.} \textbf{68} no.~24, pp. 3535--3538 (1992)}.
\newblock \href{https://zenodo.org/record/1233893}{Zenodo eprint}.
\bibAnnoteFile{krause_hihg-order_1992}

\bibitem{lhuillier_coherence_1992}
{A.~L'Huillier, P.~Balcou and L.~A. Lompr\'e}.
\newblock Coherence and resonance effects in high-order harmonic
  generation\href{http://doi.org/10.1103/PhysRevLett.68.166}{.
\newblock \emph{Phys. Rev. Lett.} \textbf{68} no.~2, pp. 166--169 (1992)}.
\bibAnnoteFile{lhuillier_coherence_1992}

\bibitem{balcou_phase-matching_1993}
{P.~Balcou and A.~L’Huillier}.
\newblock Phase-matching effects in strong-field harmonic
  generation\href{http://doi.org/10.1103/PhysRevA.47.1447}{.
\newblock \emph{Phys. Rev. A} \textbf{47} no.~2, pp. 1447--1459 (1993)}.
\bibAnnoteFile{balcou_phase-matching_1993}

\bibitem{toma_resonance-enhanced_1999}
{E.~S. Toma, P.~Antoine, A.~d. Bohan and H.~G. Muller}.
\newblock Resonance-enhanced high-harmonic
  generation\href{http://doi.org/10.1088/0953-4075/32/24/318}{.
\newblock \emph{J. Phys. B: At. Mol. Opt. Phys.} \textbf{32} no.~24, p. 5843
  (1999)}.
\bibAnnoteFile{toma_resonance-enhanced_1999}

\bibitem{figueira2002Jan}
{C.~Figueira~de Morisson~Faria, R.~Kopold, W.~Becker and J.~M. Rost}.
\newblock Resonant enhancements of high-order harmonic
  generation\href{http://doi.org/10.1103/PhysRevA.65.023404}{.
\newblock \emph{Phys. Rev. A} \textbf{65} no.~2, p. 023404 (2002)}.
\newblock \href{https://arxiv.org/abs/physics/0108010}{arXiv:physics/0108010}.
\bibAnnoteFile{figueira2002Jan}

\bibitem{Taieb2003Sep}
{R.~Ta\"{\i}eb, V.~V\'{e}niard, J.~Wassaf and A.~Maquet}.
\newblock Roles of resonances and recollisions in strong-field atomic
  phenomena. {II}. {High}-order harmonic
  generation\href{http://doi.org/10.1103/PhysRevA.68.033403}{.
\newblock \emph{Phys. Rev. A} \textbf{68} no.~3, p. 033403 (2003)}.
\bibAnnoteFile{Taieb2003Sep}

\bibitem{Ganeev2006Jun}
{R.~A. Ganeev, M.~Suzuki, M.~Baba, H.~Kuroda and T.~Ozaki}.
\newblock Strong resonance enhancement of a single harmonic generated in the
  extreme ultraviolet range\href{http://doi.org/10.1364/OL.31.001699}{.
\newblock \emph{Opt. Lett.} \textbf{31} no.~11, pp. 1699--1701 (2006)}.
\bibAnnoteFile{Ganeev2006Jun}

\bibitem{Gilbertson2008Sep}
{S.~Gilbertson, H.~Mashiko, C.~Li, E.~Moon and Z.~Chang}.
\newblock Effects of laser pulse duration on extreme ultraviolet spectra from
  double optical gating\href{http://doi.org/10.1063/1.2982589}{.
\newblock \emph{Appl. Phys. Lett.} \textbf{93} no.~11, p. 111105 (2008)}.
\bibAnnoteFile{Gilbertson2008Sep}

\bibitem{Shiner2011Jun}
{A.~D. Shiner, B.~E. Schmidt, C.~Trallero-Herrero} et~al.
\newblock Probing collective multi-electron dynamics in xenon with
  high-harmonic spectroscopy\href{http://doi.org/10.1038/nphys1940}{.
\newblock \emph{Nat. Phys.} \textbf{7}, pp. 464--467 (2011)}.
\bibAnnoteFile{Shiner2011Jun}

\bibitem{kutzner_extended_1989}
{M.~Kutzner, V.~Radojević and H.~P. Kelly}.
\newblock Extended photoionization calculations for
  xenon\href{http://doi.org/10.1103/PhysRevA.40.5052}{.
\newblock \emph{Phys. Rev. A} \textbf{40} no.~9, pp. 5052--5057 (1989)}.
\bibAnnoteFile{kutzner_extended_1989}

\bibitem{fahlman_xe_1984}
{A.~Fahlman, M.~O. Krause, T.~A. Carlson and A.~Svensson}.
\newblock Xe $5s$, $5p$ correlation satellites in the region of strong
  interchannel interactions, 28--75
  {eV}\href{http://doi.org/10.1103/PhysRevA.30.812}{.
\newblock \emph{Phys. Rev. A} \textbf{30} no.~2, pp. 812--819 (1984)}.
\bibAnnoteFile{fahlman_xe_1984}

\bibitem{becker_subshell_1989}
{U.~Becker, D.~Szostak, H.~G. Kerkhoff} et~al.
\newblock Subshell photoionization of {Xe} between 40 and 1000
  {eV}\href{http://doi.org/10.1103/PhysRevA.39.3902}{.
\newblock \emph{Phys. Rev. A} \textbf{39} no.~8, pp. 3902--3911 (1989)}.
\newblock \href{http://dx.doi.org/10.3204/PUBDB-2017-04033}{DESY eprint}.
\bibAnnoteFile{becker_subshell_1989}

\bibitem{Schmidt2012Mar}
{B.~E. Schmidt, A.~D. Shiner, M.~Gigu\`{e}re} et~al.
\newblock High harmonic generation with long-wavelength few-cycle laser
  pulses\href{http://doi.org/10.1088/0953-4075/45/7/074008}{.
\newblock \emph{J. Phys. B: At. Mol. Opt. Phys.} \textbf{45} no.~7, p. 074008
  (2012)}.
\newblock
  \href{https://nrc-publications.canada.ca/eng/view/object/?id=d64b383a-0685-47b4-91a1-79bd6b94d3e6}{NRC
  eprint}.
\bibAnnoteFile{Schmidt2012Mar}

\bibitem{huang_theoretical_1981}
{K.~N. Huang, W.~R. Johnson and K.~T. Cheng}.
\newblock Theoretical photoionization parameters for the noble gases argon,
  krypton, and xenon\href{http://doi.org/10.1016/0092-640X(81)90010-3}{.
\newblock \emph{Atom. Data Nucl. Data} \textbf{26} no.~1, pp. 33--45 (1981)}.
\bibAnnoteFile{huang_theoretical_1981}

\bibitem{Gaarde2001Jun}
{M.~B. Gaarde and K.~J. Schafer}.
\newblock Enhancement of many high-order harmonics via a single multiphoton
  resonance\href{http://doi.org/10.1103/PhysRevA.64.013820}{.
\newblock \emph{Phys. Rev. A} \textbf{64} no.~1, p. 013820 (2001)}.
\bibAnnoteFile{Gaarde2001Jun}

\bibitem{Ishikawa2003Jul}
{K.~Ishikawa}.
\newblock Photoemission and ionization of {$\mathrm{He}^{+}$} under
  simultaneous irradiation of fundamental laser and high-order harmonic
  pulses\href{http://doi.org/10.1103/PhysRevLett.91.043002}{.
\newblock \emph{Phys. Rev. Lett.} \textbf{91} no.~4, p. 043002 (2003)}.
\newblock \href{http://ishiken.free.fr/Publication/HeHHGLaser2003.pdf}{Author
  eprint}.
\bibAnnoteFile{Ishikawa2003Jul}

\bibitem{Milosevic2007Aug}
{D.~B. Milo\v{s}evi\'{c}}.
\newblock High-energy stimulated emission from plasma ablation pumped by
  resonant high-order harmonic
  generation\href{http://doi.org/10.1088/0953-4075/40/17/005}{.
\newblock \emph{J. Phys. B: At. Mol. Opt. Phys.} \textbf{40} no.~17, pp.
  3367--3376 (2007)}.
\bibAnnoteFile{Milosevic2007Aug}

\bibitem{Frolov2009Jun}
{M.~V. Frolov, N.~L. Manakov, T.~S. Sarantseva} et~al.
\newblock Analytic description of the high-energy plateau in harmonic
  generation by atoms: Can the harmonic power increase with increasing laser
  wavelengths?\href{http://doi.org/10.1103/PhysRevLett.102.243901}{.
\newblock \emph{Phys. Rev. Lett.} \textbf{102} no.~24, p. 243901 (2009)}.
\newblock \href{https://digitalcommons.unl.edu/physicsstarace/168/}{UNL
  eprint}.
\bibAnnoteFile{Frolov2009Jun}

\bibitem{Frolov2010Aug}
{M.~V. Frolov, N.~L. Manakov and A.~F. Starace}.
\newblock Potential barrier effects in high-order harmonic generation by
  transition-metal ions\href{http://doi.org/10.1103/PhysRevA.82.023424}{.
\newblock \emph{Phys. Rev. A} \textbf{82} no.~2, p. 023424 (2010)}.
\newblock \href{https://digitalcommons.unl.edu/physicsstarace/174/}{UNL
  eprint}.
\bibAnnoteFile{Frolov2010Aug}

\bibitem{Chan1991Jul}
{W.~F. Chan, G.~Cooper and C.~E. Brion}.
\newblock Absolute optical oscillator strengths for the electronic excitation
  of atoms at high resolution: Experimental methods and measurements for
  helium\href{http://doi.org/10.1103/PhysRevA.44.186}{.
\newblock \emph{Phys. Rev. A} \textbf{44} no.~1, pp. 186--204 (1991)}.
\bibAnnoteFile{Chan1991Jul}

\bibitem{Duffy2001Mar}
{G.~Duffy and P.~Dunne}.
\newblock The photoabsorption spectrum of an indium laser produced
  plasma\href{http://doi.org/10.1088/0953-4075/34/6/104}{.
\newblock \emph{J. Phys. B: At. Mol. Opt. Phys.} \textbf{34} no.~6, pp.
  L173--L178 (2001)}.
\bibAnnoteFile{Duffy2001Mar}

\bibitem{Strelkov2010Mar}
{V.~Strelkov}.
\newblock Role of autoionizing state in resonant high-order harmonic generation
  and attosecond pulse
  production\href{http://doi.org/10.1103/PhysRevLett.104.123901}{.
\newblock \emph{Phys. Rev. Lett.} \textbf{104} no.~12, p. 123901 (2010)}.
\bibAnnoteFile{Strelkov2010Mar}

\bibitem{tudorovskaya_high-order_2011}
{M.~Tudorovskaya and M.~Lein}.
\newblock High-order harmonic generation in the presence of a
  resonance\href{http://doi.org/10.1103/PhysRevA.84.013430}{.
\newblock \emph{Phys. Rev. A} \textbf{84} no.~1, p. 013430 (2011)}.
\newblock \href{https://doi.org/10.15488/1971}{LUH eprint}.
\bibAnnoteFile{tudorovskaya_high-order_2011}

\bibitem{ganeev_harmonic_2006}
{R.~A. Ganeev, H.~Singhal, P.~A. Naik} et~al.
\newblock Harmonic generation from indium-rich
  plasmas\href{http://doi.org/10.1103/PhysRevA.74.063824}{.
\newblock \emph{Phys. Rev. A} \textbf{74} no.~6, p. 063824 (2006)}.
\bibAnnoteFile{ganeev_harmonic_2006}

\bibitem{suzuki_anomalous_2006}
{M.~Suzuki, M.~Baba, R.~Ganeev, H.~Kuroda and T.~Ozaki}.
\newblock Anomalous enhancement of a single high-order harmonic by using a
  laser-ablation tin plume at 47 nm\href{http://doi.org/10.1364/OL.31.003306}{.
\newblock \emph{Opt. Lett.} \textbf{31} no.~22, pp. 3306--3308 (2006)}.
\bibAnnoteFile{suzuki_anomalous_2006}

\bibitem{suzuki_intense_2007}
{M.~Suzuki, M.~Baba, H.~Kuroda, R.~A. Ganeev and T.~Ozaki}.
\newblock Intense exact resonance enhancement of single-high-harmonic from an
  antimony ion by using {Ti}:{Sapphire} laser at 37
  nm\href{http://doi.org/10.1364/OE.15.001161}{.
\newblock \emph{Opt. Express} \textbf{15} no.~3, pp. 1161--1166 (2007)}.
\bibAnnoteFile{suzuki_intense_2007}

\bibitem{ganeev_strong_2007}
{R.~A. Ganeev, P.~A. Naik, H.~Singhal, J.~A. Chakera and P.~D. Gupta}.
\newblock Strong enhancement and extinction of single harmonic intensity in the
  mid- and end-plateau regions of the high harmonics generated in weakly
  excited laser plasmas\href{http://doi.org/10.1364/OL.32.000065}{.
\newblock \emph{Opt. Lett.} \textbf{32} no.~1, p.~65 (2007)}.
\bibAnnoteFile{ganeev_strong_2007}

\bibitem{ganeev_systematic_2007}
{R.~A. Ganeev, L.~B.~E. Bom, J.-C. Kieffer and T.~Ozaki}.
\newblock Systematic investigation of resonance-in\-duced single-harmonic
  enhancement in the extreme-ultraviolet
  range\href{http://doi.org/10.1103/PhysRevA.75.063806}{.
\newblock \emph{Phys. Rev. A} \textbf{75} no.~6, p. 063806 (2007)}.
\bibAnnoteFile{ganeev_systematic_2007}

\bibitem{Strelkov2014May}
{V.~V. Strelkov, M.~A. Khokhlova and N.~Y. Shubin}.
\newblock High-order harmonic generation and {Fano}
  resonances\href{http://doi.org/10.1103/PhysRevA.89.053833}{.
\newblock \emph{Phys. Rev. A} \textbf{89} no.~5, p. 053833 (2014)}.
\newblock \href{https://arxiv.org/abs/1307.5241}{arXiv:1307.5241}.
\bibAnnoteFile{Strelkov2014May}

\bibitem{Fano1961Dec}
{U.~Fano}.
\newblock Effects of configuration interaction on intensities and phase
  shifts\href{http://doi.org/10.1103/PhysRev.124.1866}{.
\newblock \emph{Phys. Rev.} \textbf{124} no.~6, pp. 1866--1878 (1961)}.
\bibAnnoteFile{Fano1961Dec}

\bibitem{Wahyutama2019Jun}
{I.~S. Wahyutama, T.~Sato and K.~L. Ishikawa}.
\newblock Time-dependent multiconfiguration self-consistent-field study on
  resonantly enhanced high-order harmonic generation from transition-metal
  elements\href{http://doi.org/10.1103/PhysRevA.99.063420}{.
\newblock \emph{Phys. Rev. A} \textbf{99} no.~6, p. 063420 (2019)}.
\bibAnnoteFile{Wahyutama2019Jun}

\bibitem{Flettner2003}
{A.~Flettner, T.~Pfeifer, D.~Walter} et~al.
\newblock High-harmonic generation and plasma radiation from water
  microdroplets\href{http://doi.org/10.1007/s00340-003-1329-x}{.
\newblock \emph{Appl. Phys. B} \textbf{77} no.~8, pp. 747--751 (2003)}.
\bibAnnoteFile{Flettner2003}

\bibitem{Luu2018}
{T.~T. Luu, Z.~Yin, A.~Jain} et~al.
\newblock Extreme--ultraviolet high--harmonic generation in
  liquids\href{http://doi.org/10.1038/s41467-018-06040-4}{.
\newblock \emph{Nat. Commun.} \textbf{9} no.~1, p. 3723 (2018)}.
\bibAnnoteFile{Luu2018}

\bibitem{Ghimire2011}
{S.~Ghimire, A.~D. DiChiara, E.~Sistrunk} et~al.
\newblock Observation of high-order harmonic generation in a bulk
  crystal\href{http://doi.org/10.1038/nphys1847}{.
\newblock \emph{Nat. Phys.} \textbf{7} no.~2, pp. 138--141 (2011)}.
\newblock
  \href{http://www.dimauro.osu.edu/publications/2011.Observation.pdf}{OSU
  eprint}.
\bibAnnoteFile{Ghimire2011}

\bibitem{Vampa2017}
{G.~Vampa and T.~Brabec}.
\newblock Merge of high harmonic generation from gases and solids and its
  implications for attosecond
  science\href{http://doi.org/10.1088/1361-6455/aa528d}{.
\newblock \emph{J. Phys. B: At. Mol. Opt. Phys.} \textbf{50} no.~8, p. 083001
  (2017)}.
\newblock \href{https://www.attoscience.ca/pdf/VampaBrabec_JPB_2017.pdf}{JASLab
  eprint}.
\bibAnnoteFile{Vampa2017}

\bibitem{Kuhn2012}
{T.~S. Kuhn}.
\newblock \emph{The Structure of Scientific Revolutions: 50th Anniversary
  Edition} (University of Chicago Press, Chicago, 2012).
\bibAnnoteFile{Kuhn2012}

\bibitem{Doerner2020}
{R.~D\"orner}.
\newblock Personal communication (2020).
\bibAnnoteFile{Doerner2020}

\bibitem{Mercadier2019}
{L.~Mercadier, A.~Benediktovitch, C.~Weninger} et~al.
\newblock Evidence of extreme ultraviolet superfluorescence in
  xenon\href{http://doi.org/10.1103/PhysRevLett.123.023201}{.
\newblock \emph{Phys. Rev. Lett.} \textbf{123} no.~2, p. 023201 (2019)}.
\newblock \href{https://arxiv.org/abs/1810.11097}{arXiv:1810.11097}.
\bibAnnoteFile{Mercadier2019}

\bibitem{Rohringer2020qbattles}
{N.~Rohringer}.
\newblock The role of decoherence and collisions in collective spontaneous
  emission of {FEL}-irradiated clusters.
\newblock \emph{\href{https://www.quantumbattles.com/}{Quantum Battles in
  Attoscience}} online conference (2020).
\newblock \href{https://youtu.be/MidvIsufacU}{youtu.be/MidvIsufacU}.
\bibAnnoteFile{Rohringer2020qbattles}

\bibitem{Bartlett2006}
{S.~D. Bartlett, T.~Rudolph and R.~W. Spekkens}.
\newblock Dialogue concerning two views on quantum coherence: factist and
  fictionist\href{http://doi.org/10.1142/S0219749906001591}{.
\newblock \emph{Int. J. Quantum Inform.} \textbf{04} no.~01, pp. 17--43
  (2006)}.
\newblock
  \href{https://arxiv.org/abs/quant-ph/0507214}{arXiv:quant-ph/0507214}.
\bibAnnoteFile{Bartlett2006}

\bibitem{Galilei1632}
{G.~Galilei}.
\newblock \emph{Dialogo sopra i due massimi sistemi del mondo} (1632).
\bibAnnoteFile{Galilei1632}

\end{thebibliography}

\end{document}